# Data mining, dashboards and statistics: a powerful framework for the chemical design of molecular nanomagnets

Yan Duan[†,1,2], Lorena E. Rosaleny[†,*,1], Joana T. Coutinho[†,*,1,3], Silvia Giménez-Santamarina[1], Allen Scheie[4], José J. Baldoví[1], Salvador Cardona-Serra[1], Alejandro Gaita-Ariño[*,1]

[1] Instituto de Ciencia Molecular (ICMol), Universitat de València, C/ Catedrático José Beltrán 2, 46980 Paterna, Spain

[2] Spin-X Institute, South China University of Technology, Guangzhou 510641, P. R. China

[3] Centre for Rapid and Sustainable Product Development, Polytechnic of Leiria, 2430-028 Marinha Grande, Portugal

[4] Neutron Scattering Division, Oak Ridge National Laboratory, Oak Ridge, Tennessee 37831, USA

*e-mail: Joana.t.coutinho@ipleiria.pt, rosaleny@uv.es, gaita@uv.es

## Abstract

Three decades of research in molecular nanomagnets have raised their magnetic memories from liquid helium to liquid nitrogen temperature thanks to a wise choice of the magnetic ion and coordination environment. Still, serendipity and chemical intuition played a main role. In order to establish a powerful framework for statistically driven chemical design, we collected chemical and physical data for lanthanide-based nanomagnets, catalogued over 1400 published experiments, developed an interactive dashboard (SIMDAVIS) to visualise the dataset, and applied inferential statistical analysis. Our analysis showed that the Arrhenius energy barrier correlates unexpectedly well with the magnetic memory, as both Orbach and Raman processes can be controlled by vibronic coupling. Indeed, only bis-phthalocyaninato sandwiches and metallocenes, with rigid ligands, consistently present magnetic memory up to high temperature. Analysing magnetostructural correlations, we offer promising strategies for improvement, in particular for the preparation of pentagonal bipyramids, where even "softer" complexes are protected against molecular vibrations.

## A brief history of SIMs

Molecular nanomagnets were reported for the first time at the beginning of the 1990s, when $Mn_{12}O_{12}(CH_3COO)_{16}(H_2O)_4$ was discovered to display magnetic hysteresis in analogy to classical magnets, but with a quantum tunnelling mechanism for the relaxation of the magnetisation.[1,2] This polynuclear magnetic complex was the first of a plethora of single-molecule magnets (SMMs). The term was coined for systems behaving as hard bulk magnets below a certain temperature, but where the slow relaxation of the magnetisation is of purely unimolecular origin. Their magnetic behaviour can be approximated to that of an effective anisotropic magnetic moment arising from the exchange interactions between the spins of the metal ions. The reversal of this giant anisotropic spin occurs by populating excited spin states and overcoming an energy barrier. Hence, the thermal dependence of the relaxation rate was described by the Arrhenius equation (Fig. 1b), using this effective energy barrier ($U_{eff}$) and a pre-exponential factor ($\tau_0$).[3] Both parameters were not extracted directly from the hysteresis loop (see Fig. 1c), but rather from the combined frequency- and temperature-dependence of the so-called out-of-phase component of the ac susceptibility ($\chi''$, see Fig. 1a).[3,4] The consideration of other processes in the fit, such as the Raman process and Quantum Tunneling of the Magnetisation (QTM), in principle results in more accurate values, which are denoted as $U_{eff,fit}$, $\tau_{0,fit}$. It also allows extracting additional parameters $C$, $n$ to characterise Raman, and $\tau_{QTM}$ (Fig. 1b).

The best metric for slow relaxation is hysteresis temperature ($T_{hyst}$), the highest temperature at which the system presents magnetic bistability. The first SMMs exhibited low values of $T_{hyst}$, which was attributed to their modest effective energy barrier ($U_{eff} \approx 50$ K).[5] Initial models based on effective spin Hamiltonians gave rise to the relation $U_{eff} = DS_z^2$ and concluded that the best strategy to raise $U_{eff}$ and, therefore, to improve the maximum $T_{hyst}$ is to maximise the total effective spin ($S$), rather than the magnetic anisotropy ($D$). Indeed, the latter is a less straightforward target for the synthetic chemist.[6] Despite great effort toward the synthesis of such systems and an abundance of molecules with increasing values of $S$, very little progress was made in the first decade in terms of increasing $U_{eff}$ or $T_{hyst}$.[7]

In the 2000s, a novel class of molecular nanomagnets emerged, namely bis-phthalocyaninato (Pc) "double deckers".[8] This second generation of SMMs, commonly known as Single Ion Magnets (SIMs), is based on mononuclear complexes containing a single magnetic ion embedded in a coordination environment. They constitute the smallest molecule-based magnet and their properties arise from a strong spin-orbit coupling which, combined with the crystal-field interaction with the surrounding ligands, results in an enhanced magnetic anisotropy when compared to SMMs. Identical data treatment using the Arrhenius equation led to effective energy barriers $U_{eff}$ up to an order of magnitude higher for SIMs based on rare-earth ions when compared to those of polynuclear metal complexes of the $d$-block. The characteristic maxima in the out-of-phase component of the ac susceptibility $\chi''$ also moved to higher temperatures (Fig. 1a), but $T_{hyst}$ did not increase as significantly.

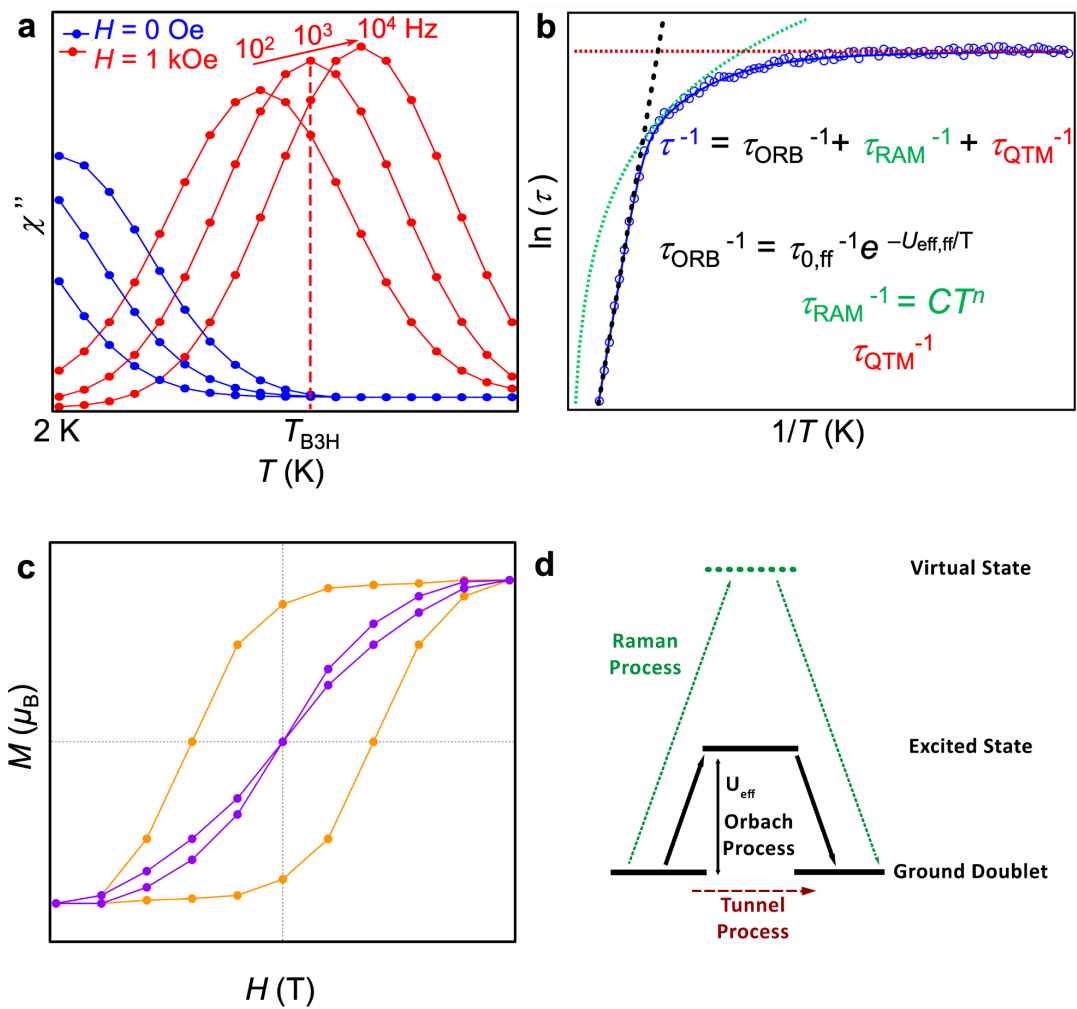

**Figure 1 | Main magnetic concepts employed in this study.** Slow relaxation of the magnetisation in SIMs can manifest in different ways. **a**, Spin blocking is often characterised by a temperature- and frequency-dependent out-of-phase ac susceptibility $\chi''$. **b**, These relaxation dynamics have most often been modelled as an Orbach process (black dots), using the Arrhenius equation. Raman (green dots) and quantum tunnelling (red dots) processes can also be relevant. **c**, Magnetic hysteresis can be full (orange) or "pinched", also known as "butterfly" (purple), signalling a fast relaxation at zero magnetic field. **d**, Scheme of different relaxation processes: tunnelling involves just the states within the ground doublet, Orbach process takes place via an excited state, and Raman process happens via a virtual state.

After the germinal LnPc$_2$, different chemical families such as polyoxometalates[9] and metallocenes[10] were shown to exhibit slow relaxation of the magnetisation of purely molecular origin (Fig. 2). Initially, oblate ions Tb and Dy, which present an equatorially expanded $f$-electron charge cloud, were the most commonly studied. Success cases were also found for lanthanide ions with axially elongated $f$-electron charge cloud (prolate ions, e.g. Er, Tm, Yb). The realisation that lanthanide SIMs were not restricted to a single chemical strategy inspired a large community of chemists. As a result, between 2003 and 2017 SIM behaviour was reported in over 600 compounds, and above a third of these compounds actually displayed magnetic hysteresis. No single chemical strategy has dominated in terms of reported examples, although many approaches have been paradigmatic (*e.g.* the use of radicals[11,12] and diketonates[13]). Recent efforts have been made to offer some perspective.[14–17] Nevertheless, anecdotal claims from proven strategies are hard to distinguish, as so many studies pursuing independent approaches have been reported. Modern techniques of data analysis and visualisation can contribute to remedy this knowledge gap. In particular, dashboards are intuitive graphical applications for dynamic data visualisation and information management, of growing popularity in different fields.[18–20]

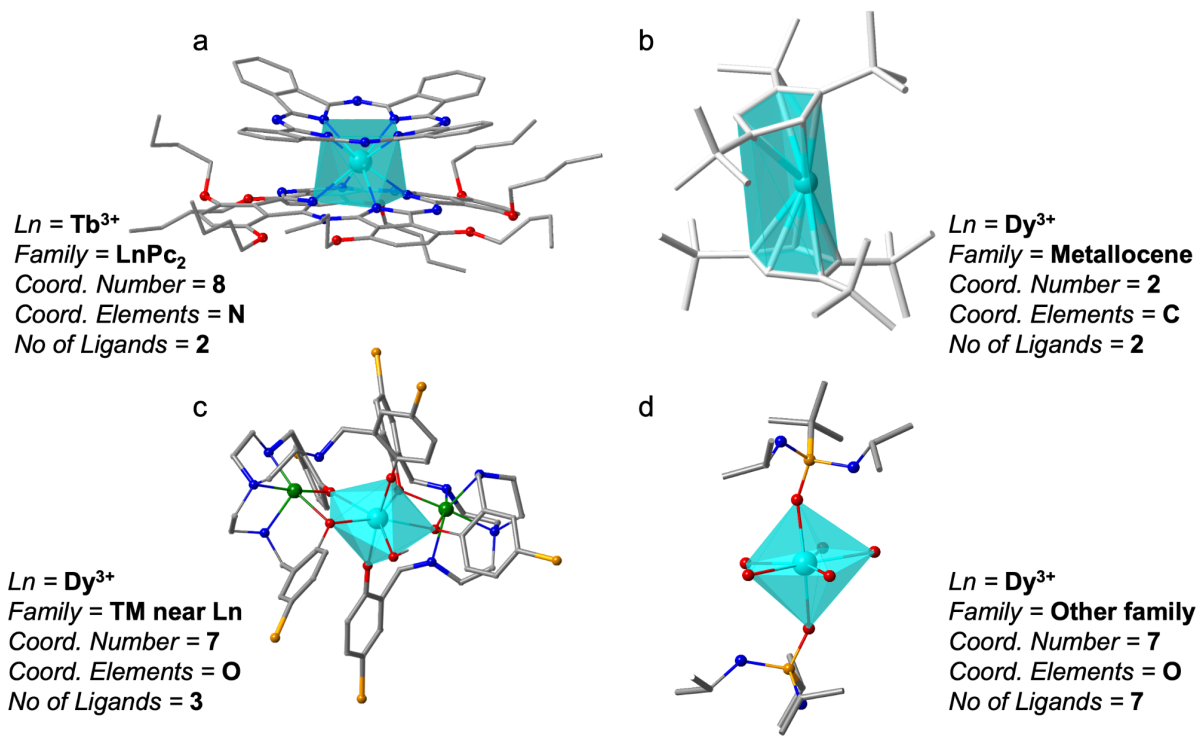

a

*Ln* = **Tb³⁺**
*Family* = **LnPc₂**
*Coord. Number* = **8**
*Coord. Elements* = **N**
*No of Ligands* = **2**

b

*Ln* = **Dy³⁺**
*Family* = **Metallocene**
*Coord. Number* = **2**
*Coord. Elements* = **C**
*No of Ligands* = **2**

c

*Ln* = **Dy³⁺**
*Family* = **TM near Ln**
*Coord. Number* = **7**
*Coord. Elements* = **O**
*No of Ligands* = **3**

d

*Ln* = **Dy³⁺**
*Family* = **Other family**
*Coord. Number* = **7**
*Coord. Elements* = **O**
*No of Ligands* = **7**

**Figure 2 | Molecular structures of some representative lanthanide-based SIMs from different chemical strategies and some of their chemical descriptors. a**, Pc "double deckers", abbreviated as LnPc₂ ($T_{hyst}$ = 2 K).[21] **b**, Metallocene complex LnCp*₂ ($T_{hyst}$ = 60 K).[22] **c**, a complex with the introduction of a diamagnetic TM ion near the lanthanide ion, [Zn₂DyL₂(MeOH)]⁻ (L is a tripodal ligand, 2,2′,2′′-(((nitrilotris(ethane-2,1-diyl))tris(azanediyl))tris(methylene))tris-(4-bromophenol)) ($T_{hyst}$ = 11 K).[23] **d**, [L₂Dy(H₂O)₅]³⁺ (L = ᵗBuPO(NHⁱPr)₂), a complex outside the main categories of the present study, which was classified as "other families'' ($T_{hyst}$ = 30 K).[24] (Color scheme for atoms: green, Zn; cyan, Tb or Dy; gray, C; blue, N; yellow, P; orange, Br; red, O. Hydrogen atoms are not shown for clarity).

The present work firstly aims to rationalise the correlations among the different physical variables involved in SIMs. A common hypothesis is that the parameters arising from the ac magnetometry (*e.g.* $U_{eff}$) are well correlated with the experimental values (*e.g.* $T_{hyst}$). This, however, has not been proven and has actually been challenged in various ways.[17,25,26] Over the years, various theoretical approaches have put the focus on the rationalisation of different physical processes and parameters.[25–28] Secondly, in order to provide the synthetic efforts with a data-driven chemical design guide, we applied the techniques of third generation computational chemistry.[29] We started by collecting a high-quality dataset from published data and represented it in an interactive dashboard. Then, we statistically analysed the correlations between molecular descriptors and physical parameters. As a second phase of the work, we expanded this dataset to rationalise the correlations found in the first phase, and analyse the influence of the shapes of coordination environments on the magnetic dynamics.

# Results

## A dataset and interactive dashboard for lanthanide SIMs

We built a dataset of the most relevant chemical and physical properties of 1411 lanthanide SIM samples collected from 448 scientific articles (Supplementary file) published between 2003 and 2019 and developed a user-friendly dashboard-style web application named SIMDAVIS (Single Ion Magnet DAta VISualisation) to host it. The dataset contains over 10000 independent pieces of chemical information, as well as over 5000 independent pieces of physical (magnetic) information. Furthermore, the dataset is hierarchically clustered into magnetostructural "taxonomies" (see Supplementary Sections 4 and 6) in order to pave the way for further analysis, including Machine Learning studies, a field that is now on the rise.[30,31] Indeed, data taxonomies are powerful tools to make sense of data, since they provide ordered representations of the formal structure of knowledge classes or types of objects within a data domain.[32]

Each chemical family that has been widely explored in this field is claimed to be promising as molecular nanomagnet, usually by citing the best reported case. However, it is crucial to avoid getting distracted by the occasional well-behaving example and instead to evaluate the general behaviour of each chemical strategy. Do members of a family generally present a slow relaxation of the magnetization, in terms of ac magnetometry and/or hysteresis? To evaluate this against a common reference, our dataset allows comparing the overall performance of samples in each family with the performance of the "Mixed Ligands" and "Others" Families, that act here as a sort of control group. Similarly, since $Tb^{3+}$ and $Dy^{3+}$ ions are oblate, as well as the cases where the record results have been obtained, it is also commonly assumed that in general complexes with oblate ions result in better SIM properties compared with prolates. Our dataset should be able to test this.

The question remains of what does one mean by "better SIM properties", or, as eloquently put recently, "How do you measure a magnet?"[33] Blocking temperature definitions in recent works include the temperature at which the relaxation time is 100 s ($T_{B2}$), the temperature at which there is a maximum in the zero-field cooled susceptibility ($T_{B1}$), and the maximum temperature at which hysteresis is observed ($T_{hyst}$). Unfortunately, only a small part of the articles provided any of these parameters, whereas older bibliography favoured only $T_{hyst}$. This potentially introduces a severe publication bias that our dataset tackles by including information about ac magnetometry, $U_{eff}$ and hysteresis (see Supplementary Section 1.1).

Qualitative and quantitative information based on the almost ubiquitous *ac* susceptibility measurements was invaluable for our analysis. Since there is an *ac* curve at (or near) 1000 Hz frequency in virtually all works in the field (given that both MPMS and PPMS magnetometers cover this range), we chose the maximum of the out-of-phase *ac* curve at this frequency as the basis for our most abundant qualitative and quantitative data source. At a given frequency, $\chi''$ does not necessarily present a maximum; an external magnetic field facilitates this effect by cancelling QTM. In this study, we register the temperature of this maximum as $T_{B3}$ ($T_{B3H}$), the blocking temperature at $10^3$ Hz in absence (in presence) of a magnetic field (Fig. 1a).

$U_{eff}$ is also very widely employed and assumed to be a good descriptor of magnetic behaviour. In contrast, in Arrhenius-type fits generally little attention is paid to $\tau_0$; according to a simplified two-phonon Orbach model, the two variables are supposed to be correlated. At the same time, $U_{eff}$ is rightfully criticised as an oversimplification that overlooks physically independent mechanisms (notably, Raman) that could be dominating the behaviour. Our dataset aimed to answer these questions.

Finally, note that a remnant magnetization at $H$=0 is a defining feature for molecular nanomagnets. If remanence is lacking, it is not feasible to store long-term information on a molecule. One can ask: is a molecule that shows out of phase magnetic susceptibility as a response to alternating current but no hysteresis, really a molecular nanomagnet? There is, however, a link between short- and long-term magnetic memory, so we included the wider definition of SIMs in the present study, as detailed above.

SIMDAVIS allows the chemical community to visualise the key relationships between chemical structures and physical properties in our catalogue of SIMs. Our interactive dashboard can be directly invoked by accessing the internet site where it is located.[34] It is organised in 6 main tabs: Home, ScatterPlots, BoxPlots, BarCharts, Data (View Data and Download Data) and About SIMDAVIS (Variables, Authors, Feedback&Bugs, Changelog and License) as we can observe in Supplementary Fig. 11.

In the SIMDAVIS dashboard, the most versatile source of graphical information is the "ScatterPlots" tab, where an example plot is explained in Supplementary Fig. 11. The next two tabs display the data in complementary ways. The "BoxPlots" tab allows to examine the distribution of each SIMs quantitative property *vs* a categorization criterion, *e.g.* we can see the distribution of $U_{eff}$ values as a function of the coordination elements. The boxplot for each category is shown, including the median and the interquartile range. The "BarCharts" tab allows the exploration of the frequency of different qualitative variables in our dataset. Stacked bar graphs allow the simultaneous analysis of two qualitative variables, *e.g.* we can display, for each chemical family, the number of samples which present magnetic hysteresis. The "Data" tab is a powerful interface to browse the dataset, featuring the possibility to choose the data columns to show, ordering in ascending or descending order, and filtering by arbitrary keywords; it also permits downloading all data, including links to the CIF files, when available. Finally, the "About SIMDAVIS" tab contains information about the variables contained in the dataset.

## Statistically driven chemical design of SIMs

SIMDAVIS allows the visualisation of the relationships between chemical and physical variables in SIMs, and thereby enables determining the main variables that the synthetic chemist needs to consider to obtain the desired physical properties. We will first analyse this qualitatively employing a series of boxplots, violin plots and bar charts (see Fig. 3 and Supplementary Figs. 11.1-11.6, 12.1-12.5, 13.1-13.2). The full statistical analysis is presented in Supplementary Sections 4, 5 and 6.

First, let us focus on the effective energy barrier $U_{eff}$ and the blocking temperature $T_{B3}$ (the temperature for maximum out-of-phase ac susceptibility $\chi''$ at $10^3$ Hz, see Fig. 1). From Fig. 3 and Supplementary Figs. 11.1-11.4, we can see that the chemical families with a distinctly good behaviour are LnPc$_2$ and metallocenes, with median values of $U_{eff} > 200$ K and $T_{B3} > 30$ K. Equivalently, one can see that $Dy^{3+}$ and $Tb^{3+}$ present somewhat higher $U_{eff}$, $T_{B3}$ than the rest (Supplementary Figs. 11.3c, 11.4a) and that, in general, oblate ions perform better than prolate ions, for both properties. In addition, non-Kramers ions present higher median $T_{B3}$ but similar $U_{eff}$ values compared with Kramers ions.

Now, let us analyse the maximum hysteresis temperature $T_{hyst}$. The only chemical family with a distinct positive behaviour is the metallocene family. More surprisingly, $Er^{3+}$ complexes have distinctly high hysteresis temperatures, markedly with a higher median than $Dy^{3+}$ or $Tb^{3+}$ complexes. This is in sharp contrast with their relative $T_{B3}$ values which are consistently much lower in the case of $Er^{3+}$ complexes. This not only indicates that searching for equatorial environments, precisely the ones that favour good magnetic properties in $Er^{3+}$ complexes,[27] often results in more rigid ligands, but also indicates an underexplored territory. It is certainly possible that chemical modifications of $[Er(COT)_2]^-$ (or other $Er^{3+}$ record-bearing complexes) designed to optimise the detrimental effect of molecular vibrations may achieve records that are competitive with $DyCp_2$. Prolate ions are consistently -and surprisingly- better than the oblate ones, having a higher median value for $T_{hyst}$. This is again in contrast with the opposite behaviour which is observed for $T_{B3}$ and $U_{eff}$, and possibly again due to the influence of $Er^{3+}$ complexes with their more rigid equatorial environments. This behaviour of $Er^{3+}$ is unexpected after a recent theoretical contribution,[35] which calculated the electronic structure of $Er[N(SiMe_3)_2]_3$ variants, concluding that no geometrical optimization can significantly improve $U_{eff}$ for $Er^{3+}$. Nevertheless, all the high $U_{eff}$ cases involving $Er^{3+}$ in our database are based on the COT ligand, meaning our differing conclusions stem from different chemical strategies.

Finally, the coordination number and the number of ligands do have an influence on the statistically expected hysteresis temperature, with the best ones being 2 and 7 in the case of the coordination number and just 7 for the number of ligands. As we will discuss below, there are chemical insights to be gained from this if one analyses the influence of the coordination environment shape.

To put all these trends into perspective, it is important to numerically analyse the connection between the different variables and the clustering of our data. A lognormal analysis (see Supplementary Section 4.3) shows that the three main chemical variables, namely the chemical family, the lanthanide ion and the coordination elements, are sufficient to reasonably explain the variation of values of the others. This means there is a limit on the information one can independently extract from the rest of the chemical variables. Multiple correspondence analysis (see Supplementary Sections 4.1, 4.2) suggests a chemical clustering that consists in three small groups, namely $Gd^{3+}$ complexes, metallocenes and $LnPc_2$ double deckers, and two much larger groups with a large overlap with oblate and prolate ions respectively. A factorial analysis of mixed data considering also all magnetic information available (see Supplementary Section 6) simplifies the clustering to three groups. Again, the two distinct families present a large overlap with metallocenes and $LnPc_2$ double decker chemical families, both of them presenting significantly better properties than the other kinds of samples. Finer clustering categorisations are possible and indeed available in the dataset. These "taxonomies" can serve to guide future theoretical work. In the current stage, they mainly serve to confirm that, in layman's terms, all chemical families within our current dataset present basically indistinguishable magnetic behaviours, except metallocenes and double deckers.

Further insight is provided representing the reported behaviour of magnetic hysteresis and *ac* magnetic susceptibility as a function of different chemical variables, namely (i) chemical family, (ii) metal ion, (iii) coordination number and (iv) coordination elements (Fig. 3, Supplementary Figs. 12.1 to 12.5). Remember that for hysteresis we are limited by the minority of the samples where hysteresis or its absence is reported; in the vast majority of the cases this information is lacking. Nevertheless, it is apparent that certain chemical families such as $LnPc_2$ and metallocenes tend to display hysteresis, with diketonates being in a distant third position. In contrast, complexes based on POMs or on Schiff bases seldomly report hysteresis, and actually tend to not even present out of phase *ac* signals (Supplementary Fig. 12.2). In terms of metal ions, $Dy^{3+}$, $Tb^{3+}$ and $Er^{3+}$ are clearly the best behaved (Fig. 3 and Supplementary Fig. 12.3).

The suggestion of future directions to guide synthetic efforts requires a more detailed study of the physical mechanisms of relaxation and the predictive power of the different parameters, as well as taking into account the shape of the coordination environment. This information is not immediately available from the literature. We performed these tasks and here we present the results in the following sections.

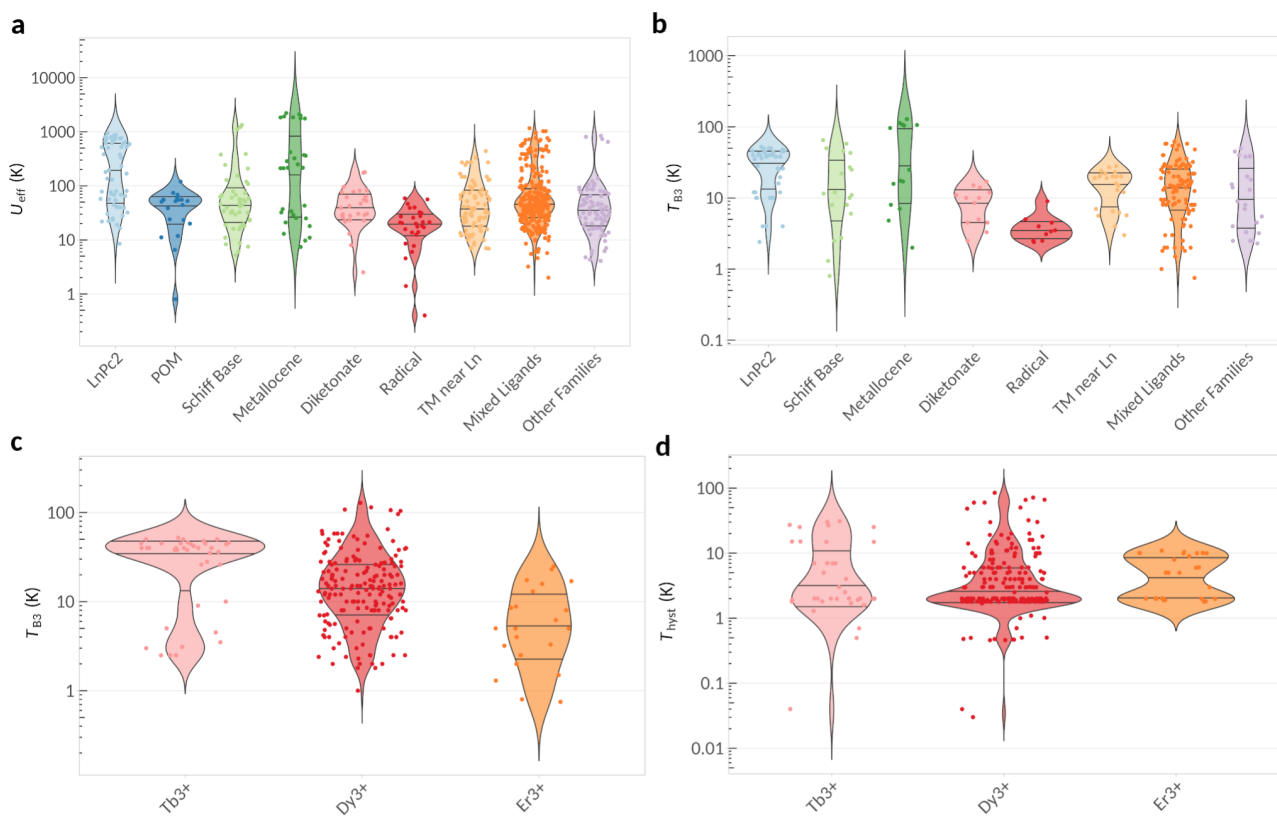

**Figure 3 | Violin plots and bar charts relating magnetic relaxation behaviour with the main chemical parameters.** The width of each violin plot is proportional to the density of data points for this range of values, and the horizontal stripes mark the quartiles. **a**, Values of $U_{eff}$ for samples in each chemical family. **b**, Values of $T_{B3}$ for samples in each chemical family. **c**, Values of $T_{B3}$ for samples containing $Tb^{3+}$, $Dy^{3+}$, $Er^{3+}$. **d**, Values of $T_{hyst}$ for samples containing $Tb^{3+}$, $Dy^{3+}$, $Er^{3+}$.

## Orbach mechanism: oversimplified, predictive… a function of vibronic coupling?

A key question is: how much have the analyses in this field been affected by the simplified assumption that SIMs relax via an Orbach mechanism, characterised by $\tau_0$ and $U_{\mathrm{eff}}$? We strived to quantify up to what level the value of $U_{\mathrm{eff}}$ is well correlated with the slow relaxation of the magnetisation, or to determine whether one would need to employ $U_{\mathrm{eff,ff}}$ instead. A visual inspection (Fig. 4a), a categorical analysis (Figs. 4b, 4c), an in-depth statistical analysis of all physical parameters based on the Akaike Information Criterion (Supplementary Section 5.3) and a factorial analysis of mixed data (Supplementary Section 6) conclude that $U_{\mathrm{eff}}$ derived from a simple Arrhenius plot is, by itself, an excellent predictor for magnetic behaviour. Supplementary Section 5.4 presents the full discussion of this question. As an alternate approach to evaluate the validity of the Orbach mechanism, it has been pointed out that frequently, as $U_{\mathrm{eff}}$ increases, $\tau_0$ decreases, leaving relaxation times essentially constant.[5] An approximate relation between $\tau_0$ and $U_{\mathrm{eff}}$ can be derived for two-phonon Orbach process within a Debye phonon model. Fitting experimental data to it results in a large dispersion but only moderate deviations from the expected parameter range, meaning these approximations can be useful (Supplementary Section 5.5 for details).[34]

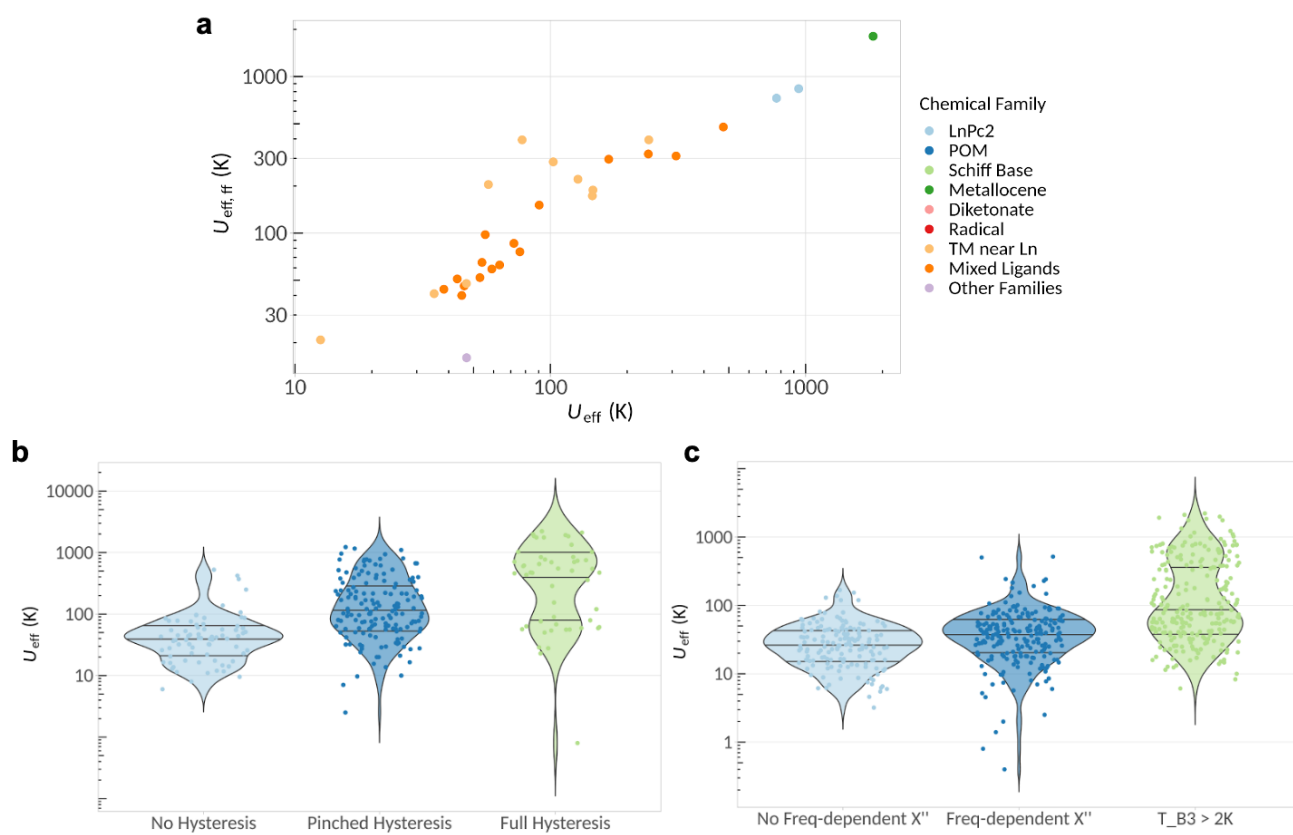

**Figure 4 | Main dependencies between the physical variables. a**, Dependence between $U_{\mathrm{eff}}$ and $U_{\mathrm{eff,ff}}$. **b**, Distribution of $U_{\mathrm{eff}}$ for samples depending on their qualitative hysteresis behaviour, **c**, Distribution of $U_{\mathrm{eff}}$ for samples depending on their qualitative ac $\chi''$ susceptibility behaviour. For a more complete analysis, see Supplementary Sections 5.1 and 6.

Having shown that $U_{eff}$ is very successful -perhaps unreasonably so- at predicting the magnetic behaviour, we turn our attention to the other relaxation pathways, notably Raman and quantum tunnelling of the magnetization processes. The former is characterised by a prefactor $C$ and an exponent $n$, whereas the latter by a time $\tau_{QTM}$. Since fits including this information are relatively scarce, one needs to note that this phase of the analysis has much less statistical power. We extracted $C$, $n$, $\tau_{QTM}$ from all samples which presented $U_{eff,ff}$, and explored possible correlations among the different parameters. We found that $U_{eff}$ seems to correlate quite well with $C$ (see Supplementary Figure 34, top), and this correlation is perhaps more clear with $U_{eff,ff}$ (see Figure 5). In particular, cases with high $U_{eff,ff}$ (>200K) present a low $C$ (<10$^{-3}$ s$^{-1}$) and vice versa. While the number of points is limited, this correlation persists remarkably for different ways of categorising the points (e.g. for different coordination elements, see Figure 5). But these are supposed to be two fully independent mechanisms, so what could be the reason behind this apparent coincidence?

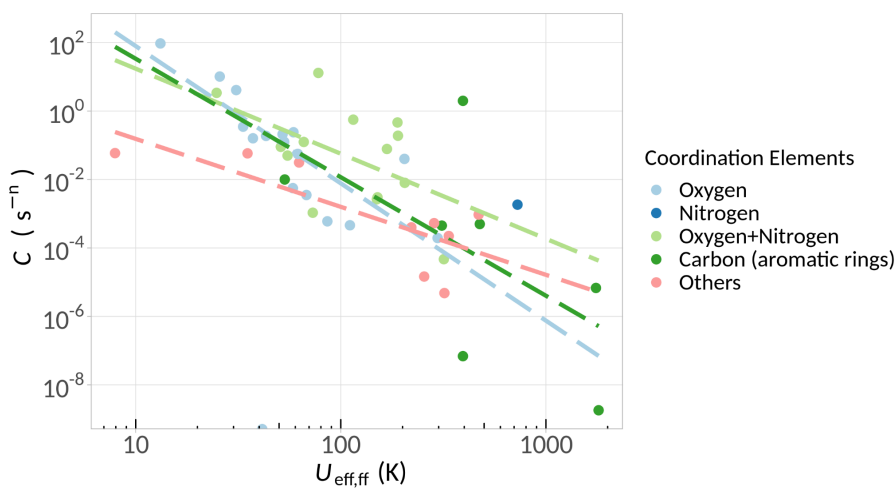

**Figure 5 | Relation between $U_{eff,ff}$ and Raman relaxation parameter $C$.** The dependence between $C$ and $U_{eff,ff}$ is robust enough to be preserved also when considering samples grouped by different categories, here for illustration we group the complexes by their coordination elements..

First we need to be aware of the fact that even a fit considering different relaxation mechanisms is a simplification. Indeed it has been shown that the anomalous Raman exponents so often found in these fits come from this fact.[36] These fits assume that one is studying an experiment with 3 physical processes that can be independently parameterised. Instead, many more processes are simultaneously taking place, including alternate multi-phonon Orbach mechanisms, competing Raman mechanisms dominated by different vibrational frequencies and the direct process. This means it is not surprising that, when fitting an overly complex process with a few parameters, some of them are "unphysically" correlated.

However, there is also a possible physical reason behind the correlation we found. It is the coupling between spin states and vibrations: a common factor for both relaxation pathways. Spin-vibration coupling plays a vital role in both Raman and Orbach mechanisms. This means that a strong crystal field is not a sufficient condition for a high $U_{eff}$. One also needs a low vibronic coupling so that the effective barrier is closer to the total crystal field, rather than the first excited state. This hypothesis is consistent with the interpretation of their own results in the record-setting dysprosoceniums with hysteresis up to 60 K[22] and 80 K[37]. Supporting this interpretation is also the fact that this correlation with $U_{\text{eff,ff}}$ is apparently absent in the case of $\tau_{\text{QTM}}$ (see Supplementary Figure 35). Furthermore, this vibrationally-controlled $U_{\text{eff}}$ would contribute to explain the typically weak correspondence between predicted (or even experimentally determined) energy levels and the $U_{\text{eff}}$ extracted from the spin relaxation experiments, a problem which is often minimised and sometimes justified by a role of QTM. Crucially, according to this idea, $LnPc_2$ and metallocenes would behave as exceptionally good nanomagnets not just because they provide exceptionally strong crystal field, but because they additionally provide exceptionally weak spin-vibrational coupling due to their rigidity, blocking Orbach and Raman processes simultaneously.

We have now obtained a likely rationalisation of why the controversial, oversimplified $U_{\text{eff}}$ is such a good predictor for the magnetic behaviour, and why a parameter that, resulting from a simplified fit, effectively summarises other relaxation mechanisms and correlates so unexpectedly well with the true $U_{\text{eff,ff}}$. The thermal dependence of the spin relaxation depends on Orbach+Raman, but $U_{\text{eff}}$ is correlated with $C$ and, as can be rationalised from the current understanding of spin relaxation,[36] $U_{\text{eff}}$, $C$ are heavily controlled by the spin-vibrational relaxation. Whether the spin levels are real or virtual, to exchange energy with the thermal bath the spin needs to couple to vibrations. Thus a high $U_{\text{eff}}$ can be understood as acting as a witness for a weak spin-vibrational coupling.

## Influence of the coordination environment shape

Since we have established that $U_{\text{eff}}$ is a good predictor for magnetic behaviour and also rationalised how and why it correlates with Raman relaxation, let us now turn our attention to rational chemical design strategies. A key question is: are there coordination polyhedra that are intrinsically well suited to produce high effective barriers? Plotting all $U_{\text{eff}}$ values vs the closest polyhedron for each complex, this seems to be the case. In contrast to what a cursory review of claims in literature would suggest, preparing lanthanide complexes that present a coordination environment close to $D_{4d}$ is not the best path. A more detailed analysis can be read in the Supporting Information Sections 7-8, including a critical assessment of data scarcity. Let us focus here on a salient case constituted by pentagonal bipyramids (with CN=7), which present a striking distribution of $U_{\text{eff}}$, with consistently high values compared with any of the other common polyhedra (see Fig. 6a) as well as a high success rate both in terms of presenting a peak in $\chi''$ at $10^3$ Hz and magnetic hysteresis (see Supplementary Fig. 32.2). Indeed, pentagonal bipyramids, much like square antiprisms, present no extradiagonal crystal field terms therefore minimising spin mixing. Additionally, all of their diagonal terms are in first approximation protected from low-energy vibrations, minimising vibronic coupling (for a longer discussion of this see Supplementary Section 7). Their barriers can be maximised by vertical compression (see Supplementary Fig. 33).

The natural follow-up question would be how to chemically favour this kind of coordination rather than, for example, capped trigonal prisms, which also present CN=7. From the dataset, it is obvious that employing non-chelating ligands massively favours the formation of pentagonal bipyramids (see Fig. 6b). The greatest synthetic competitor would seem to be octahedra, that also forms most often from ligands coordinating via a single atom, whereas most other shapes with CN $\geq$ 8 tend to result from chelating ligands. Similarly, a combination of oxygens and nitrogens is to be avoided, since for CN=7 this tends to result in capped trigonal prisms; to obtain pentagonal bipyramids, an all-oxygen coordination sphere is often employed instead.

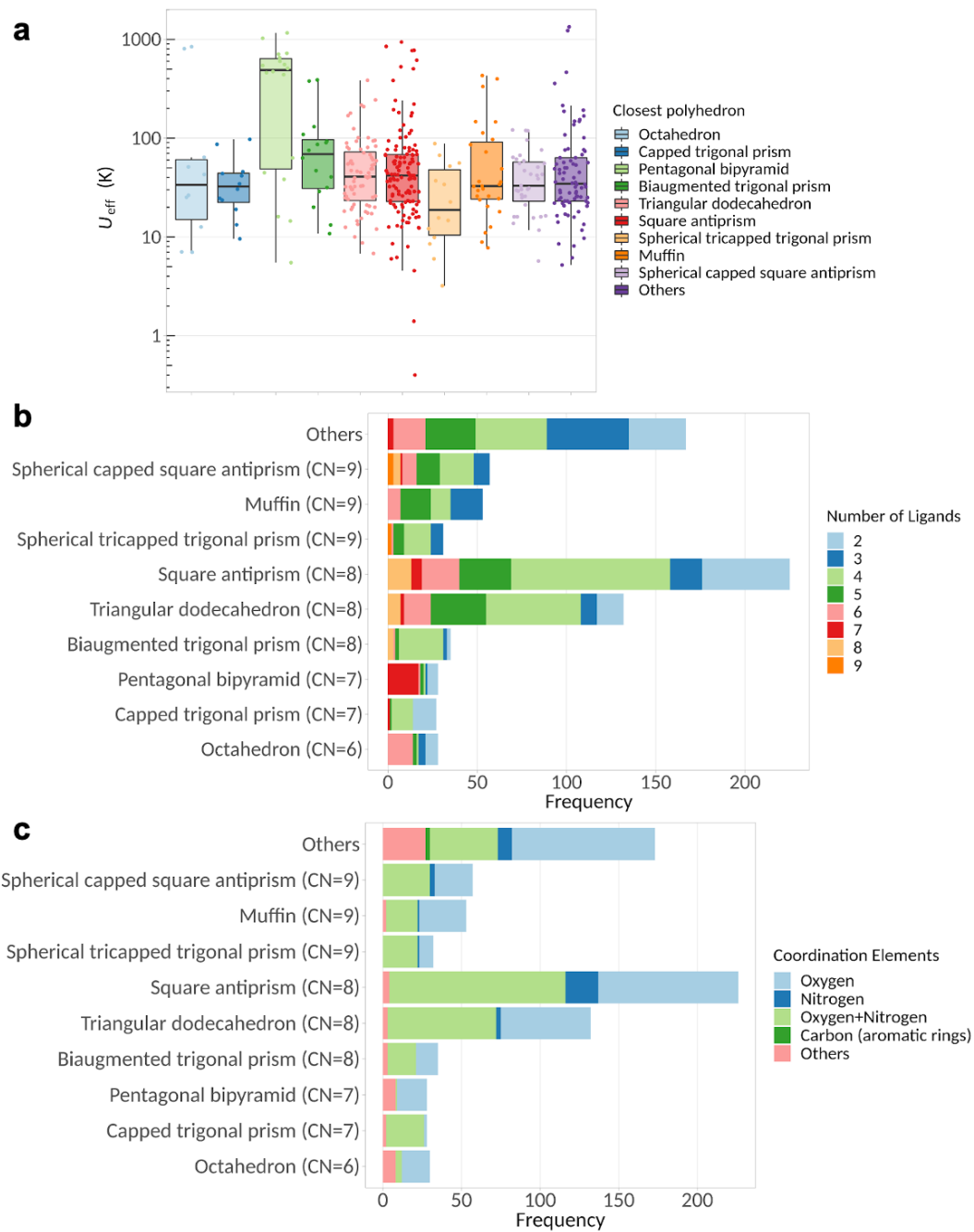

**Figure 6 | Relation between $U_{\text{eff}}$ and chemical design. a,** Dependence between $U_{\text{eff}}$ and coordination polyhedra. **b,** Numbers of ligands that are present in the different coordination polyhedra. **c,** Elements in the coordination sphere that are present in the different coordination polyhedra.

# Discussion

We have systematically analysed over 450 articles to collect information from over 1400 samples reported over the first 17 years of the field of lanthanide-based SIMs and built a user-friendly dashboard for the visualisation of all the collected data. Moreover, we carried out an in-depth statistical analysis that allowed extracting trends, distinguishing the most relevant variables and grouping the data in clusters based on their chemical and physical properties. From this study, we can highlight two main pieces of information.

In the first place, from the point of view of the parametric characterisation, the simple Arrhenius fit assuming an Orbach process has been proven to be surprisingly meaningful, with the expected approximate relation between $\tau_0$ and $U_{\text{eff}}$. One can therefore perform this oversimplified theoretical fit knowing that the effective energy barrier $U_{\text{eff}}$ has been proven to present a consistently good correlation with SMM behaviour, as well as with Raman parameters $C$, $n$. Crucially, we have also shown the different nature of short-term magnetic memory in the form of the blocking temperature $T_{\text{B3}}$ at $10^3$ Hz and its long-term counterpart in the form of maximum hysteresis temperature $T_{\text{hyst}}$. The best strategies that optimise the former are not necessarily the best for the latter.

In the second place, the chemical roadmap for the preparation of lanthanide complexes with higher $T_{\text{hyst}}$ becomes now a little clearer. Generally, oblate ions are superior to prolate in terms of ac and $U_{\text{eff}}$, but not in $T_{\text{hyst}}$. So far, there has been a single chemical strategy to consistently and prolifically produce good magnetic memories, namely sandwiching an oblate ion between two rigid, planar, aromatic ligands; furthermore, the ion should be chosen to result in the most favourable $M_J$ structure, given the electron distribution offered by the ligand. Up to now, only two chemical families are well adapted to this strategy, namely TbPc$_2$ complexes and dysprosium metallocenes. Optimization is ongoing within these two families, for example TbPc$_2$ complexes featuring a radical Pc display enhanced properties,[38] and the reduced (divalent) analogues of DyCp$_2$.[39] We find comparatively little value in further pursuing chemical strategies that have been amply explored and never yielded hysteresis above 10 K, like polyoxometalates, Schiff bases, diamagnetic transition metals placed near the magnetic lanthanide, or radical ligands, except when acting as a bridge or as a part of a TbPc$_2$ complex. On the other hand, we also evidence that there is, of course, value in chemical ingenuity and exploration, in the quest for another successful strategy, which according to our results might well be based on equatorial erbium complexes, since these display consistently high $T_{\text{hyst}}$ values. Note that a few complexes included in our data fall into ample families such as "mixed ligands" or "other families", and yet present excellent hysteresis temperatures. It is possible that the next family of record-setters is related to one of the promising candidates in Fig. 7. Two axial phosphine oxide ligands with bulky substituents seem to function in a similar way as metallocenes, despite the five equatorial H$_2$O molecules.[24,40] This strategy is not restricted to phosphine oxides and deserves to be explored further: complexes with 7 ligands have median values of $T_{\text{hyst}}$ close to 10 K, as high as those with 2 ligands. Indeed, and as pointed out above, axially compressed pentagonal bipyramids are a most promising yet underexplored strategy, and monodentate oxygen-based ligands seem to be a consistent path to achieve them.

At the same time, here we provide a catalogue of lanthanide SIMs, together with SIMDAVIS, a dashboard that allows its interactive navigation; this is a type of tool utterly missing in the field of molecular nanomagnets. Perhaps more importantly in the wider perspective of design of new materials[41–43] and new molecules.[29,44] The dataset curated in this work will serve for Machine Learning studies and can also be employed as an annotated training data set for the development of new web scraping systems to retrieve chemical data,[45,46] or even word embeddings,[47] from the scientific literature. Finally, this work constitutes a step towards the availability of findable, accessible, interoperable, and reusable (FAIR) data in Chemistry.[48]

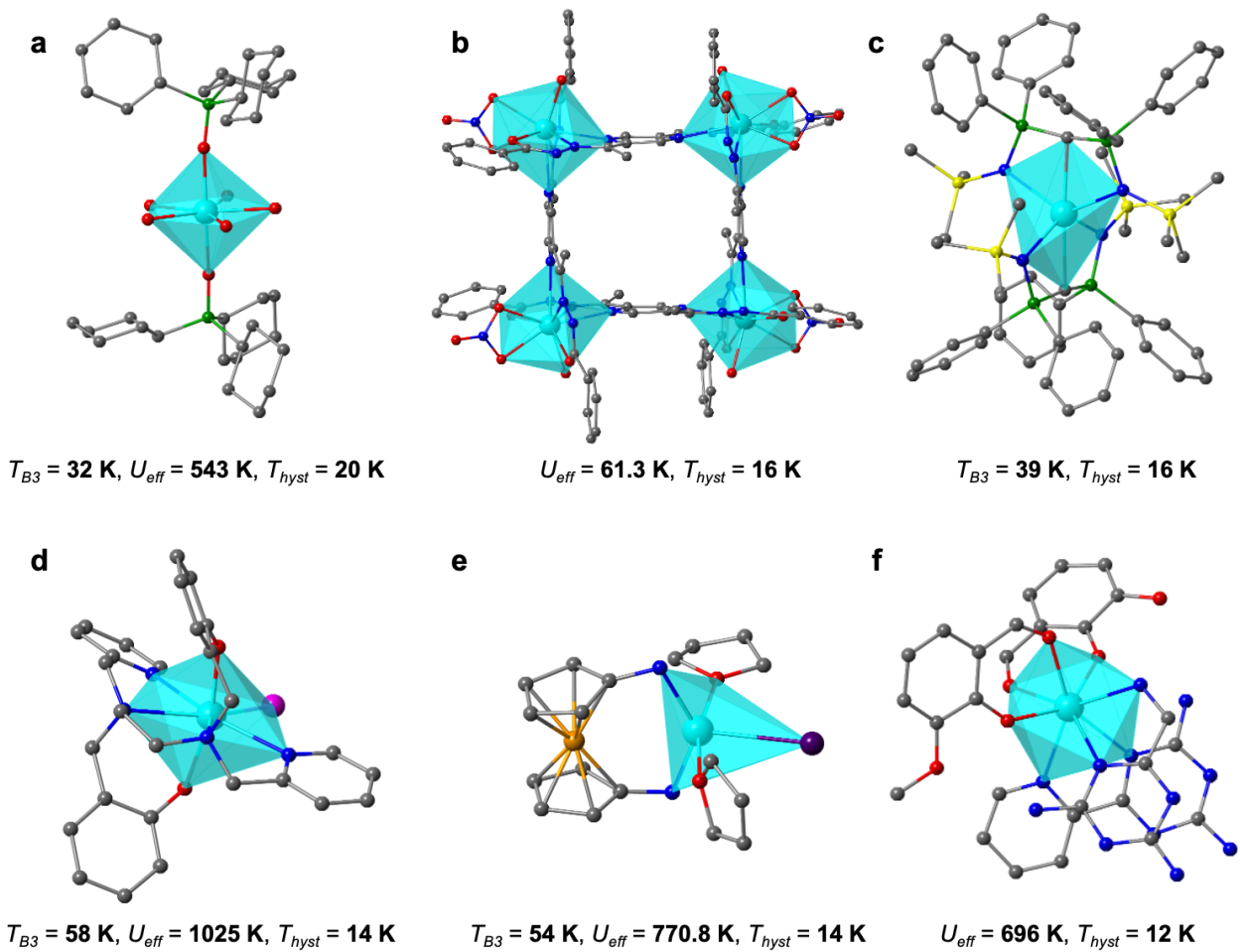

**a**
$T_{B3}$ = 32 K, $U_{eff}$ = 543 K, $T_{hyst}$ = 20 K

**b**
$U_{eff}$ = 61.3 K, $T_{hyst}$ = 16 K

**c**
$T_{B3}$ = 39 K, $T_{hyst}$ = 16 K

**d**
$T_{B3}$ = 58 K, $U_{eff}$ = 1025 K, $T_{hyst}$ = 14 K

**e**
$T_{B3}$ = 54 K, $U_{eff}$ = 770.8 K, $T_{hyst}$ = 14 K

**f**
$U_{eff}$ = 696 K, $T_{hyst}$ = 12 K

**Figure 7 | Promising systems for the development of new high-$T_{hyst}$ SIMs, all chemically distinct from each other and from the TbPc$_2$ and metallocene categories.**[40,49–53] **a**, [Dy(Cy$_3$PO)$_2$(H$_2$O)$_5$]$^{3+}$ (Cy$_3$PO = tricyclohexyl phosphine oxide);[40] **b**, [Dy$_4$(bzhdep-2H)$_4$(H$_2$O)$_4$(NO$_3$)$_4$] (bzhdep = pyrazine-2,5-diyl-bis(ethan-1-yl-1-ylidene)-di-(benzohydrazide));[49] **c**, [Dy(BIPM$^{TMS}$)$_2$]$^-$ (BIPM$^{TMS}$ = {C(PPh$_2$NSiMe$_3$)$_2$}$^{2-}$);[50] **d**, [Dy(bbpen)Br] (H$_2$bbpen = N,N′-bis(2-hydroxybenzyl)-N,N′-bis(2-methylpyridyl)ethylenediamine);[51] **e**, (NN$^{TBS}$)DyI(THF)$_2$ (NN$^{TBS}$ = fc(NHSi$^t$BuMe$_2$)$_2$, fc = 1,1′-ferrocenediyl);[52] **f**, [DyLz$_2$(o-vanilin)$_2$]$^+$ (Lz =

6-pyridin-2-yl-[1,3,5]triazine-2,4-diamine).[53] (Color scheme for atoms: green, P; cyan, Dy; gray, C; blue, N; yellow, Si; orange, Fe; red, O; magenta, Br; purple, I. Hydrogen atoms are not shown for clarity.)

# Methods

**Data gathering**. This process started with the collection and organisation of literature data. The following search criterion was applied for the manuscript: articles are searched via Web of Science, employing this code:

TOPIC: TS=((lanthan* OR 4$f OR "rare$earth") AND ((single NEAR/1 magnet*) OR "slow relaxation"))
Timespan: 2003-2019

For an article to be included in the study, it needs to contain data on at least one compound with certain requirements as follows: (a) contain one trivalent lanthanide ion from the set Ln = {Pr, Nd, Sm, Gd, Tb, Dy, Ho, Er, Tm, Yb} and (b) contain no other paramagnetic entity with the only accepted exception being the presence of a single radical in the coordination sphere and (c) present no strong Ln-Ln interaction, in particular meaning the Ln-Ln distance needs to be larger than 5 Å and more than 3 bridging atoms between neighbouring Ln centres, and there cannot be a radical in the bridge. Additionally, the data needs to include at least one of the following information: (a) whether $\chi''$ presents a maximum as a function of $T$, or a mere frequency-dependence, or neither; (b) $\chi''$ vs $T$ with at least one frequency ($f$) in the window 0.9 kHz $\leq f \leq$ 1.1 kHz and at a field ($B$) in the window $0 \leq B \leq 2$ T; (c) $U_{eff}$; (d) the presence or absence of hysteresis; (e) $T_{hyst}$ at sweep speeds ($v$) below 0.3 T/s. The compounds were classified in chemical families: $LnPc_2$, polyoxometalates, Schiff base, metallocenes, diketonates, radicals, TM near Ln, mixed ligands, and other families. Furthermore, we registered for each sample (when available), the lanthanide ion, its concentration, the coordination number and number of ligands coordinated to the lanthanide ions, the coordination elements, the presence of a field-dependent $\chi''$ or a maximum, the temperature of said maximum in presence or absence of an external magnetic field, the external magnetic field, the extracted effective energy barrier and relaxation time, either from a simplified Arrhenius fit or from a model considering all relaxation processes, whether these were extracted from the maxima of $\chi''$ vs T at different frequencies or from an Argand fit, the presence of hysteresis in the magnetisation, and the maximum temperature at which it was recorded. Additionally the DOI, the full reference to the original article, and a link to a CIF file were recorded for each sample. The question of publication bias is addressed at the end of Supplementary Section 1.1. Further details including the classification in chemical families and the criteria for data extraction are provided in Supplementary Sections 1 and 2.

**SIMDAVIS dashboard.** We programmed the dashboard using R language[54,55] and shiny,[56] an open source R package to create the interactive web app. The design aimed to obtain a clean and simple user interface that adapts automatically to any screen size. The R packages readr,[57] dplyr,[58] DT,[59] ggplot2[60] and rcrossref[61] were also employed in the development of the dataset or the app. The dashboard-style web application is available at https://go.uv.es/rosaleny/SIMDAVIS. This interface allows for variables in the analysis, and subsets of the data, to be adjusted and chosen in real time.

**Statistical analysis.** The statistical analysis was also based on R, a widely used software environment for statistical computing and graphics, and included the Gifi system for Multiple Correspondence Analysis[62] (R homals package,[63] ade4 package,[64] see details in Supplementary Section 4.1), hierarchical clustering studies (FactoMineR,[65] see details in Supplementary Section 4.2), lognormal modelling (Poisson's distribution, see Supplementary Section 4.3), factorial analysis of mixed data (FactoMineR and factoextra,[66] see details in Supplementary Section 6) as well as Pearson's product-moment correlation and the Akaike information criterion (AIC)[67] (see details in Supplementary Section 5.3). The analysis was repeated and verified an overall excellent qualitative and quantitative consistency in all results between the period 2003-2017 (1044 samples) and 2003-2019 (1405 samples).

**SHAPE analysis.** We employed a modified version of the pyCrystalField code[68] to extract the coordination environments of samples with a cif file in either the COD or the CCDC databases. We employed the SHAPE program to compare these with the reference polyhedra. Elongated and compressed versions of the reference polyhedra were also evaluated. We searched for correlations between the new data and the rest of the dataset.

# Data and code availability

The dataset collected and analysed during this study is freely available for download at https://go.uv.es/rosaleny/SIMDAVIS.
All custom code generated and employed for this study, namely the SIMDAVIS app version 1.1.9, is freely available for download at https://bitbucket.org/rosaleny/simdavis/src/issue-6/.

# Acknowledgements

This work has been supported by the COST Action MolSpin on Molecular Spintronics (Project 15128), H2020 (FATMOLS project), the European Research Council (ERC) under the European Union's Horizon 2020 research and innovation programme (grant agreement No 78822for AdG "MOL2D" and ERC-2021-StG-101042680 "2D-SMARTiES"), the Spanish MINECO (grants PID2020-117264GB-I00 and PID2020-117177GB-I00 co-financed by FEDER and Excellence Unit María de Maeztu CEX2019-000919-M), the Fundação para a Ciência e a Tecnologia (projects UIDB/04044/2020 and UIDP/04044/2020), and the Generalitat Valenciana (Prometeo Program of Excellence/2019/066, CDEIGENT/2019/022, CIDEGENT/2021/018). This research used resources at the Spallation Neutron Source, a DOE Office of Science User Facility operated by the Oak Ridge National Laboratory. The statistical analysis was performed by Raquel Gavidia Josa with the Statistical Section of the S.C.S.I.E. (Universitat de València). TOC and Supplementary Fig. 1.1 were created with BioRender.com.

# Author contributions

All authors contributed to the different stages of the work plan as detailed below.

A.G.A. suggested the starting point of the analysis, with contributions from J.J.B. and S.C.S.

J.C. and Y.D. designed the whole procedure for raw data extraction and classification.

Y.D., J.C., A.G.A., S.C.S, S.G.S. and J.J.B. did the manual data-mining. Y.D., J.C., A.G.A, L.E.R. and S.C.S. double-checked the raw data.

A.S. adapted pyCrystalField and extracted the coordination environments. A.S and S.G.S. performed all calculations.

L.E.R. and A.G.A. cleaned and organised the raw data into a tidy dataset.

L.E.R. and A.G.A. conceived and A.G.A. supervised the statistical data analysis.

L.E.R. conceived and programmed the dashboard-style interactive web application for data visualisation and analysis.

All authors contributed to the preparation of the manuscript.

# Competing interests

The authors declare no competing interests.

# Additional information

Supplementary information is available for this paper.

# Supplementary Information

# Data mining, dashboards and statistics: a powerful framework for the chemical design of molecular nanomagnets

Yan Duan[†,1,2], Lorena E. Rosaleny[†,*,1], Joana T. Coutinho[†,*,1,3], Silvia Giménez-Santamarina[1], Allen Scheie[4], José J. Baldoví[1], Salvador Cardona-Serra[1], Alejandro Gaita-Ariño[*,1]

[1] Instituto de Ciencia Molecular (ICMol), Universitat de València, C/ Catedrático José Beltrán 2, 46980 Paterna, Spain
[2] Spin-X Institute, South China University of Technology, Guangzhou 510641, P. R. China
[3] Centre for Rapid and Sustainable Product Development, Polytechnic of Leiria, 2430-028 Marinha Grande, Portugal
[4] Neutron Scattering Division, Oak Ridge National Laboratory, Oak Ridge, Tennessee 37831, USA

**Supplementary Section 1. Construction of the dataset**

Data extraction was restricted to variables that can be systematically extracted from articles included in the present study. Our objectives for the data analysis were twofold: the correlation between different variables of the same (physical or chemical) category and the correlation between two variables from different (chemical and physical) categories. In the first case, the goal is to determine whether the variables are closely correlated with each other, and thus to simplify our analysis and to avoid false correlations. In the second case, the goal is to determine which chemical variables are proven to be the most influential on the physical performance.

For physical variables, we focus on the magnetic hysteresis and the behaviour of the out-of-phase component of the ac susceptibility. These two kinds of experimental observations are the most basic experimental tell-tale signs for SIM behaviour. For both, we extract from the articles qualitative and quantitative information. For ac susceptibility, we extract as qualitative information whether the out-of-phase component of the ac susceptibility $\chi''$ vs the temperature has a maximum when there is no external dc magnetic field, or whether this maximum is absent but there is a frequency-dependent behaviour of $\chi''$ with the temperature. As quantitative information, we extract the temperature of said $\chi''$ vs T maximum. If the maximum of the out-of-phase component of the ac susceptibility appears in the presence of an external dc magnetic field, we extract the temperature at which the said maximum value appears and the applied external field. From reported magnetisation vs the magnetic field experiments, we extract as qualitative information the presence of full hysteresis (with remnant magnetisation and/or a coercive field), or at least a pinched (butterfly) hysteresis, and as quantitative information the maximum hysteresis temperature reported. In addition, a series of variables from the more extended theoretical analysis of the experimental data, namely the effective energy barrier $U_{\mathrm{eff}}$ and the relaxation time $\tau_0$, as well as relevant information on what kind of fit gave rise to these parameters are also included.

In the case of the chemical variables, the information we collected and analysed is as follows: (a) the chemical family of the complex; (b) the lanthanide (Ln) ion; (c) whether the Ln ion is oblate, prolate or isotropic; (d) whether the Ln ion is Kramers or not; (e) the concentration of the sample (if the diamagnetic dilution is studied); (f) the coordination number of the Ln ion; (g) the number of coordinated ligands; (h) the coordination elements.

A full list of the variables, including the number of data points and the percentage of samples with valid values for each variable in the dataset, can be found in Supplementary Figs. 1.1 and 1.2. As a consequence of this sparseness in the data, not all samples will be present in all graphs: any (x vs y) plot can only include samples for which x and y are simultaneously present in the dataset.

| Chemical Variables | N |
| --- | --- |
| Chemical Family | 1411 |
| Ln ion | 1411 |
| Ln anisotropy | 1411 |
| Ln Kramers | 1411 |
| Coordination Number | 1337 |
| Number of Ligands | 1379 |
| Coordination Elements | 1411 |
| Concentration | 1402 |
| Closest Polyhedron | 795 |
| CSM | 795 |
| Axial Distortion | 622 |
| Molecular Cluster | 1404 |

| Physical Variables | N |
| --- | --- |
| $\chi_{max}''$ : Presence of maximum in $\chi''(T)$ | 1215 |
| $T_{B3}$ : Temperature of $\chi_{max}''$ at zero-field | 251 |
| $T_{B3H}$ : Temperature of $\chi_{max}''$ at field H | 347 |
| $H$ : External Applied Magnetic Field | 348 |
| $Fit$ : $U_{eff}$ fitted by $T_{B3}$ or Argand | 472 |
| $U_{eff}$ : Effective barrier with Orbach process | 668 |
| $U_{eff2}$ : $U_{eff}$ for a second relaxation process | 24 |
| $U_{eff,ff}$ : $U_{eff}$ for all processes (full fit) | 76 |
| $\tau_0$ : Attempt time with Orbach process | 621 |
| $\tau_{0,ff}$ : Attempt time with all relaxation processes | 68 |
| $C$ : Raman prefactor | 54 |
| $n$ : Raman exponent | 56 |
| $\tau_{QTM}$ : Quantum tunneling of the magnetisation time | 41 |
| $Hyst$ : Presence of pinched or full hysteresis | 380 |
| $T_{hyst}$ : Maximum hysteresis temperature | 283 |
| Mag. Struct.Cluster : Clustering by chem. & phys. props. | 608 |

**Supplementary Figure 1.1 | Chemical and physical variables included in the dataset.** Correspondence between variables and symbols and number (N) of samples in the dataset containing that information.

Let us start by defining the chemical variables and explaining the different values they can take. When appropriate a numerical labelling equivalence for each value is given in square brackets, this is used in some of the statistical plots in later sections.

-The parameter "Chemical family" is categorical and takes one of the following 9 values for each sample: {LnPc$_2$ [1]; polyoxometalate [2]; Schiff base [3]; metallocene [4]; diketonate [5]; radical [6]; TM near Ln [7]; mixed ligands [8]; other families [9]}. Details on this classification are given in Supplementary Section 2.

-The parameter "Ln ion" is categorical and takes one of the following 10 values for each sample: {Pr$^{3+}$ [1]; Nd$^{3+}$ [2]; Sm$^{3+}$ [3]; Gd$^{3+}$ [4]; Tb$^{3+}$ [5]; Dy$^{3+}$ [6]; Ho$^{3+}$ [7]; Er$^{3+}$ [8]; Tm$^{3+}$ [9]; Yb$^{3+}$ [10]}.

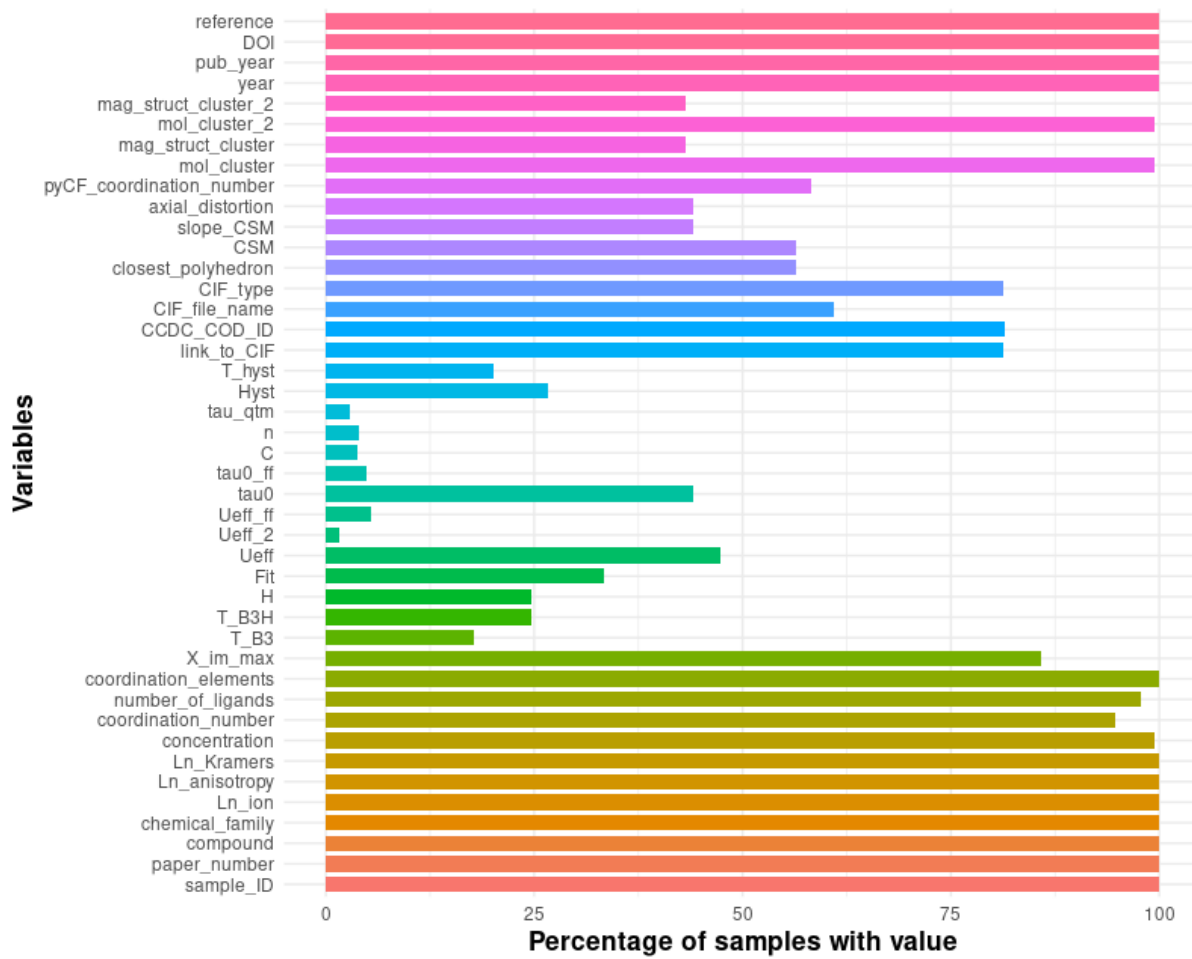

**Supplementary Figure 1.2 | Percentage of samples with data included in the dataset.** Percentage of samples containing valid values for each variable.

-The parameter "Ln anisotropy" is categorical and takes one of the following 3 values for each sample: {prolate [0]; oblate [1]; isotropic [2]}. This is determined directly by the Ln ion.

-The parameter "Ln Kramers" is categorical and takes one of the following 2 values for each sample: {non Kramers [0]; Kramers [1]}. Like the anisotropy, this is determined directly by the Ln ion.

-The parameter "concentration" takes the percentage value, e.g. concentration=1 is read as 1% concentration of the magnetic Ln ion in a diamagnetic matrix where 99% of the molecules are *e.g.* the $Y^{3+}$ analog.

-The parameter "Coordination Number" (or CN) is an integer number between 2 and 9. Note that, for $LnPc_2$ we assigned CN = 8, as corresponding to the 8 N donor atoms; and for metallocenes we assigned CN = 2, assuming that the electron density is delocalized within each aromatic ring.

-The parameter "Number of ligands" is an integer number between 2 and 9, corresponding to the total number of ligands contributing donor atoms. The absolute number of ligands is registered, not the number of chemically different ligands: N identical ligands count as N.

-The parameter "Coordination Elements" is categorical and takes one of the following 5 values for each sample: {Oxygen [1]; Nitrogen [2]; Oxygen+Nitrogen [3]; Carbon [4]; Others [5]}. Any combination of oxygens and nitrogens is counted as "Oxygen+Nitrogen", and complexes with coordination elements different from O, N or C in the coordination sphere of Ln ions are in the category of "Others".

-The parameters "Closest polyhedron", "CSM" and "Axial distortion" are extracted from calculations as defined in Supplementary Sections 7, 8. "Closest polyhedron" is categorical, while "CSM" and "Axial distortion" are continuous numbers.

-The parameters of the type "Molecular cluster" are categorical and are extracted from statistical data processing as defined in Supplementary Section 4.2.

Let us continue by defining the physical variables.
-The parameter labelled as $\chi''_{max}$ (in plots), or $\chi_{im,max}$ (in data table), takes one of these possible values:
·[0]: Freq-independent $\chi''$ (neither $T_{B3} > 2$ K reported, nor frequency-dependence in $\chi''$ vs T),
·[1]: Freq-dependent $\chi''$ (no $T_{B3} > 2$ K, but frequency-dependence in $\chi''$ vs T measured),
·[2]: $T_{B3} > 2$ K, and
·[3]: Not Measured (no available data to assign the sample into one of the previous three categories).

-$T_{B3}$ ($T_{B3H}$) is the temperature at which one finds the maximum value of $\chi''$ vs T at $10^3$ Hz, in absence (in presence) of an external magnetic field; H is the magnetic field, if present. It can be understood as the maximum temperature for which the system maintains short-term (millisecond) magnetic memory. For articles that provide $\chi''$ vs T with a curve for each different frequency, we simply chose the curve corresponding to the frequency $10^3$ Hz (or the closest one) and registered the temperature for the maximum $\chi''$, or the absence of a maximum. However, if the articles represent $\chi''$ vs frequency as isothermal curves for each different temperature, the same information is accessible indirectly by reading the points in the graph vertically at the abscissa value corresponding to the frequency $10^3$ Hz and checking in consecutive temperature curves whether $\chi''$ values present a non-monotonic evolution with respect to temperature, and therefore a maximum.

-"Fit" registers whether the parameters to determine $U_{eff}$ and $\tau_0$ were obtained from $\chi''(T)$ maxima at different frequencies or from an Argand plot.

-$U_{eff}$, $U_{eff,2}$, $U_{eff,ff}$ are the effective energy barriers and $\tau_0$, $\tau_{0,ff}$ are the attempt times, which means the pre-exponential factors. The values of the effective energy barrier $U_{eff}$ and of the attempt time $\tau_0$ are recorded if they are determined from a fit considering a single Orbach process. In the cases where a second Orbach process is considered, we register (besides $U_{eff}$, $\tau_0$ for the first process) its effective energy barrier $U_{eff,2}$. If a more complete model for relaxation is employed including an Orbach process as well as the Raman process, Quantum Tunnelling of the Magnetization and/or a direct process, we consider this a "full fit"(in short: ff), and record the value of the effective energy barrier $U_{\text{eff,ff}}$ and the attempt time or pre-exponential factor $\tau_{0,\text{ff}}$.

-The parameter labelled as "*Hyst*" takes one of the four values as follows:

·[0]: No hysteresis above 2 K reported,
·[1]: Pinched/butterfly Hysteresis (magnetic hysteresis above 2 K reported, but no magnetic coercivity field or remnant magnetisation can be determined; see Supplementary Fig. 1.3),
·[2]: Full Hysteresis (magnetic hysteresis above 2 K reported, and additionally either magnetic coercive field or remnant magnetisation can be determined; see Supplementary Fig. 1.3),
·[3]: Not Measured (no available data to assign the sample into one of the previous three categories).

To standardise data as far as possible, instead of taking the hysteresis temperatures as reported in the main text by the different researchers (that often employ different criteria) we examined all figures available ourselves and employed a uniform criterion to extract the data. As a consequence, in many cases our data do not coincide with the author's explicit claims.

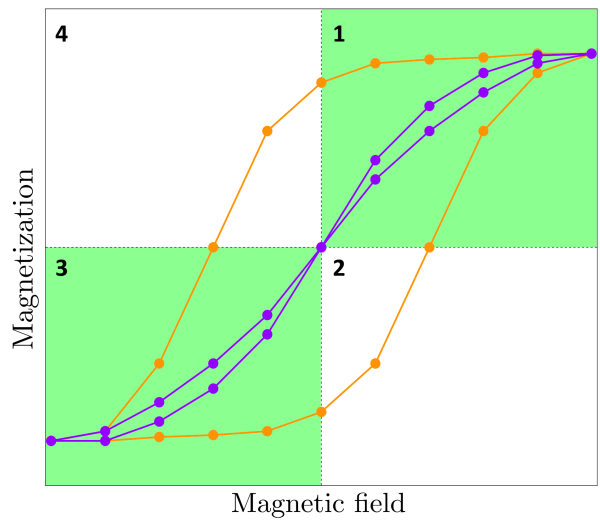

**Supplementary Figure 1.3 | Full vs pinched (butterfly) hysteresis.** Full hysteresis curves (orange) present at least one point either in quadrants 2,4 (*i.e.* different signs for Magnetic field and Magnetization) or in the x and/or y axes. Pinched (butterfly) hysteresis curves (violet) present points only in quadrants 1,3 (*i.e.* same signs for *H* and *M)* and, sometimes, also at the origin of coordinates.

-The related parameter "$T_{\text{hyst}}$" takes the highest temperature value at which hysteresis is reported. In contrast with $T_{\text{B3}}$, this quantifies the temperature up to which the system maintains long-term magnetic memory.

-The parameters of the type "Magneto Structural cluster" are categorical and are extracted from statistical data processing as defined in Supplementary Section 6.

One of the main problems for the data extraction during the construction of the dataset was that different criteria are chosen by different groups to characterise the hysteresis. For example, the hysteresis is measured only at 2 K in many studies, resulting in an overrepresentation of $T_{hyst} = 2$ in the dataset; while in many other cases the hysteresis is measured up to the highest possible temperature. Therefore, the very same compound could then present different $T_{hyst}$ depending on an arbitrary choice by the researchers, decreasing the quality of the dataset. Another difficulty is that the information of the applied magnetic field sweep rate which directly affects $T_{hyst}$ is missing in some articles. Note however that the works where data from more than a single sweep rate are reported demonstrate that the order of magnitude of $T_{hyst}$ is very robust unless ultra-fast magnetic field pulses are employed. These extreme cases are discarded from our dataset (we only include sweep speeds (v) below 0.3 T/s), meaning that even when mixing data from different sweep rates within this range, the data points will consistently be in the right region of the graphs (e.g. cases with $T_{hyst} = 3K$ will not be mistaken for $T_{hyst} = 13K$ or $T_{hyst} = 30K$, nor vice versa). In addition, in some articles, not only a full hysteresis, presenting coercive field and magnetic remanence, is measured up to a certain temperature but also a pinched (butterfly) hysteresis is measured up to a higher temperature. This introduces some noise in the dataset, but is less problematic since both cases are going to be registered as a SIM with good properties.

Finally, it was not practical to include in the dataset other descriptive parameters such as $\tau_{switch}$ ,[1] the temperature at which the relaxation time is 100 s or the temperature at which there is a maximum in the zero-field cooled susceptibility because, while they will hopefully be the standard in the field, the number of publications reporting these parameters within our studied period (2003-2019) is too reduced to extract trustworthy statistical information from them.

## 1.1 The question of publication bias

Note that published results are generally biassed towards positive results, and this is generally considered as a problem in meta-studies. Publication bias may invalidate the conclusions unless they are robust to possible non-random selection mechanisms.[2] In the present case there are two important questions about this.

In the first place: how did we address the bias in available data towards positive results, and how abundant negative data are within our dataset? We addressed this in the design phase of our study by (a) performing an automated search for articles based on certain keywords (related to the topic, not necessarily with the result), (b) recording all negative data, which we found to be very abundant, especially in lower-impact journals and (c) distinguishing between different categories of negative results. For this it was key to record not just hysteresis data, where an absence of hysteresis is rarely acknowledged explicitly, but also the indirect information provided by ac magnetometry. By following this strategy, we found that the bibliography in this field is in practice very rich in negative results, whether these take the form of absolutely no ac signal, or of just a frequency dependence in ac magnetometry but no out-of-phase peak above 2 K. Even ac peaks at low temperature can be understood as negative results, since typically the compounds with hysteresis also behave well in ac (see Supplementary Figure 21).

Note that it is frequent that a series of compounds is studied in the same publication, changing either the Ln metal or doing systematic modifications in the ligands, and among them only some present good SMM properties, and in the vast majority of cases this is explicitly acknowledged in the text, so it is possible to extract negative data. There are also many cases of studies focused on other properties (e.g. optical), where the magnetic behaviour is recorded but does not determine publication. As we will see, from over 1400 samples, about 600 present no frequency-dependent out-of-phase magnetic susceptibility $\chi''$, compared with 200 with no data, and about 300 each for a maximum in $\chi''$ above 2K or no maximum but a frequency-dependent $\chi''$. In that sense, the problematic data are related to hysteresis, where the vast majority of the samples (>1000) contain no information. While we did not rely on this for our analysis, it would be reasonable to assume that in most cases a lack of information on hysteresis means that the sample presents no hysteresis.

In the second place: what are the consequences? In the present study we were careful not to ask absolute questions which would be affected by publication bias, such as how often, out of a novel 100 complexes, Schiff base ligands are expected to produce certain results in terms of magnetic behaviour, and we do not extract conclusions from how many Dy SIMs have been reported, in absolute numbers, compared with Er SIMs. Instead, we always compare relative frequencies. One can safely assume that the effect of publication bias will affect equally different lanthanides, different chemical families, etc, meaning that relative comparisons should be safe from publication bias. From the dataset we want to answer relative questions such as:

-are LnPc$_2$, or metallocenes, distinctly promising as SIMs, compared with any other chemical families?

-the same for several other chemical families: are {Schiff bases, polyoxometalates, diketonates, radicals, TM near Ln} promising as SIMs, in relative terms? (see below for the definition of the families)

-are complexes of oblate ions (Dy,Tb…) better SIMs than prolate ions (Er…) in terms of ac susceptibility, higher $U_{\text{eff}}$, magnetic hysteresis?

-is $U_{\text{eff}}$ as good a predictor for $T_{\text{hyst}}$ as often assumed? Is it correlated with $\tau_0$ and/or with Raman?

-are there any coordination polyhedra with high relative frequency of good magnetic behaviour?

**Supplementary Section 2. Classification in chemical families**

Lanthanides are a group of *f*-block elements with atomic numbers ranging from 57 (lanthanum) to 71 (lutetium). Most of the Ln elements exhibit the oxidation state of +3. Our dataset only includes the trivalent Ln$^{3+}$ (Pr, Nd, Sm, Gd, Tb, Dy, Ho, Er, Tm and Yb) ions containing complexes. Ln ions possess large coordination numbers (CNs) due to their large ionic radii. The geometrical arrangement around these trivalent ions basically depends on the steric properties of the coordinated ligands; thus a suitable design of the ligand molecules leads to an easy tuning of the CNs. In particular, CNs between 2 and 12 are documented for Ln ions. Note that in this work we consider one rigid aromatic ring as equivalent to a contribution to CN/ring = 1 when it is of the cyclopentadienyl/cyclooctatetraenyl kind, whereas we consider a contribution of CN/ring = 4 when it is of the phthalocyaninato kind. In the low CN cases, the coordination ligands are usually bulky ligands, e.g. bis(trimethylsilyl) amine gives CN = 3; whereas cyclopentadienyl ligands need to be smartly substituted to achieve the same steric impediment. In contrast, in the case of complexes with high CN, the ligands are usually small bidentate ligands, such as nitrate and/or macrocyclic ligands. In the present work, we found that the most frequent are CN = 8 and CN = 9. This coincides with what is known for Ln ions, namely, Ln ions tend to spontaneously favour these CNs, typically with distorted square antiprismatic coordination (CN = 8) or distorted tricapped trigonal prism coordination (CN = 9).

Ln-based SIMs are interesting because the 4$f$ electrons are less exposed to ligand field effects and exhibit larger spin-orbital coupling if compared with the $d$-shell. The first Ln-based mononuclear single molecule magnets (SMMs) were generated by Ishikawa and co-workers in 2003 using two macrocyclic ligands to sandwich the $Ln^{3+}$ ion in a double-decker fashion.[3] They can also be prepared by using a range of acyclic ligands, such as polyoxometalates (POMs),[4–6] Schiff bases,[7,8] radicals,[9–17] and ketones.[18–20] Between 2003 and 2019, several hundreds of articles referring to Ln-based SIMs have been published. Among them, the vast majority focused on the chemical approaches in designing lanthanide-based SIMs with superior properties. A fundamental key parameter of the magnetic properties of SIMs is the molecular symmetry which can be controlled by: (a) the ligand design and modification, (b) the substitution of the coordination elements as a means to alter electrostatic potential and/or Ln to coordination atom bond lengths, and (c) the peripheral ligand functionalization/substitution. Here, we classify the collected complexes into 9 categories according to the type of coordination ligands or the chemical strategy used for the design of the magnetic complex. These 9 categories (Chemical Family) are listed below and will be briefly described in this section.

1) LnPc$_2$ family
2) POM family
3) Schiff Base family
4) Metallocene family
5) Diketonate family
6) Radical family
7) TM near Ln family
8) Mixed ligands family
9) Other families

Note that endohedral metallofullerenes, nowadays a very promising SIM family, have not been classified as a separate family in the present study merely because at the point where we started designing the data collection, metallofullerenes were still quite scarce and not yet established as a SIM family. They could be included in a future update of the dataset. Radical-bridge dimers are now recognized to be strong candidates but were not included in the study at all because we considered that they introduce extra degrees of freedom that would only apply to a minority of cases. They would require their own study, which at the present time cannot be statistics-based.

## 2.1. LnPc₂ family

The first category is constituted by "double-decker complexes" related to the classical LnPc₂ family, namely, the Ln ion in the complex is octa-coordinated by nitrogen atoms from two Pc (or their related functionalized complexes, or porphyrin-like, or even tetraaza[14]annulenes) ligands. As we will see below, this criterion has priority over the presence of a spin S = 1/2 (radical ligand, in this case corresponding to oxidised or reduced Pc ligands) and also over the presence of diamagnetic transition metal ions in the vicinity (in this case often corresponding to multiple deckers which coordinate $Cd^{3+}$). In both cases, these complexes are classified as the LnPc₂ family. Complexes composed of phthalocyanine ligands or porphyrins with nitrogen-based donating atoms have shown very important roles in Ln-based SIMs. There are several reasons for choosing phthalocyanines and porphyrins for SIM design: a) these tetrapyrrole macrocyclic ligands containing four isoindole or pyrrole nitrogen atoms have the ability to strongly coordinate to Ln ions; b) special features such as intramolecular $\pi$-$\pi$ stacking interaction and the intrinsic nature of their macrocyclic rotation; and c) their structural characteristics of those sandwich-type complexes since the ligand field constructed by this type of ligands with a $C_4$-symmetric axis (pseudo-$D_{4d}$ symmetry) is very important for the zero-field splitting of the ground state into the magnetic sublevels. The combination of the large magnetic anisotropy with strong spin-orbital interactions leads to the SIMs behaviours.

The first examples of Ln-based SIMs reported are from the LnPc₂ family, which was proposed in 2003 by Ishikawa and co-workers.[3] They successfully demonstrated that slow magnetic relaxation could occur in mononuclear lanthanide complexes, such as those in which a Ln ion is sandwiched between two Pc ligands, formulated as $(Bu_4N)[LnPc_2]$ ($Ln^{3+}$ = $Tb^{3+}$ or $Dy^{3+}$, $Bu_4N$ = tetrabutylammonium) (Supplementary Fig. 2a). Later on, a massive synthetic effort has led to an ever-increasing number of compounds from the LnPc₂ family, which includes the introduction of a wide range of substituents at the periphery of the Pc macrocycles without significantly interfering with the metal binding properties of the ligands.[21–25] Some structure representations of examples from this family studied in this work are shown below (Supplementary Fig. 2), as the sandwich complex $[Bu_4N][DyPc(OTBPP)]$, (Supplementary Fig. 2b) in which one of the nitrogen atoms of one porphyrin pyrrole is replaced by an oxygen atom. Compared with the typical LnPc₂ complex, the atom replacement significantly enhances the effective energy barrier of the SIMs.[25] Another example is the use of tetraazaporphyrins (or porphyrazines) in place of the bulkier Pc ligands, giving rise to a series of neutral double-decker complexes that show analogous magnetic features as their Pc counterparts (Supplementary Fig. 2c).[26]

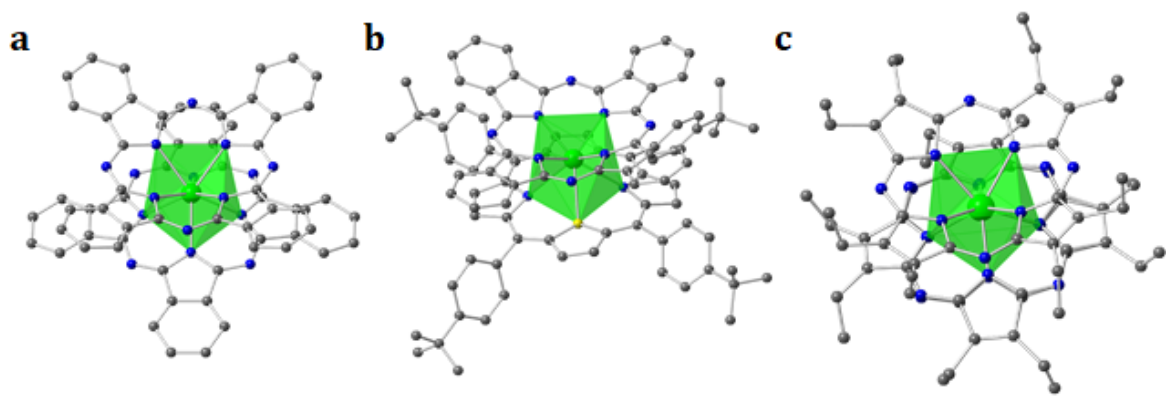

**Supplementary Figure 2 | Combined polyhedral and ball-and-stick models of the coordination spheres around Ln ions of some cases from the LnPc₂ family. a**, [Pc₂Ln]⁻ from reference [3]. **b**, The sandwich-type mixed phthalocyaninato with core-modified porphyrinato double-decker complexes [DyPc(OTBPP)]⁺ or [Dy(Pc)(STBPP)]⁺ from reference [25]. **c**, [Ln(OETAP)₂]⁺, where OETAP is octa(ethyl)tetraazaporphyrin.[26] (Color code: grey sphere, C; green sphere and polyhedron, Ln; blue sphere, N.)

## 2.2. POM family

The second representative family consists of polyoxometalates (POMs). This family contains all compounds where $Ln^{3+}$ ions coordinate with POM ligands, including the cases where the coordination sphere is completed with other ligands. POMs are molecular metal-oxo clusters with early transition metals (W, Mo, Nb, Ta or V) in their highest oxidation states. The ability of these inorganic species to incorporate almost any kind of metal or non-metal addenda heteroatoms, together with their enormous molecular and electronic structural diversity, makes them of relevance in the molecular magnetism field. One relevant feature of POM ligands is that their diamagnetic structures can encapsulate Ln ions with coordination geometries similar to those of bis(phthalocyaninato)lanthanide complexes from LnPc₂ family.[3] More recently, POMs were used as extremely versatile inorganic building blocks for the construction of SMMs based either on 3*d* or 4*f* metal ions.[27] Some representative cases of complexes included in this study are shown below in Supplementary Fig. 3. The first example from the POM family exhibiting SIM behaviour is [ErW₁₀O₃₆]⁹⁻ (Supplementary Fig. 3a).[4] Later on, two families of POM-based SIMs with formula [Ln(W₅O₁₈)₂]⁹⁻ and [Ln(β₂-SiW₁₁O₃₉)₂]¹³⁻ (Supplementary Fig. 3b) are reported in 2009, both of which show slow relaxation of the magnetisation, typical of the SIM-like behaviour.[5] Another well-known series of complexes is [LnP₅W₃₀O₁₁₀]¹²⁻ (Supplementary Fig. 3c), in which its unusual $C_5$ axial symmetry allows the study of new SIMs having 5-fold symmetry. The $Dy^{3+}$ and $Ho^{3+}$ derivatives exhibit SIM behaviour.[6]

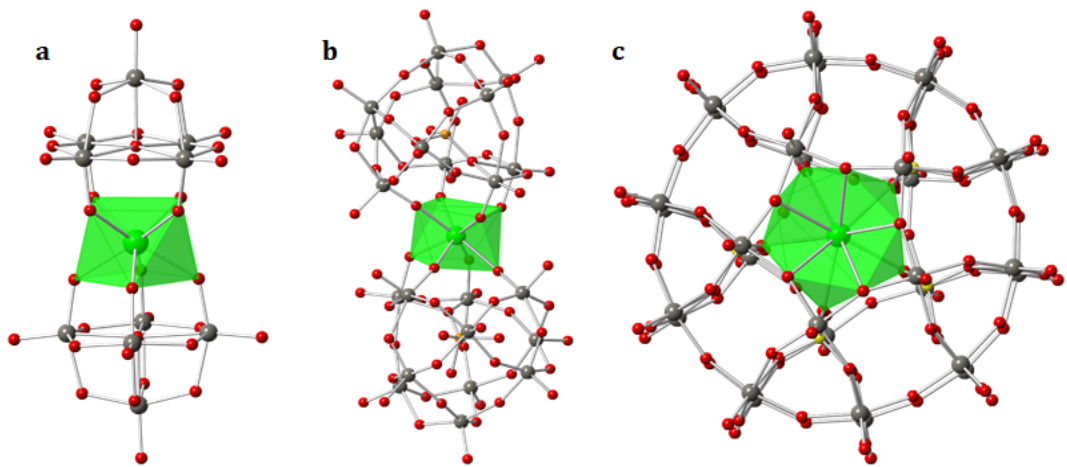

**Supplementary Figure 3 | Combined polyhedral and ball-and-stick models of the coordination spheres around Ln ions of three representative cases from the POM family. a**, $[ErW_{10}O_{36}]^{9-}$ from reference [4]. **b**, $[Ln(\beta_2\text{-}SiW_{11}O_{39})_2]^{13-}$ from reference [5]. **c**, $[LnP_5W_{30}O_{110}]^{12-}$ from reference [6]. (Color code: grey sphere, W; green sphere and polyhedron, Ln; red sphere, O; yellow sphere, P; orange sphere, Si.)

## 2.3. Schiff base family

The third family is based on Schiff base ligands. This includes all samples where the $Ln^{3+}$ ion coordinates to only Schiff base ligands; in addition, we included the cases where the strategy pursued by the authors (as stated in the title) relies on Schiff base ligand, even if other small ligands are used to complete the coordination sphere. Schiff base ligands are polydentate macrocyclic or macro-acyclic ligands, which typically contain both nitrogen and oxygen donors. However, the donor atom can be varied between sulfur, phosphorus, nitrogen, and oxygen. Due to their facile synthesis, Schiff base ligands are considered to be "privileged ligands", which can easily make a coordination bond with many different metal ions and stabilize them in various oxidation states. In addition, when two equivalents of salicylaldehyde are combined with a diamine, a particular chelating Schiff base is produced, which is called salen ligands. Salen ligands present four coordinating sites (tetradentate) and two axial sites that are open to ancillary ligands, thus similar to porphyrins but with an easier preparation process.

Schiff bases derived from condensation reactions of aromatic aldehydes with primary amines have been the subject of extensive research because of their enormous versatility with respect to the formation of metal complexes with sophisticated discrete or expanded architectures and functional properties. The choice of initial reagents for the condensation determines the ligand coordination fashion and allows one to utilize both chelate and bridging functions of the obtained Schiff base. Schiff base complexes continue to intrigue chemists regarding their structure and reactivity. Their geometries are strongly influenced by the ligands and tend to be five- or six-coordinate. The first case listed here comprises two mono-deprotonated Schiff base $[LH]^-$ ligands, showing SIM behaviour and with a $U_{eff}$ of 44.4 K in presence of a dc field (Supplementary Fig. 4a).[28] Another case from this family is the $Dy^{3+}$ complex with tridentate NNO ligands of N-[(imidazol-4-yl)methylidene]-DL-alanine (Supplementary Fig. 4b), which shows an out-of-phase signal with frequency-dependence in ac susceptibility under a dc bias field of $10^3$ Oe, indicative of field induced SIM.[29] One other representative case from this family is the Salen-type mononuclear Ln$^{3+}$ complex [Ln(3-NO$_2$-salen)$_2$]$^-$ (Supplementary Fig. 4c), which shows slow magnetisation relaxation processes associated with SIM behaviour.[30]

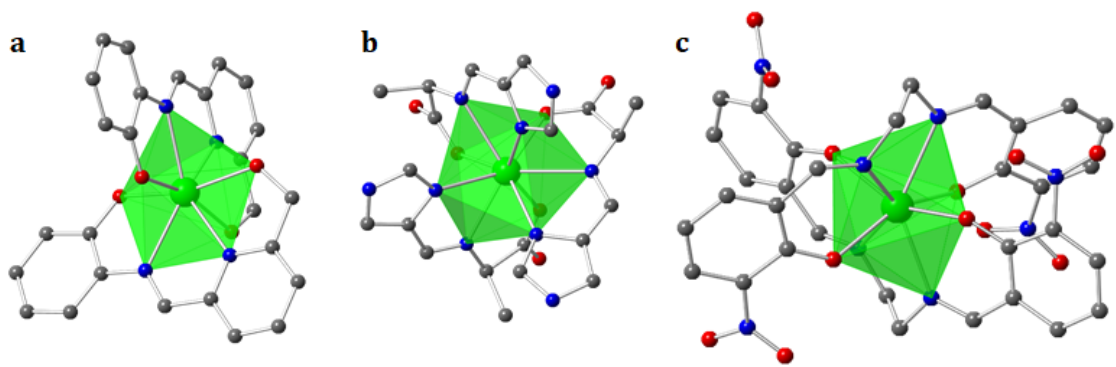

**Supplementary Figure 4 | Combined polyhedral and ball-and-stick models of the coordination spheres around Ln ions of some cases from the Schiff base family. a**, [Ln(LH)$_2$]$^-$ where H$_2$L = 2-((6-(hydroxymethyl)pyridin-2-yl)-methyleneamino)phenol.[28] **b**, '*fac*'-[Dy$^{III}$(HL$^{DL-ala}$)$_3$], where H$_2$L$^{DL-ala}$ is N-[(imidazol-4-yl)methylidene]-DL-alanine.[29] **c**, [Ln(3-NO$_2$-salen)$_2$]$^-$, where Ln can be Dy, Er or Yb, and 3-NO$_2$-salen$^{2-}$ = N,N'-bis(3-nitro-salicylaldehyde)ethylenediamine dianion.[30] (Color code: grey sphere, C; green sphere and polyhedron, Ln; red sphere, O; blue sphere, N.)

## 2.4. Metallocene family

The fourth family is based on the small aromatic ligands derived from conjugated hydrocarbon ligands, typically cyclopentadienyl or cyclooctatetraene anions. We only include in this classification the complexes where the coordination sphere is completed by this kind of ligands, in contrast with cases with an extra "equatorial" coordination site. Compared with heteroatomic donor atoms such as oxygen and nitrogen, which have limited orbital overlap with the shielded 4$f$ orbitals, the aromatic ligands allow the perturbation of the crystal field of the lanthanide ions through the use of an electron $\pi$-cloud. Thus, it can further control over the anisotropic axis and induction of *f-f* interactions, making donor atoms as conjugated hydrocarbons.[31] Here we list some examples by employing delocalized ligands to design SIMs with prominent uniaxial anisotropy. An Er$^{3+}$ ion sandwiched by two aromatic ligands, pentamethylcyclopentadienide anion (C$_5$Me$_5^-$, Cp*) and cyclooctatetraenide dianion (C$_8$H$_8^{2-}$, COT) (Supplementary Fig. 5a), displays a butterfly-shaped hysteresis loop at 1.8 K up to even 5 K.[32] Another example is a bis-monophospholyl Dy$^{3+}$ SIM, [Dy(Dtp)$_2$][Al{OC(CF$_3$)$_3$}$_4$] (Supplementary Fig. 5b), which shows an effective energy barrier to magnetisation reversal of 1760 K (1223 cm$^{-1}$) and magnetic hysteresis up to 48 K.[33] The use of planar cyclooctatetraenide (COT''$^{2-}$) ligands allows the access to the sandwich type complex [Dy(COT'')$_2$]Li(DME)$_3$ (Supplementary Fig. 5c), which exhibits slow relaxation of the magnetisation indicating its SIM behaviour.[34]

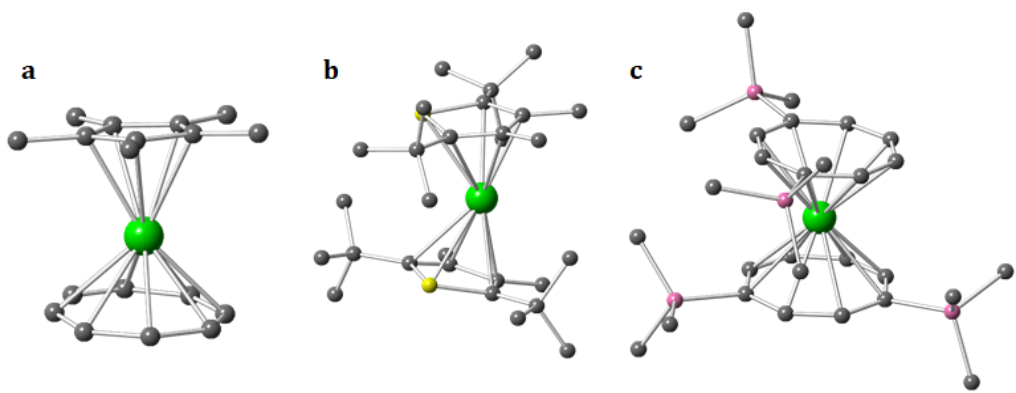

**Supplementary Figure 5 | Ball-and-stick models of the coordination spheres around Ln³⁺**

**Supplementary Figure 5 | Ball-and-stick models of the coordination spheres around Ln³⁺ ions of some cases from the metallocene family. a**, (Cp\*)Er(COT), where Cp\* = $C_5Me_5^-$ and COT = $C_8H_8^{2-}$, from reference [32]. **b**, [Dy(Dtp)₂][Al{OC(CF₃)₃}₄], where Dtp = {P(C′BuCMe)₂}.[33] **c**, [DyCOT″₂]⁻, where COT″²⁻ = cyclooctatetraenide rings.[34] (Color code: grey sphere, C; green sphere, Ln; pink sphere, Si; yellow sphere, P.)

## 2.5. Diketonate family

The fifth family is the diketonate family of complexes, it includes those samples with Ln³⁺ ions coordinated with diketonate ligands and diketonate ligands mixed with other molecules which are not defined in the classification. The diketonate ligands are bidentate and bond through delocalized chelate rings formed through two oxygen atoms. β-diketone SIMs have received much attention in recent years, since β-diketone can provide a stable bidentate chelating mode to afford eight-coordinated mononuclear lanthanide complexes. There are two different polyhedron coordination geometries for the β-diketone complexes, square antiprism with $D_{4d}$ symmetry and triangular dodecahedron with $D_{2d}$ symmetry. After the SMM behaviour of a simple acetylacetonate complex has been reported on several β-diketone complexes, much effort is devoted to the synthesis and investigation of β-diketone SIMs. In addition to the coordination geometry, the stability of the SIMs upon heating is also an important topic. Lanthanide β-diketone complexes with fluorides as substituent groups, such as hexafluoroacetylacetone (hfac), can make the complexes stable upon heating. By using the β-diketonate ligand dibenzoylmethane (DBM) anion, mononuclear Dy complex [Hex₄N][Dy(DBM)₄] (Supplementary Fig. 6a) was obtained, in which slow magnetic relaxation is observed.[18] A typical compound of β-diketone is formulated as (cation)[Ln(β-diketone)₄], in which the Ln³⁺ ion is surrounded by four β-diketone forming a LnO₈ environment. The complex shown in Supplementary Fig. 6b, using hfac ligand, exhibits field-induced slow magnetization relaxation.[19] Another case is the use of a sulfonyl amidophosphate (SAPh), acting as a β-diketone homologue for the complexation of Ln ion, which gives rise to complex LnL₃Phen (L = C₆H₅SO₂NP(O)[N(CH₃)(C₆H₅)]₂) with in-field SIM behaviours (Supplementary Fig. 6c).[35]

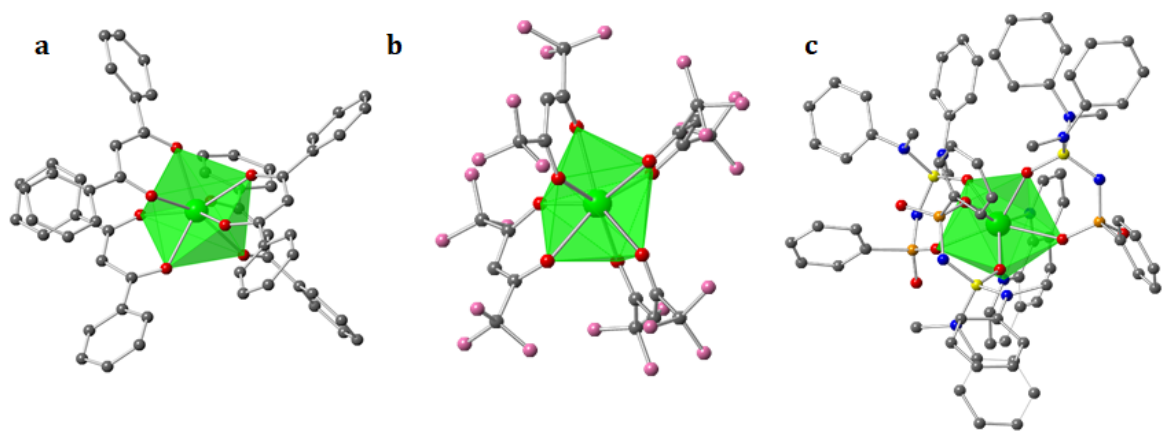

**Supplementary Figure 6 | Combined polyhedral and ball-and-stick models of the coordination spheres around Ln ions of some cases from the diketonate family. a**, [Dy(DBM)₄]⁻, where DBM = dibenzoylmethane anion ligand.[18] **b**, {Dy(hfac)₄}, where hfac = hexafluoroacetylacetone.[19] **c**, [LnL₃(phen)], where Ln can be Dy or Er, and L is deprotonated bis(methyl(phenyl)amino)phosphoryl)-benzenesulfonamide, C₆H₅SO₂NP(O)[N(CH₃)(C₆H₅)]₂, and Phen = phenanthroline.[35] (Color code: grey sphere, C; green sphere and polyhedron, Ln; red sphere, O; yellow sphere, P; pink sphere, F; blue sphere, N; orange sphere, S.)

## 2.6. Radical family

The sixth family is composed of complexes in which Ln³⁺ ion is coordinated with radical-based ligand(s), such as nitronyl nitroxide and semiquinones. Radical ligands are one of the most efficient bridging ligands for the design of molecular magnetic materials.[36] The radical systems are relatively abundant in our dataset, being more numerous than any other family presented so far. The reason of choosing radical ligands is that they possess 2*p* diffuse spin orbitals that can potentially penetrate the core electron density of the lanthanide ions to reach deeply buried 4*f* orbitals, whose shielded magnetic orbitals are usually a drawback for their use in extended magnetically coupled structures.[37] The strong 2*p*-4*f* heterospin exchange coupling effectively shifts degenerated $m_J$ sublevels to different energies and, furthermore, significantly reduces the probability of resonant quantum tunnelling and lengthens the relaxation time. We will show some examples from this family (Supplementary Fig. 7). The first case is the nitronyl nitroxide radical complex [Ln(tfa)₃(NIT-BzImH)], (Supplementary Fig. 7a) in which Ln³⁺ ion is 8-coordinated to one NIT-BzImH and three trifluoroacetylacetonate (tfa) ligands. It shows slow magnetic relaxation suggesting that they behave as SIMs.[12] The second case is a dinuclear Ln³⁺ compound with its formula as {Cp₂Co}{[Dy(tmhd)₃]₂(bptz)} in radical anion form (Supplementary Fig. 7b), in which the rare earth ions are isolated by an organic ligand bridged species. It exhibits out-of-phase ac susceptibility signals below 4 K.[17] Another relevant case is a cyclic dimer structure, in which each pyridine substituted radical links two different metal ions through the oxygen of a nitroxide group and the pyridine nitrogen (Supplementary Fig. 7c). It shows frequency-dependent ac magnetic susceptibility, indicating SIM behaviour.[11]

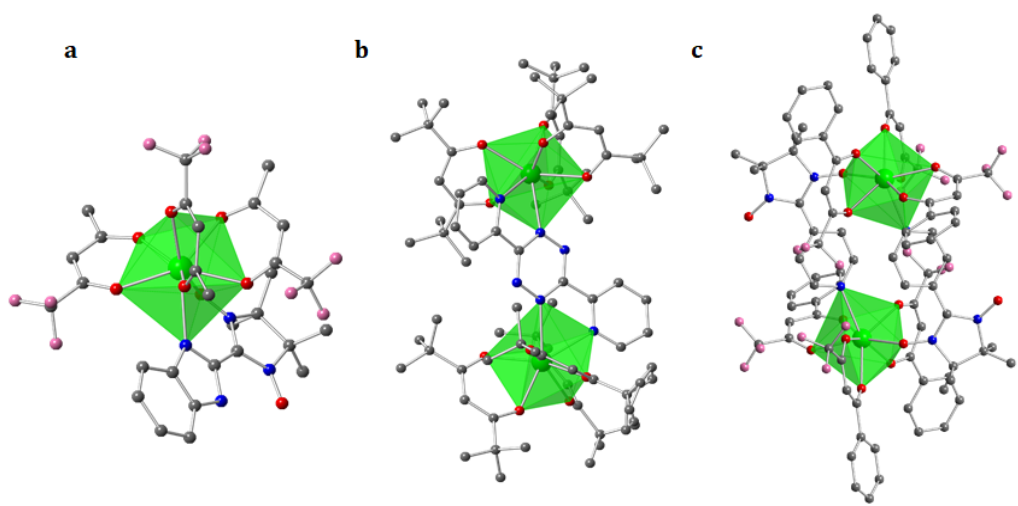

**Supplementary Figure 7 | Combined polyhedral and ball-and-stick models of the coordination spheres around Ln ions of some cases from the radical family. a**, [Ln(tfa)₃(NIT-BzImH)], where tfa = trifluoroacetylacetonate; NIT-BzImH = 2-(2′-benzimidazolyl)-4,4,5,5-tetramethylimidazolyl-1-oxyl-3-oxide.[12] **b**, {Cp₂Co}{[Dy(tmhd)₃]₂(bptz)}, where tmhd = 2,2,6,6-tetramethyl-3,5-heptane dionate and bptz = 3,6-bis(2-pyridyl)-1,2,4,5-tetrazine.[17] **c**, [Ln(Phtfac)₃(NITpPy)]₂, where HPhtfac = 4,4,4-trifluoro-1-phenylbutane-1,3-dione and NITpPy = 2-(4-pyridyl)-4,4,5,5-tetramethyl-4,5-dihydro-1H-imidazolyl-1-oxyl-3-oxide.[11] (Color code: grey sphere, C; green sphere and polyhedron, Ln; red sphere, O; pink sphere, F; blue sphere, N.)

## 2.7. TM near Ln family

This family of complexes is defined when a diamagnetic transition metal (TM) ion exists in the coordination sphere of $Ln^{3+}$ ion. There are several Ln-based SIMs containing one or more 3*d* metal ions.[38–41] Most of this type of complexes contain Schiff base ligand or a diketone ligand.[38] For example, Yamashita *et al.* reported an Er-based SIM,[42] where the $Er^{3+}$ ion is coordinated with a Schiff base ligand, which in turn is connected to the diamagnetic transition metal $Zn^{2+}$ through oxygen. Macrocyclic ligands provide discrete metal binding pockets and, therefore, offer a more predictable cluster nuclearity and structure than acyclic analogues can.[38] For example, the [3+3] macrocycle provides three $N_2O_2$ pockets for 3*d* metal ions and one central $O_6$ pocket for a Ln ion, making mixed-metal $M_3Ln$ tetrametallic macrocyclic complexes predictable.[42;43] Macrocycles usually provide enhanced stability, solubility and fine-tunability (vary the choice of M and Ln, whilst retaining the $M_3Ln$ core) over acyclic analogues. It's documented that the $U_{eff}$ of Ln-based SIMs can be enhanced by introducing diamagnetic metal ions in the coordination sphere. The diamagnetic ion may induce large electrostatic interaction between the $Ln^{3+}$ ion and coordination atoms, giving rise to the destabilization of excited states and increasing the gap between the ground state and the first excited state.[44–46] There are many compounds that fall into the "diamagnetic TM near the Ln center" category. For instance, the pentagonal-bipyramid (quasi-$D_{5h}$) [Zn–Dy–Zn] complex (Supplementary Fig. 8a) exhibits a large thermally activated barrier with long relaxation times.[46] Other cases are {[Zn(Me$_2$valpn)]$_2$Dy(H$_2$O)Cr(CN)$_6$}$_2$ (Supplementary Fig. 8b)[41] and [Zn($\mu$-L)($\mu$-OAc)Er(NO$_3$)$_2$] (Supplementary Fig. 8c)[47] , both exhibit SIM behaviour.

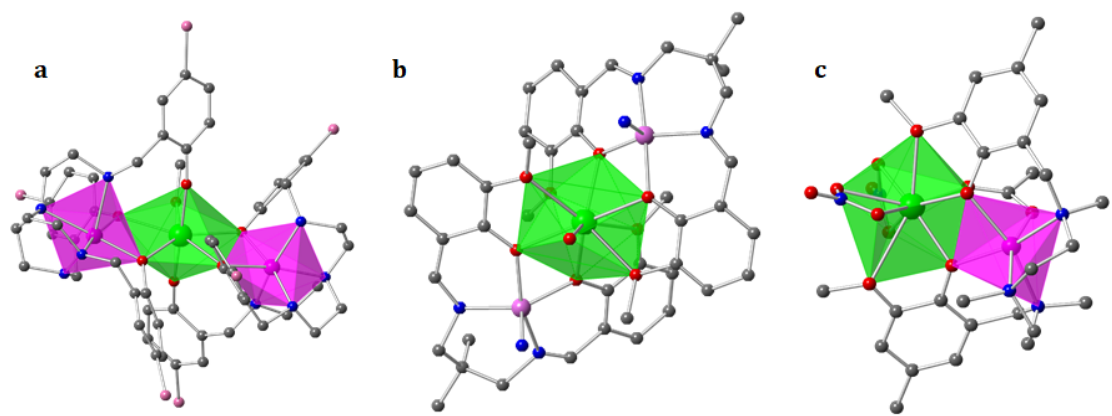

**Supplementary Figure 8 | Combined polyhedral and ball-and-stick representations of the coordination spheres around Ln ions of examples from the TM near Ln family. a**, [Zn$_2$DyL$_2$(MeOH)]$^-$, where L is 2,2',2''-(((nitrilotris(ethane-2,1-diyl))tris(azanediyl))tris(methylene))tris-(4-bromophenol).[46] **b**, {[Zn(Me$_2$valpn)]$_2$Dy(H$_2$O)Cr(CN)$_6$}$_2$, where Me$_2$valpn$^{2-}$ is dianion of N,N'-2,2-dimethylpropylenebis(3-methoxysalicylideneimine).[41] **c**, [Zn($\mu$-L)($\mu$-OAc)Er(NO$_3$)$_2$], where H$_2$L is N,N',N''-trimethyl-N,N''-bis(2-hydroxy-3-methoxy-5-methylbenzyl)diethylenetriamine.[47] (Color code: grey sphere, C; green sphere and polyhedron, Ln; magenta sphere and polyhedron, Zn; red sphere: O; pink sphere, Br; blue sphere: N.)

## 2.8. Mixed ligands family

The eighth category is defined as mixed ligands. It contains all cases where the Ln$^{3+}$ ion is coordinated with one kind of ligands defined above together with another ligand not defined, thus, mixed ligands. The design strategy of using mixed ligands for high performance SIMs is promising. There are many complexes from this category which possess SIM behaviour. One example is using N,N'-bis(2-hydroxybenzyl)-N,N'-bis(2-methylpyridyl)ethylenediamine and Cl (or Br) as ligands for synthesis of the seven-coordinate complex [Dy(bbpen-CH$_3$)X], which produces high performance SIMs (Supplementary Fig. 9a).[48] Another representative case is the half-sandwich organometallic complex [Cp*Dy(DBM)$_2$(THF)] (Supplementary Fig. 9b) with a Janus structural motif, where the ligands are composed of THF, DBM$^-$ and [Cp*]$^-$. It displays slow magnetic relaxation in the absence of an applied magnetic field, indicating SIMs properties.[49] By combination of $\beta$-diketonate with 6-pyridin-2-yl-[1,3,5]triazine-2,4-diamine ligands, a series of SIMs were obtained and investigated (Supplementary Fig. 9c).[50]

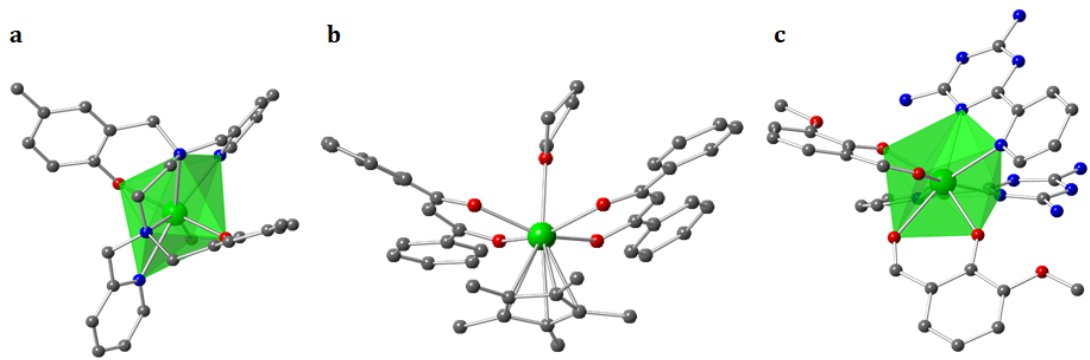

**Supplementary Figure 9 | Combined polyhedral and ball-and-stick models of the coordination spheres around Ln ions of some cases from the mixed ligands family. a**, [Dy(bbpen-CH₃)X], where X = Cl or Br and H₂bbpen = N,N'-bis(2-hydroxybenzyl)-N,N'-bis(2-methylpyridyl)ethylenediamine.[48] **b**, [Cp*Dy(DBM)₂(THF)], where Cp*=C₅Me₅⁻ and DBM⁻=dibenzoylmethanoate anion.[49] **c**, [DyLz₂(o-vanilin)₂]⁺, where Lz = 6-pyridin-2-yl-[1,3,5]triazine-2,4-diamine and X = Br⁻, NO₃⁻, CF₃SO₃⁻, from reference [[50]]. (Color code: grey sphere, C; green sphere and polyhedron, Ln; red sphere, O; blue sphere, N.)

## 2.9. Other families

The last category is named as "other families". It includes all complexes which fall into the criterion of complex selection but the coordination ligands of Ln ions are not in the ligand families previously defined. Large numbers of complexes included in this work are from this category. For instance, the octahedral dysprosium aluminate complex [Dy(AlMe₄)₃] shows fast relaxation of the magnetisation via quantum tunnelling (Supplementary Fig. 10a).[51] Also, the alkoxide cage complexes [DyY₃K₂O(OᵗBu)₁₂] and [DyY₄O(OⁱPr)₁₃] (Supplementary Fig. 10b) incorporate a small amount of DyCl₃ in the synthesis of [Dy₄K₂O(OᵗBu)₁₂] and [Dy₅O(OⁱPr)₁₃] to produce {DyY₃K₂} in a {Y₄K₂} matrix, or {DyY₄} in {Y₅}. These complexes show a single dominant relaxation process with very high $U_{eff}$ values.[52] Another relevant cases are the five-coordinate complexes Ln(NHPhⁱPr₂)₃(THF)₂, (Ln = Dy and Er), with trigonal bipyramidal geometry, both of which exhibit slow magnetic relaxation under a zero/non-zero dc applied magnetic field (Supplementary Fig. 10c).[53]

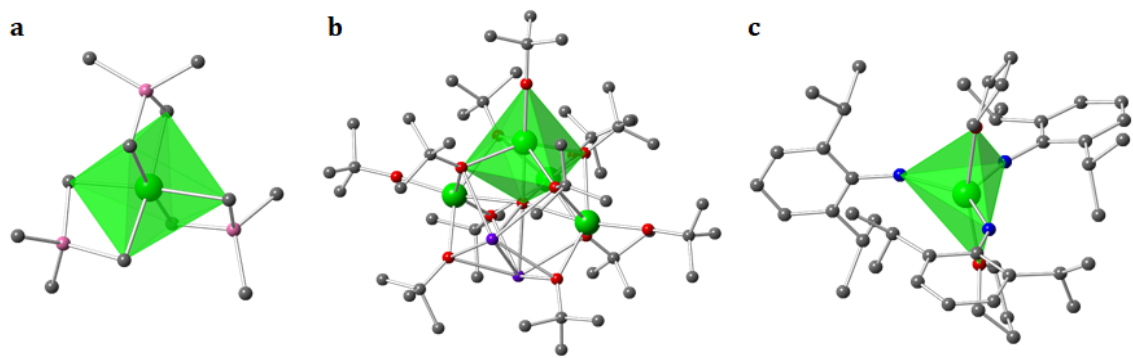

**Supplementary Figure 10 | Combined polyhedral and ball-and-stick models of the coordination spheres around Ln ions of three examples from the other families. a**, [Dy(AlMe$_4$)$_3$] from reference [51]. **b**, [DyY$_3$K$_2$O(O$^t$Bu)$_{12}$] from reference [52]. **c**, Ln(NHPh$^i$Pr$_2$)$_3$(THF)$_2$, in which Ln$^{3+}$ can be Dy$^{3+}$ or Er$^{3+}$, from reference [53]. (Color code: grey sphere, C; green sphere and polyhedron, Ln or Y$^{3+}$; red sphere, O; blue sphere, N; pink sphere, Al.)

## Supplementary Section 3. A graphical, interactive, browsable App

To facilitate a broader use by the chemical community of the data collected in the present study, we developed the tool SIMDAVIS (Single Ion Magnet DAta VISualization): a graphical, interactive, browsable online database of over 1400 samples. Employing SIMDAVIS, any user can study the data in different and complementary ways. The four modes of operation, accessible in different tabs within the program, are "ScatterPlots", "BoxPlots", "BarCharts" and "Data" table. There is also an information subtab denoted as "Variables" within the "About SIMDAVIS" tab in which the definition of each variable can be found.

The basic use of the "ScatterPlots" tab is the representation of quantitative data against each other, *e.g.* the maximum hysteresis temperature ($T_{hyst}$) *vs* the effective energy barrier ($U_{eff}$). This allows a visual estimate on the relation between different experimental and theoretical descriptors of the magnetic behaviour. Other relevant numerical variables in the dataset include $T_{B3}$, $T_{B3H}$, the alternate estimate for the effective energy barrier ($U_{eff,ff}$), or the pre-exponential factors $\tau_0$, $\tau_{0,ff}$, for either the simplistic equation or the full fit (see details about the variables in Supplementary Section 1). Furthermore, the "ScatterPlots" tab allows to distinguish the data points plotted according to a number of qualitative (categorical) variables, which can be of chemical nature, such as the chemical family, or which lanthanide ion was employed. Also, you may distinguish the points by some categorical variables of physical nature, such as presence or absence of magnetic memory above 2 K, in form of hysteresis or maximum in the $\chi''$ (categorical variable $\chi''_{max}$ in our dataset). It also allows the user to select or deselect the represented data depending on these qualitative variables, to help distinguish quantitative correlations that might be different for different classes of compounds. Finally, there is also an option to fit linear regressions between the two represented quantitative variables for each of the categorical classes. These variations can combine to hundreds of thousands of distinct meaningful plots.

The "BoxPlot" tab allows a different type of representation. One can plot the values of any of the quantitative variables *vs* any of the qualitative variables, for a total of 108 possible variable pairs producing distinct representations. The distribution of a single quantitative variable (*e.g.* $U_{eff}$) is represented, showing the data points, the median, the low (first or Q1) quartile, upper (third or Q3) quartile, and whiskers. The upper whisker extends from the hinge to the largest value no further than 1.5xIQR from the hinge (where IQR is the interquartile range, or distance between the first and third quartiles). The lower whisker extends from the hinge to the smallest value at most 1.5xIQR of the hinge. This representation is done in parallel for different values of a qualitative variable, *e.g.* "Coordination Elements". An advantage of these boxplots *vs* the more sophisticated scatterplots is a larger amount of data to be represented at any given time. Note that there is virtually no paper that contains simultaneously all the kinds of information recorded in the dataset. For example, only a minority of the papers have historically performed a full fit considering Orbach, Raman, quantum tunneling and/or direct mechanism of relaxation. This means that the scatter plots, by being restricted to samples where two particular quantitative data kinds are well defined, effectively work with less data, so while they enable us to extract more nuanced dependencies, inevitably some information is lost.

In the "BarCharts" tab, the different qualitative data types can be represented vs each other, for a total of 144 variable pairs producing distinct representations. Since qualitative information is available for almost all samples, bar charts contain almost all data points and allow for a quick frequency check of frequencies of different values in the dataset. Again, they provide a complementary mode of analysis of correlations. In our case, rather than the standard bar chart, SIMDAVIS employs stacked bar graphs meaning we can analyze the covariation of two variables, *e.g.* $Tb^{3+}$ is more common in the SIM literature than $Er^{3+}$, but whereas this is especially true for the chemical families of "$LnPc_2$" and "radicals", the reverse trend is found for the metallocene family.

The "Data" tab contains a mini-menu with two options: it allows the user to download the raw dataset, and it allows the user to browse the data set. The browsing is interactive in different, complementary ways. First, it allows the user to select the columns to show, *e.g.* by default each entry just shows 7 columns of data, namely the sample ID, formula of the compound, its chemical family, the Ln ion, the coordination elements, $U_{eff}$, and the DOI of the article where the information was obtained, while the other 24 columns are hidden. Second, it allows the user to arrange the information by ascending or descending order of the chosen variable. Finally, it includes a search tool that filters for text strings in real time.

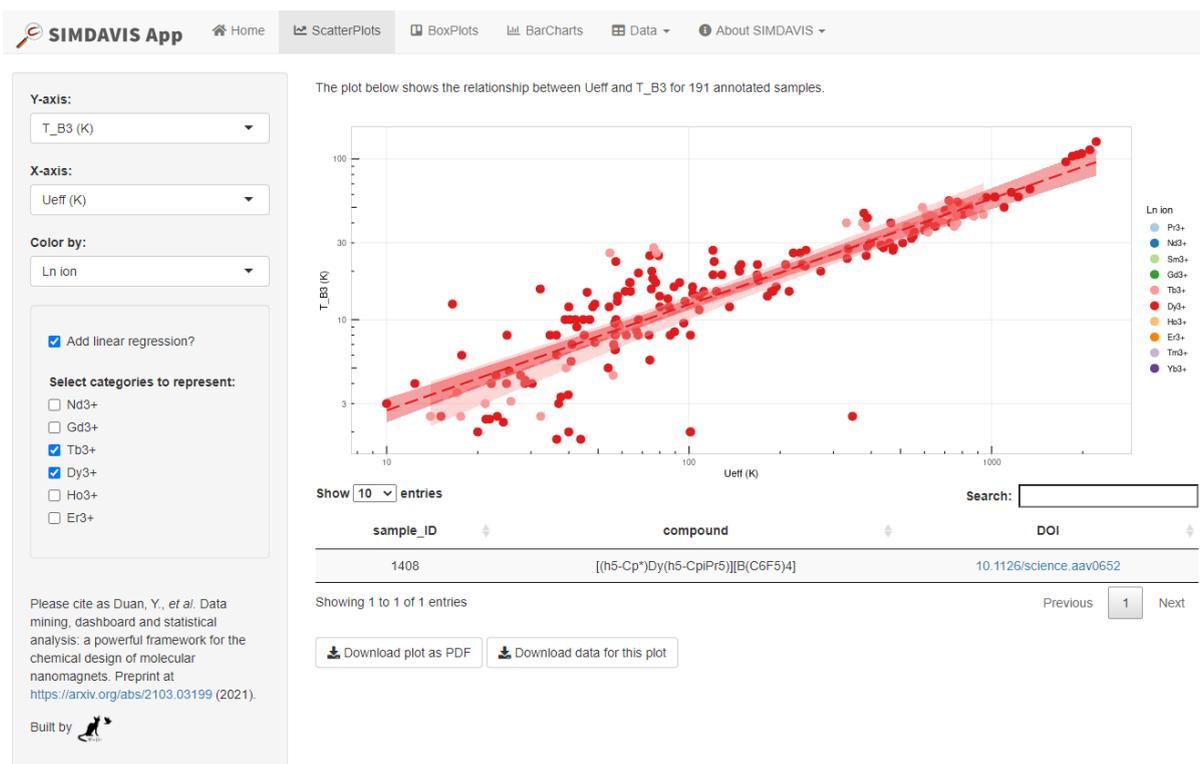

**Figure 11 | Screenshot of the SIMDAVIS dashboard.** It shows an example plot employing the "ScatterPlots" tab, where the users can represent 9 quantitative physical properties versus another in logarithmic scales, as well as a chemical qualitative variable from a dropdown menu, which contains 12 qualitative categorization possibilities; each data point is identified by a colour corresponding to its category. This permits the interactive exploration of hundreds of potential magnetostructural correlations between chemical variables, measured experimental values and parameters fitted from physical measurements. In the example, $T_{B3}$ *vs* $U_{eff}$ presented. Checkboxes were used to add a linear regression for each category and to hide all metal ions except for $Tb^{3+}$ and $Dy^{3+}$. This visual estimate on the relation between descriptors of the magnetic behaviour may uncover trends for specific qualitative variables. At any time, the chosen plots can be downloaded as vectorial PDF files. In the example, the data point with the highest $T_{B3}$ was clicked to display its sample ID, compound name and DOI linking to the article, facilitating further analysis.

## 3.1. Gallery of graphs: chemical variables to optimise the physical properties

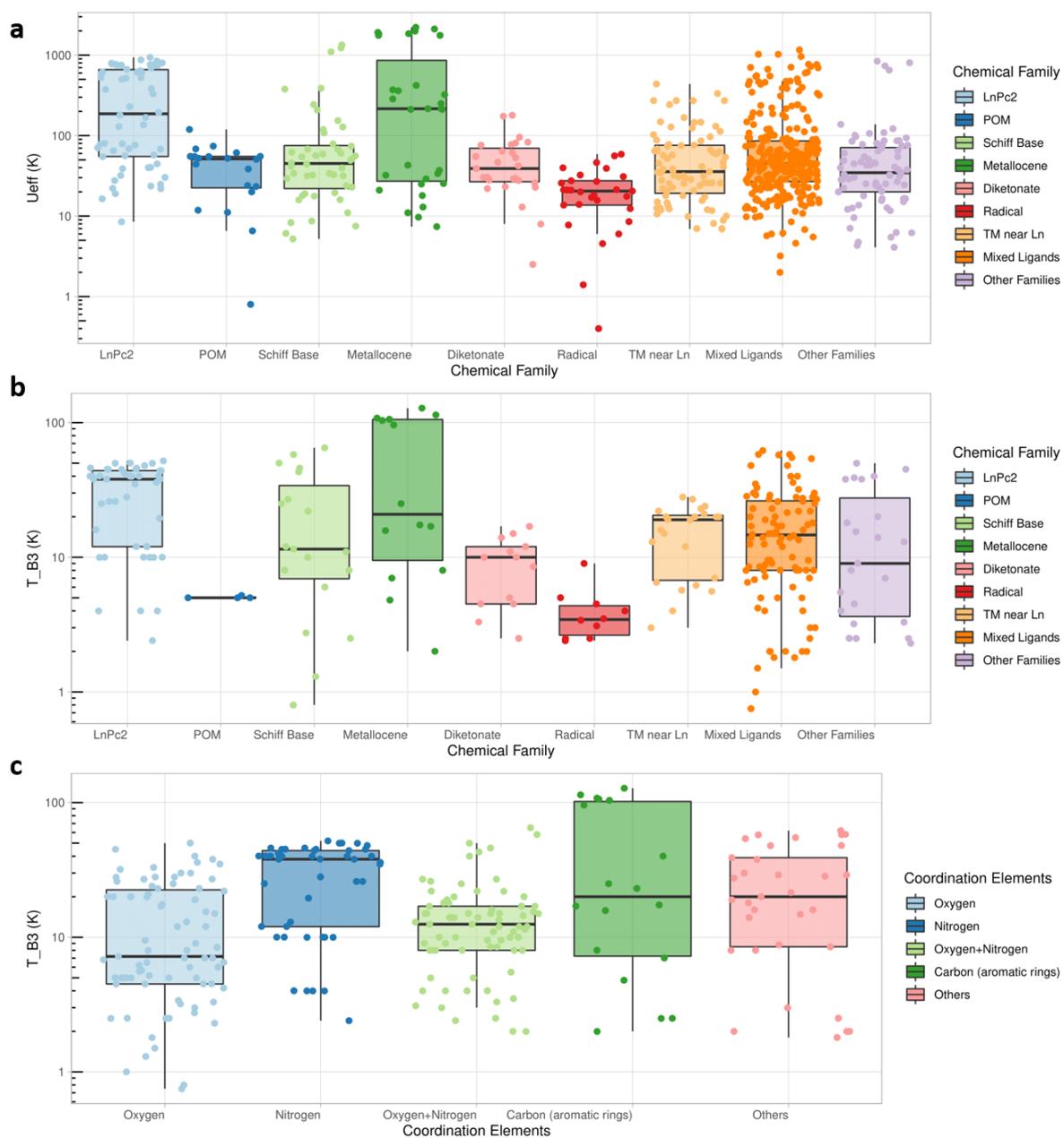

**Supplementary Figure 11.1 | Boxplots of physical variables *vs* chemical variables. a**, $U_{\text{eff}}$ *vs* chemical family. **b**, $T_{\text{B3}}$ *vs* chemical family. **c**, $T_{\text{B3}}$ *vs* coordination elements.

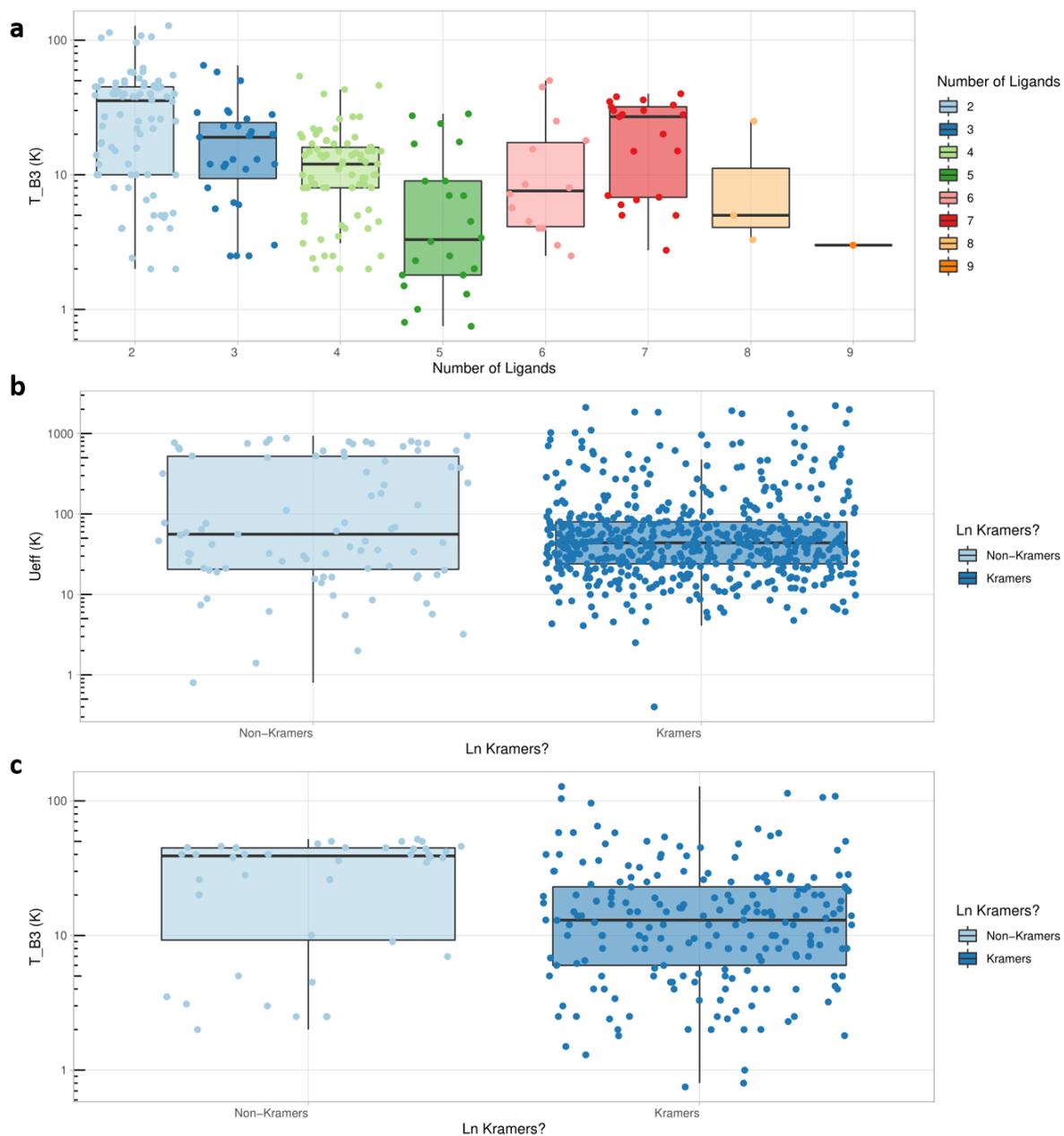

**Supplementary Figure 11.2 | Boxplots of physical *vs* chemical variables. a**, $T_{B3}$ *vs* number of ligands. **b**, $U_{eff}$ *vs* spin parity of the metal ion. **c**, $T_{B3}$ *vs* spin parity of the lanthanide ion.

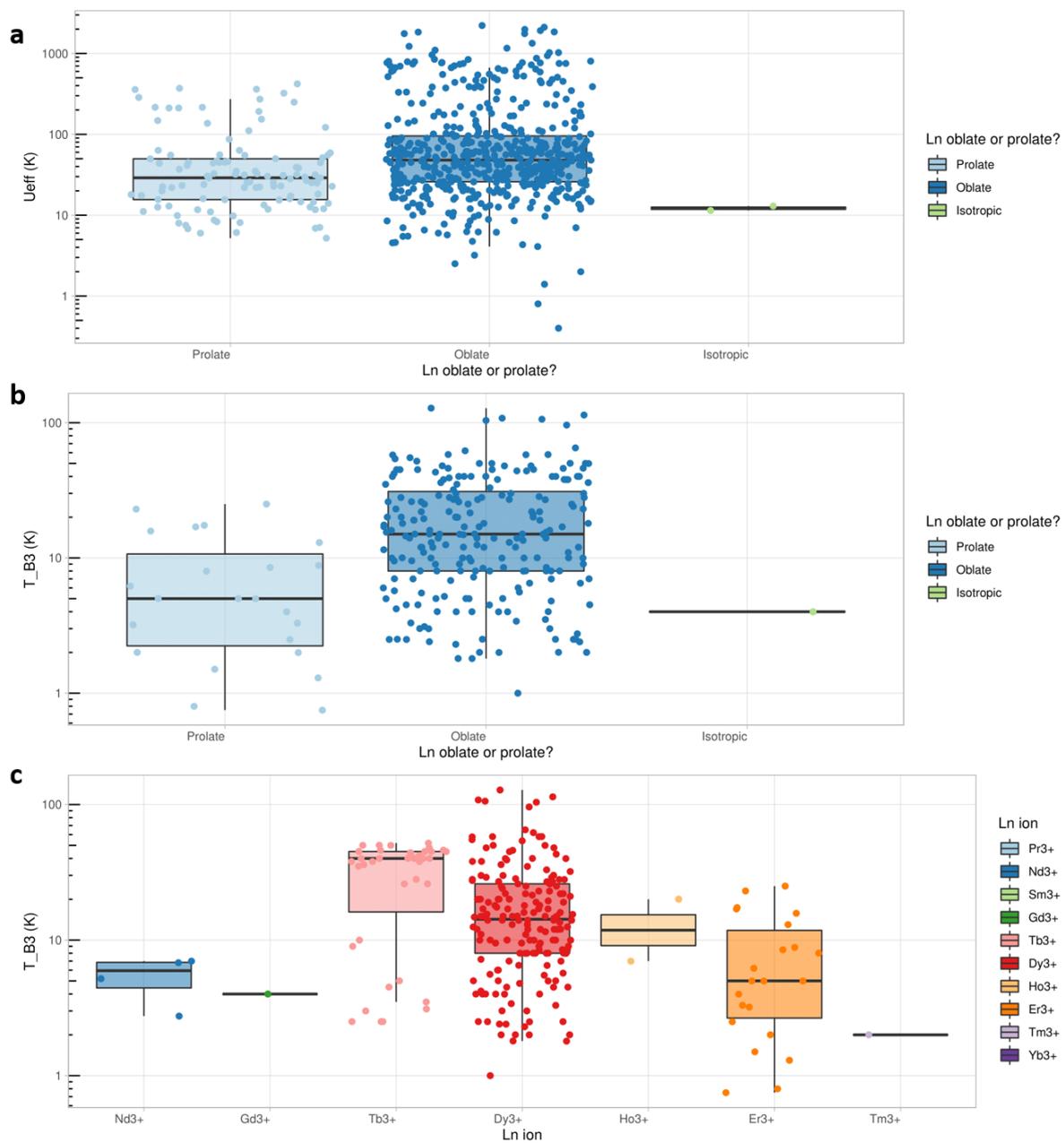

**Supplementary Figure 11.3 | Boxplots of physical vs chemical variables. a**, $U_{\text{eff}}$ *vs* anisotropy of the lanthanide ion. **b**, $T_{\text{B3}}$ *vs* anisotropy of the lanthanide ion. **c**, $U_{\text{eff}}$ *vs* lanthanide ion.

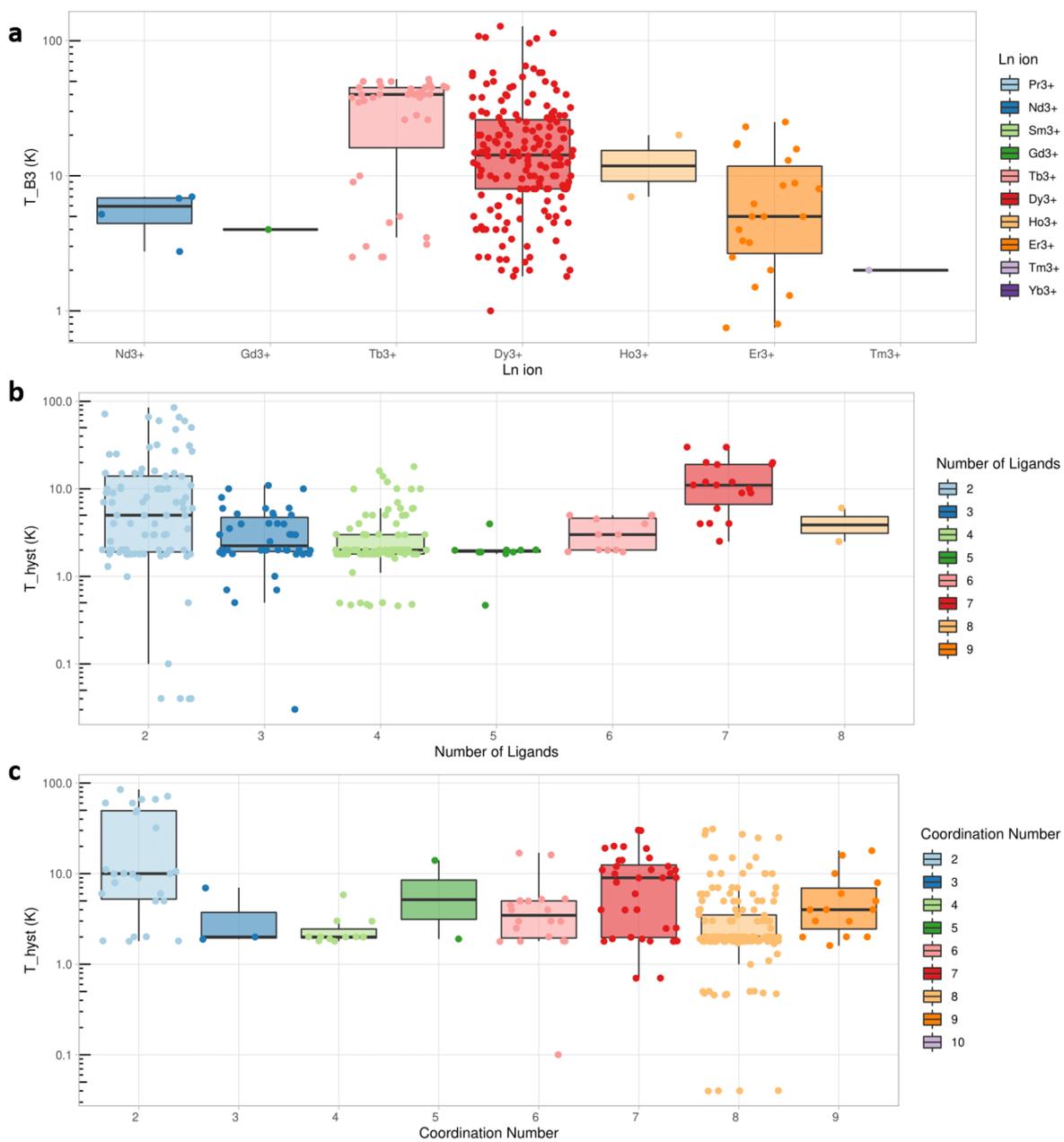

**Supplementary Figure 11.4 | Boxplots of physical vs chemical variables. a**, $T_{B3}$ *vs* lanthanide ion. **b**, $T_{hyst}$ *vs* number of ligands. **c**, $T_{hyst}$ *vs* coordination number.

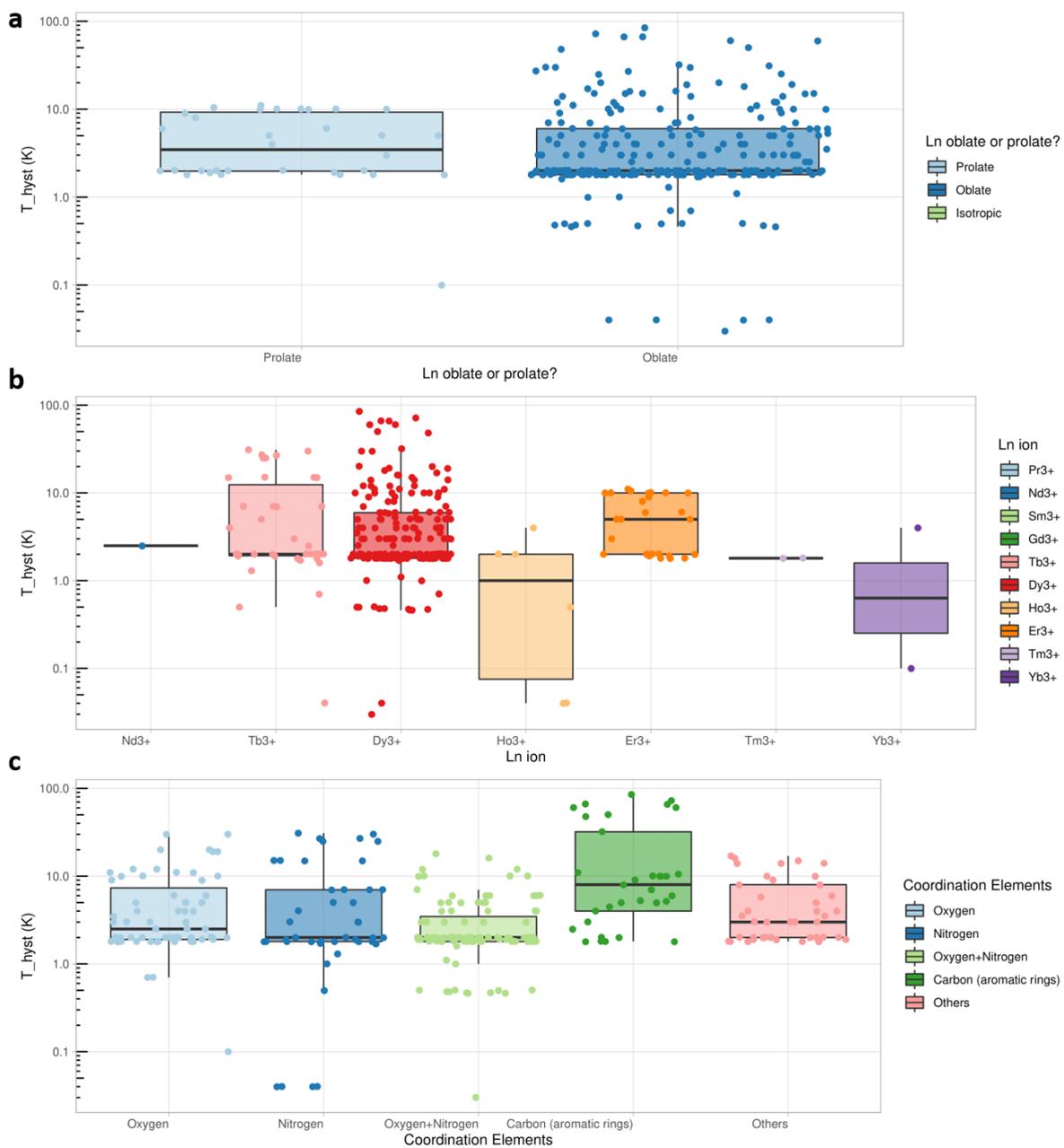

**Supplementary Figure 11.5 | Boxplots of physical vs chemical variables. a**, $T_{hyst}$ *vs* anisotropy of the lanthanide ion. **b**, $T_{hyst}$ *vs* lanthanide ion. **c**, $T_{hyst}$ *vs* coordination elements.

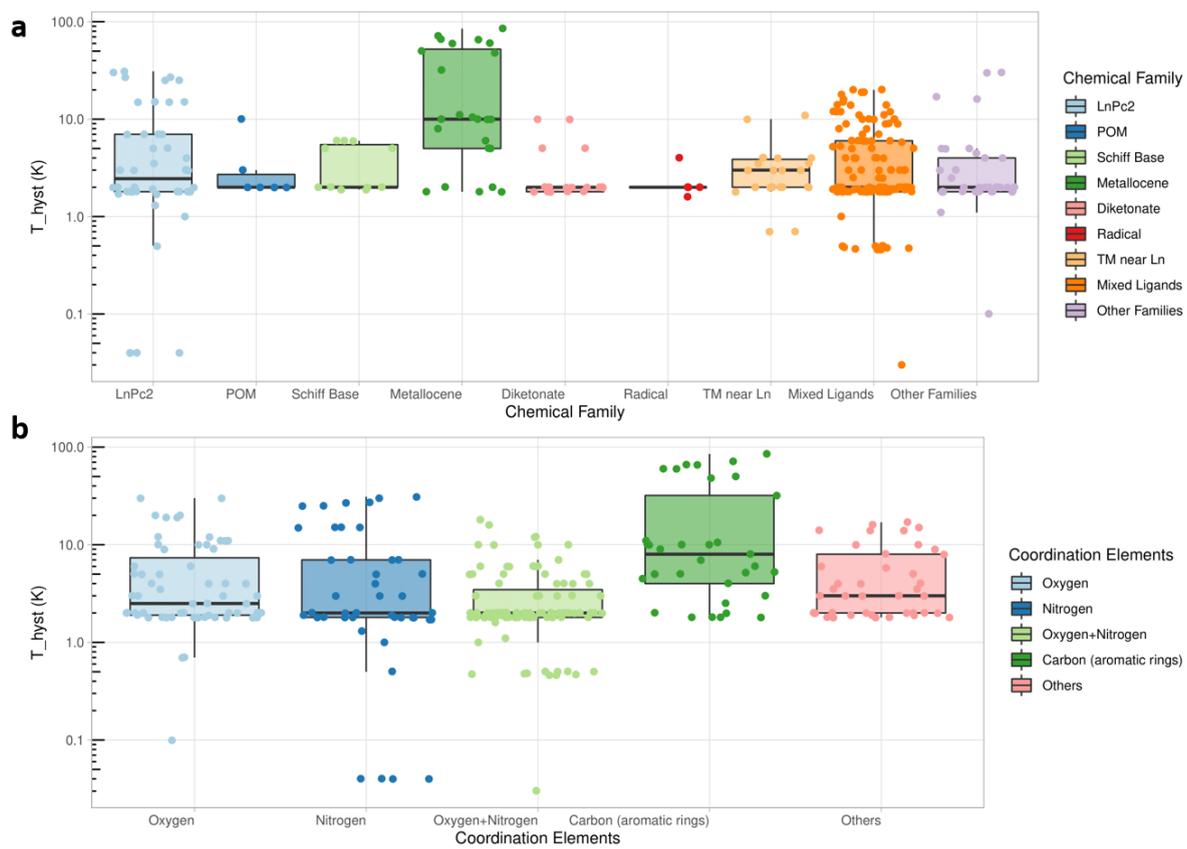

**Supplementary Figure 11.6 | Boxplots of physical vs chemical variables. a**, $T_{\text{hyst}}$ *vs* chemical family. **b**, $U_{\text{eff}}$ *vs* coordination elements.

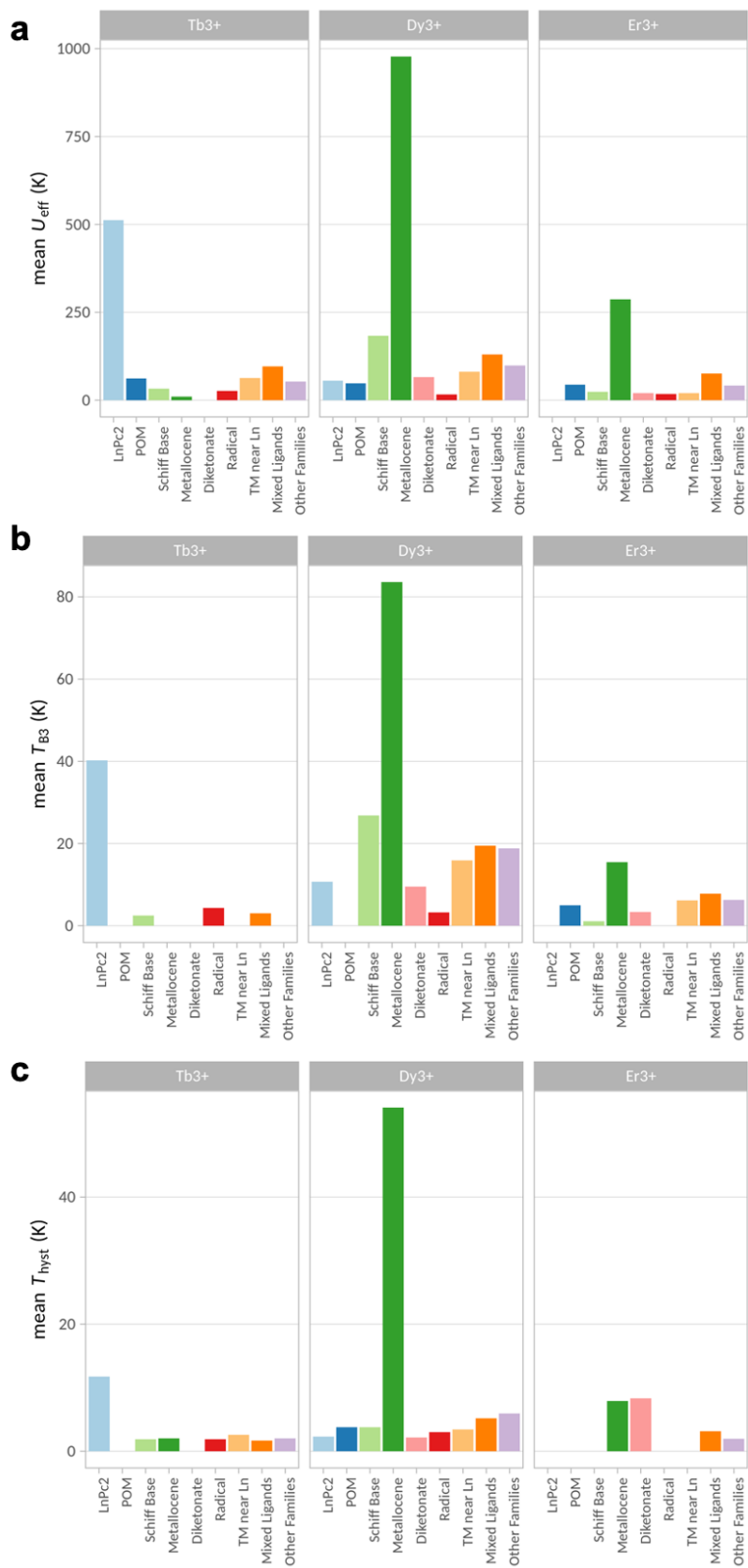

**Supplementary Figure 12.1 | Bar chart representations of the relation between Ln ion, chemical family and arithmetic mean of $U_{eff}$, $T_{B3}$ and $T_{hyst}$.** Bar charts showing the mean

values for every combination of categories between the main metal ions {$Tb^{3+}$, $Dy^{3+}$, $Er^{3+}$} and all chemical categories. **a**, $U_{eff}$, **b**, $T_{B3}$, **c**, $T_{hyst}$.

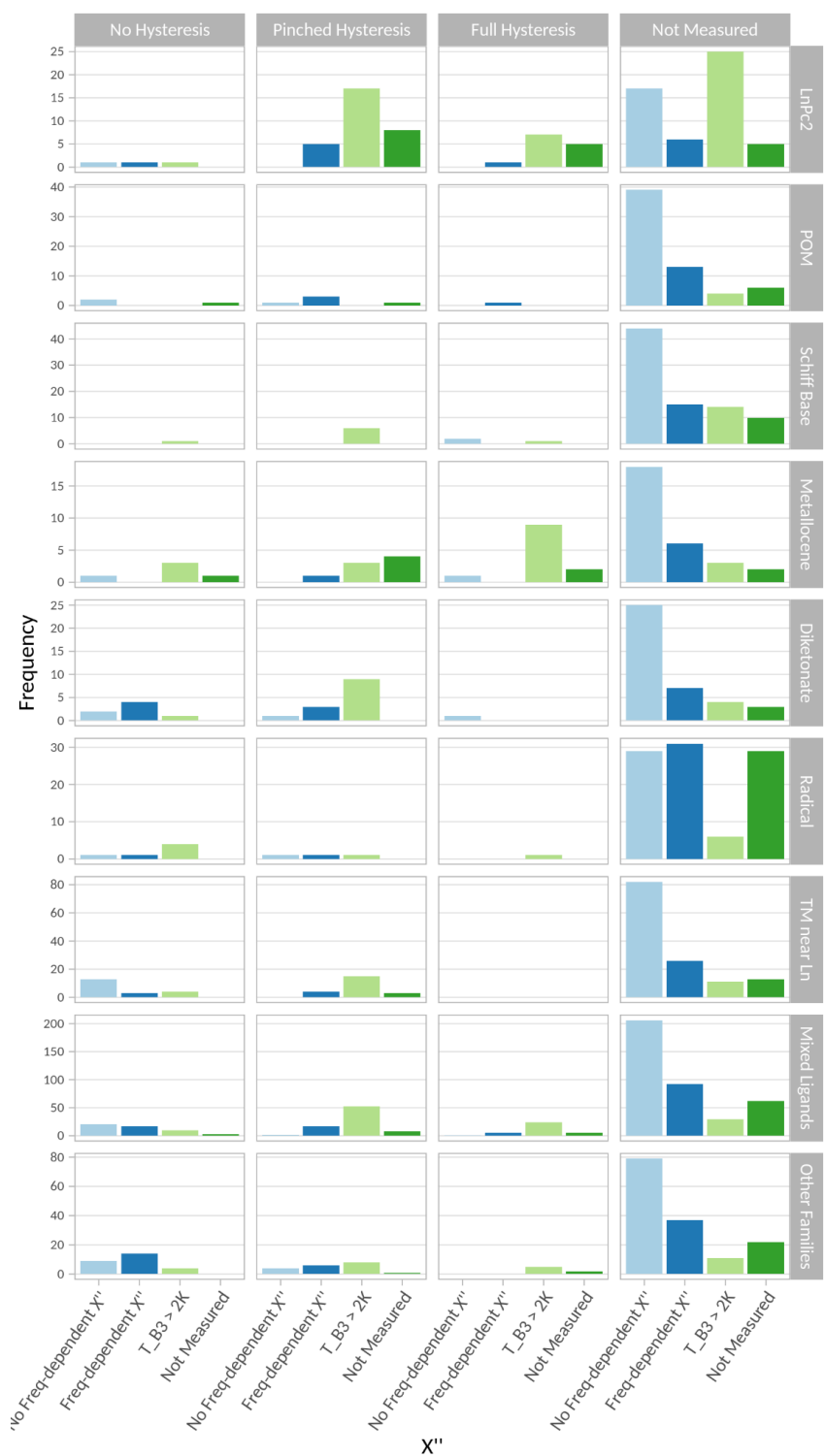

**Supplementary Figure 12.2 | Bar chart representations of the relation between chemical family and magnetization dynamics.** Bar charts showing the frequency of samples for every combination of categories between the categorical variables "chemical family", "Hyst" and "$\chi''_{max}$". Graphs are normalised to the maximum frequency in each chemical family.

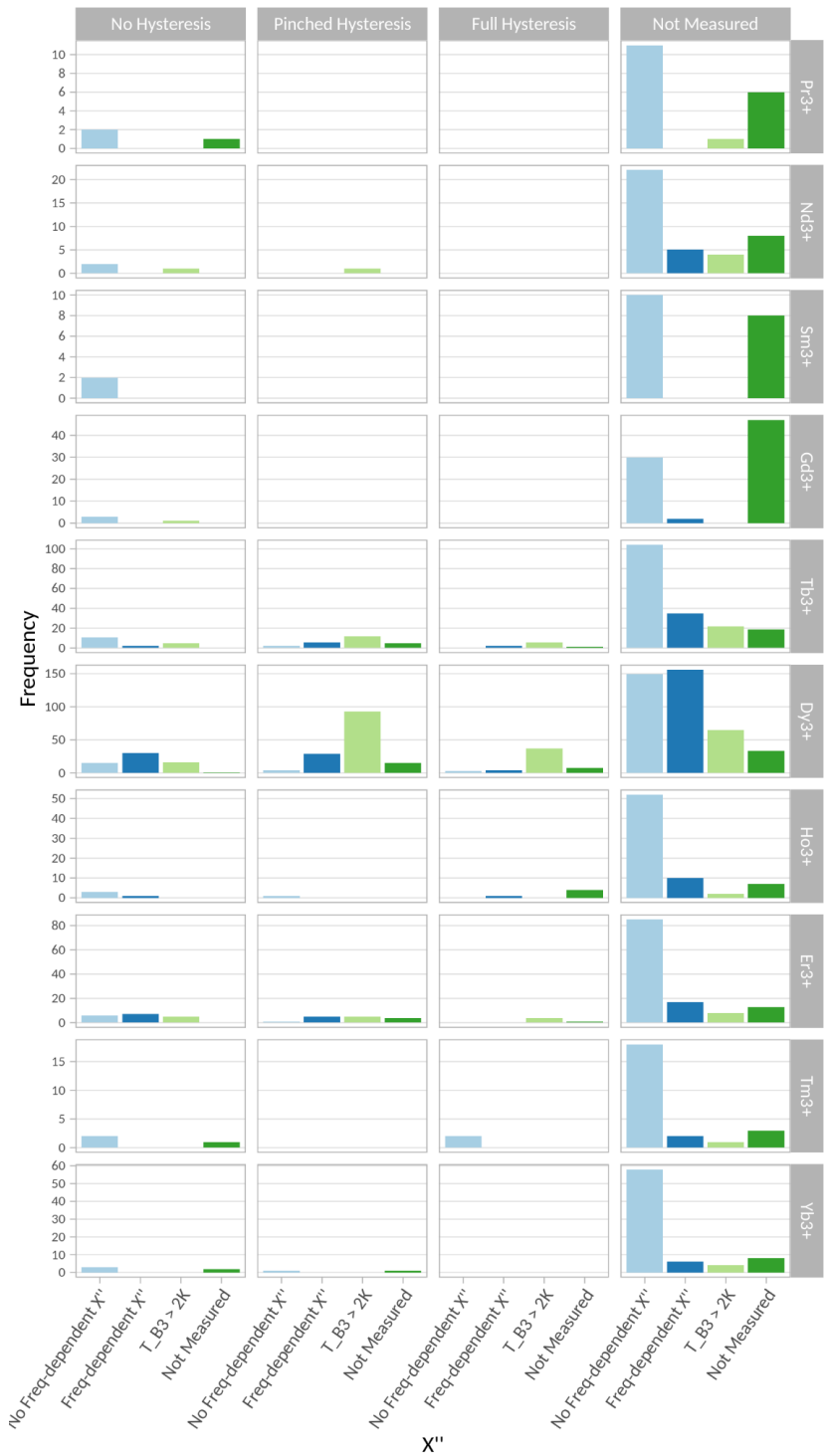

**Supplementary Figure 12.3 | Bar chart representations of the relation between lanthanoid ion and magnetization dynamics.** Bar charts showing the frequency of samples for every combination of categories between the categorical variables "lanthanoid ion", "Hyst" and "$\chi''_{max}$". Graphs are normalised to the maximum frequency in each chemical family.

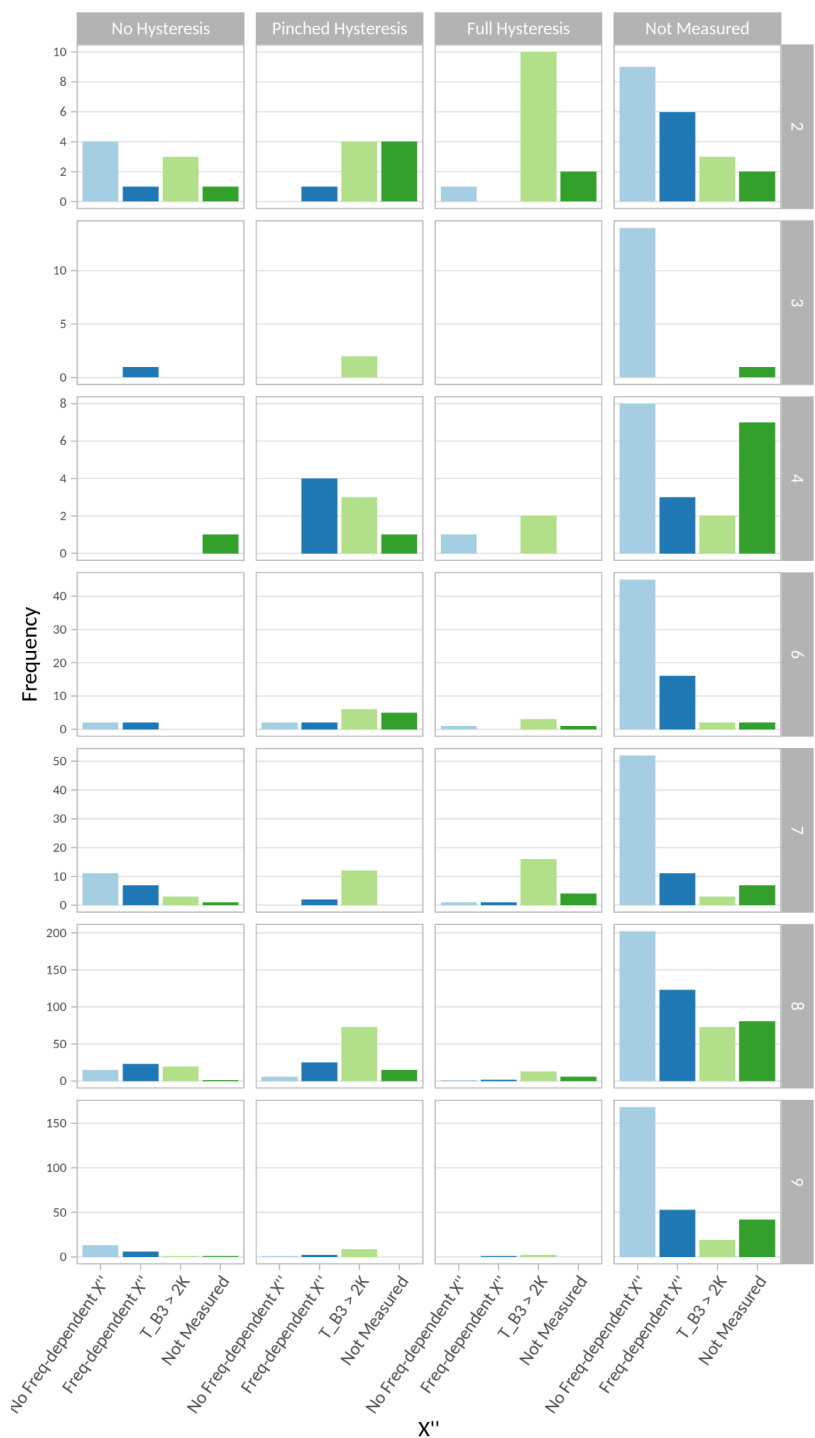

**Supplementary Figure 12.4 | Bar chart representations of the relation between coordination number and magnetization dynamics.** Bar charts showing the frequency of samples for every combination of categories between the categorical variables "coordination number", "Hyst" and "$\chi''_{max}$". Graphs are normalised to the maximum frequency in each chemical family.

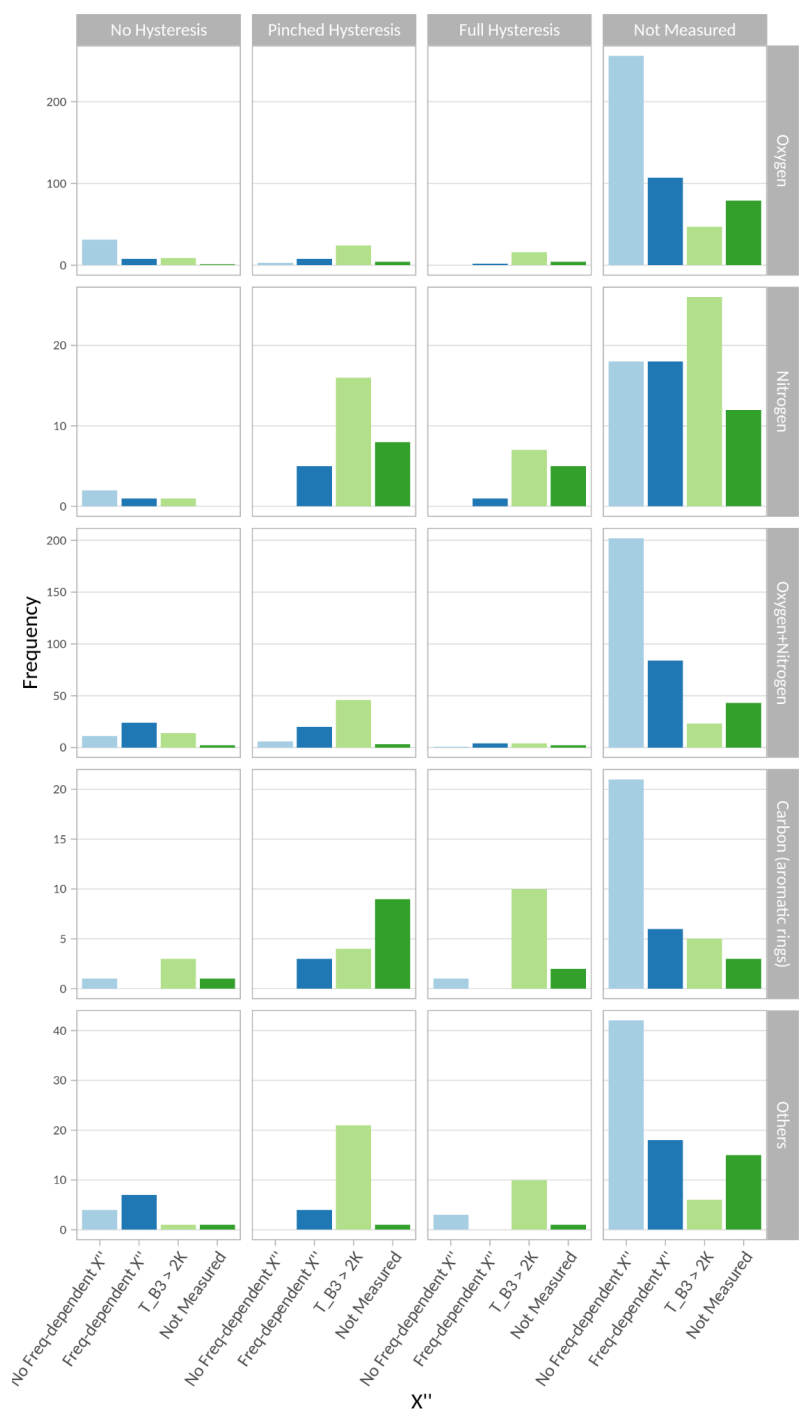

**Supplementary Figure 12.5 | Bar chart representations of the relation between coordination elements and magnetization dynamics.** Bar charts showing the frequency of samples for every combination of categories between the categorical variables "coordination elements", "Hyst" and "$\chi''_{max}$". Graphs are normalised to the maximum frequency in each chemical family.

## 3.2. Extended gallery of SIMDAVIS graphs: Arrhenius equation parameters

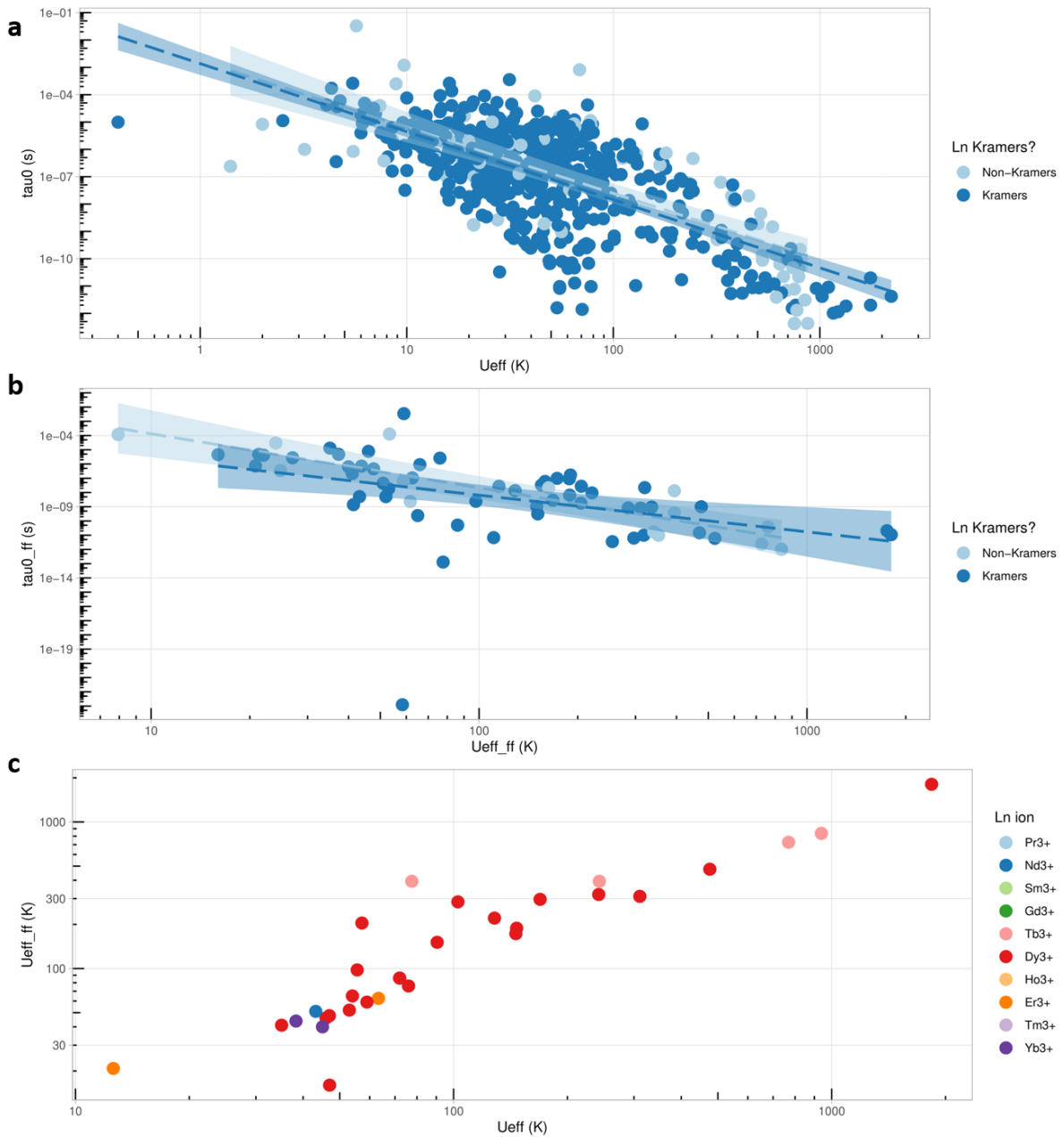

**Supplementary Figure 13.1 Scatterplot representations of the relation between $U_{eff}$, $U_{eff,ff}$, $\tau_0$, $\tau_{0,ff}$.** **a**, $\tau_0$ vs $U_{eff}$, or Kramers and non-Kramers ions, with linear regressions; **b**, $\tau_{0,ff}$ vs $U_{eff,ff}$, for Kramers and non-Kramers ions, with linear regressions; **c**, $U_{eff,ff}$ vs $U_{eff}$ for different lanthanide ions, with a visible linear behaviour.

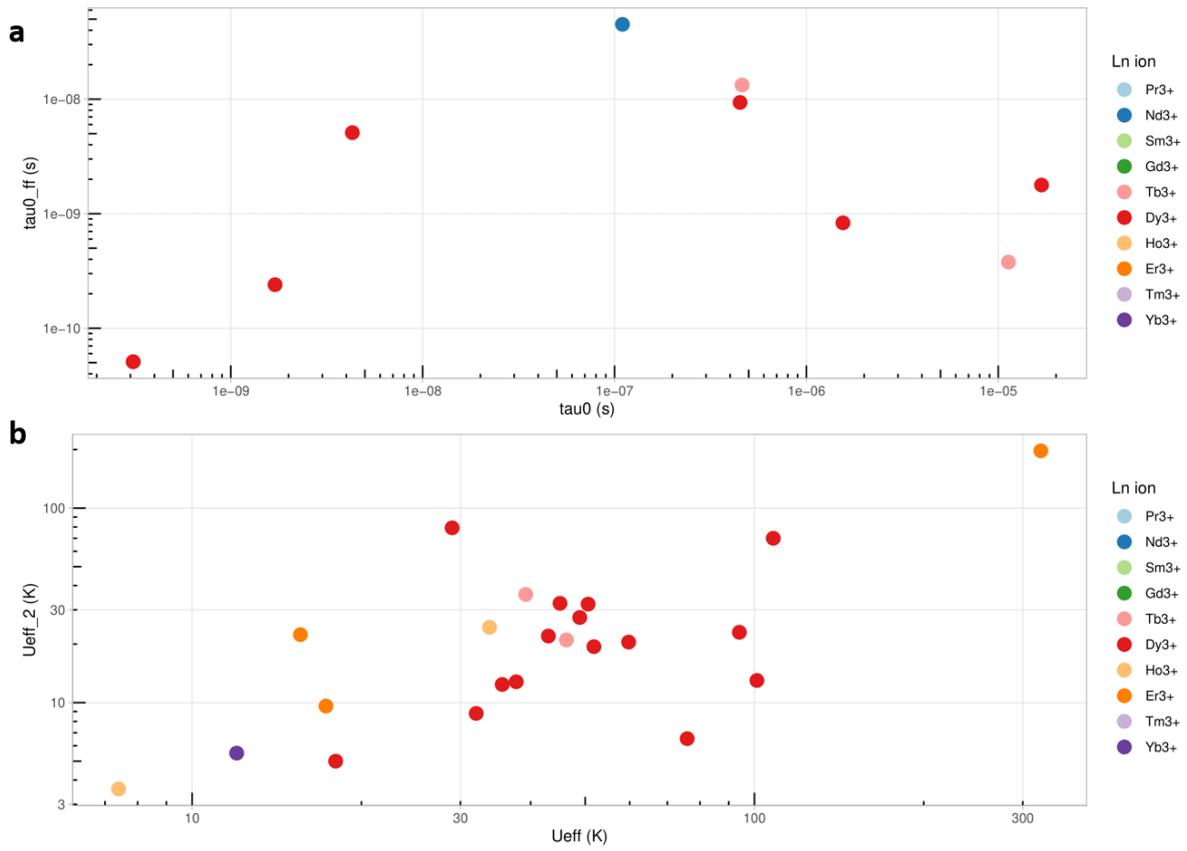

**Supplementary Figure 13.2 | Scatterplot representations of $\tau_{0,ff}$, $U_{eff,2}$ vs $\tau_0$, $U_{eff,ff}$. a,**: $\tau_{0,ff}$ *vs* $\tau_0$ for different lanthanide ions, with no discernible relation between the parameters; **b**, $U_{eff,2}$ *vs* $U_{eff}$ for different lanthanide ions, with no discernible relation between the parameters.

**Supplementary Section 4. Statistical analysis of the chemical variables**

All qualitative results presented in this section considered the whole dataset containing data from literature from 2003 to 2019 for the analyses of chemical variables that follow. Additionally, we repeated the study with the data subset in the timeframe 2003-2017 (~1000 samples instead of ~1400), and the conclusions were robust, with no difference resulting from whether one considers the whole data set from 2003-2019 or only 2003-2017 subset. All quantitative numbers given herein are also consistent between the two studies, within a 5% difference. We can therefore conclude that the relations among the chemical variables are stable, *i.e.* no new trend has been revealed since 2017.

**4.1. Initial multiple correspondence analysis**

Here we are striving to determine the existing statistical correlations among the chemical variables in the studied sample. This is necessary in order to avoid being misled later on by meaningless correlations between chemical design variables and physical behaviour. For example, we find that the variables "number of ligands" and "coordination elements" happen to be strongly correlated with each other, then it is likely that they will both display the same correlations with a given physical behaviour. In particular, one can expect that an "all-nitrogen coordination environment" will be strongly correlated with "number of ligands = 2", and with "chemical family = LnPc$_2$", and relatively few other complexes in the dataset present only nitrogens as donor atoms. Therefore, if one obtains a correlation between a desirable physical behaviour and "chemical family = LnPc$_2$", it would be unwarranted to deduce that this behaviour can be obtained solely by employing an all-nitrogen coordination environment, or solely by preparing complexes with 2 ligands.

Correspondence Analysis (CA) or reciprocal averaging is a multivariate statistical technique that is employed for the graphical analysis of the dependence or independence of a set of categorical variables from data in a contingency table. It consists in summing up the information in the rows and columns so that it can be projected on a reduced subspace, and represent simultaneously the row and the column data, allowing to obtain conclusions about each pair of variables.

CA only requires data to be organised in categories. Since in our case there are more than two variables, we employed Multiple Correspondence Analysis (MCA). Different approaches for MCA have been proposed; we employed the widely used Gifi system.[54] This system consists of a set of multivariate methods developed around the Alternating Least Squares (ALS) algorithm. Among these methods, Homogeneity Analysis provides a model that is equivalent to MCA. ALS's solution for Homogeneity Analysis is known as HOMALS. We employed the R homals package to obtain the following graphical representations.[55] Results are plotted in Supplementary Figs. 14, 15 and 16.

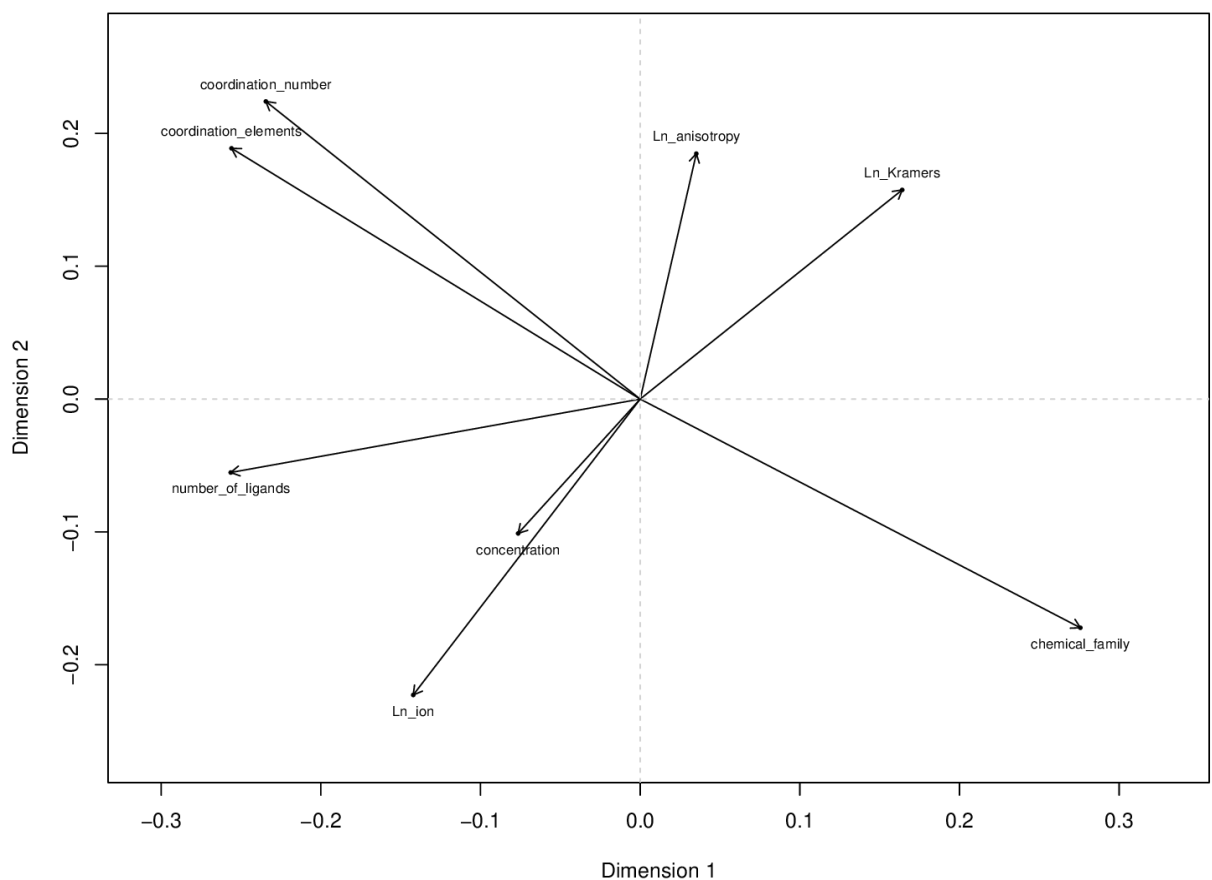

**Supplementary Figure 14 | Multiple Correspondence Analysis: minimal representation of each of the variable loadings on the two main dimensions.**

The graph in Supplementary Fig. 14 is read as follows: (i) the length of the vector approximates the variation within each variable, (ii) the cosine between two vectors approximates the correlation between two variables, *i.e.* parallel vectors correspond to perfectly correlated variables, (iii) the distance between the endpoints of two vectors approximates the dissimilarity between the two variables, (iv) the projection of each vector allows to order the data points for that variable. These two MCA dimensions will be employed to understand further analysis, in particular clustering studies. Supplementary Figures 15 and 16 are complementary to Supplementary Fig. 14, and allow for a more complete understanding.

We employed the R ade4 package[56] for MCA, which only allows the analysis of categorical variables, and obtained the results collected in Supplementary Table 1 for the correlation ratio of each variable.

**Supplementary Table 1 | Correlation ratio of each chemical variable with the two MCA dimensions.**

|  | RS1 | RS2 |
|---|---|---|
| chemical_family | 0.81096423 | 0.6147058 |
| Ln_ion | 0.20607706 | 0.5364464 |
| Ln_anisotropy | 0.05097696 | 0.3121107 |
| Ln_Kramers | 0.16243450 | 0.2918366 |
| coordination_number | 0.62057925 | 0.3528815 |
| number_of_ligands | 0.47815893 | 0.1020615 |
| coordination_elements | 0.73301849 | 0.5353323 |

To achieve some clarity in the following data-rich representations, we assigned numerical labels to all categorical values as indicated in Supplementary Section 1. Employing these labels and within the same package we obtained the following complementary boxplot representation (Supplementary Fig. 15) that allows us to see the distribution of values, for each variable, along the axis defined by the MCA dimension 1. It is easy to see that the most extreme negative values of MCA dimension 1 are displayed by samples that have carbon or nitrogen as coordination elements, 2 ligands, CN = 2 or 3, as well as those of the metallocene chemical family. As significant positive values in MCA dimension 1, one needs to highlight Gd$^{3+}$, as well as isotropic complexes which are obviously the same ones.

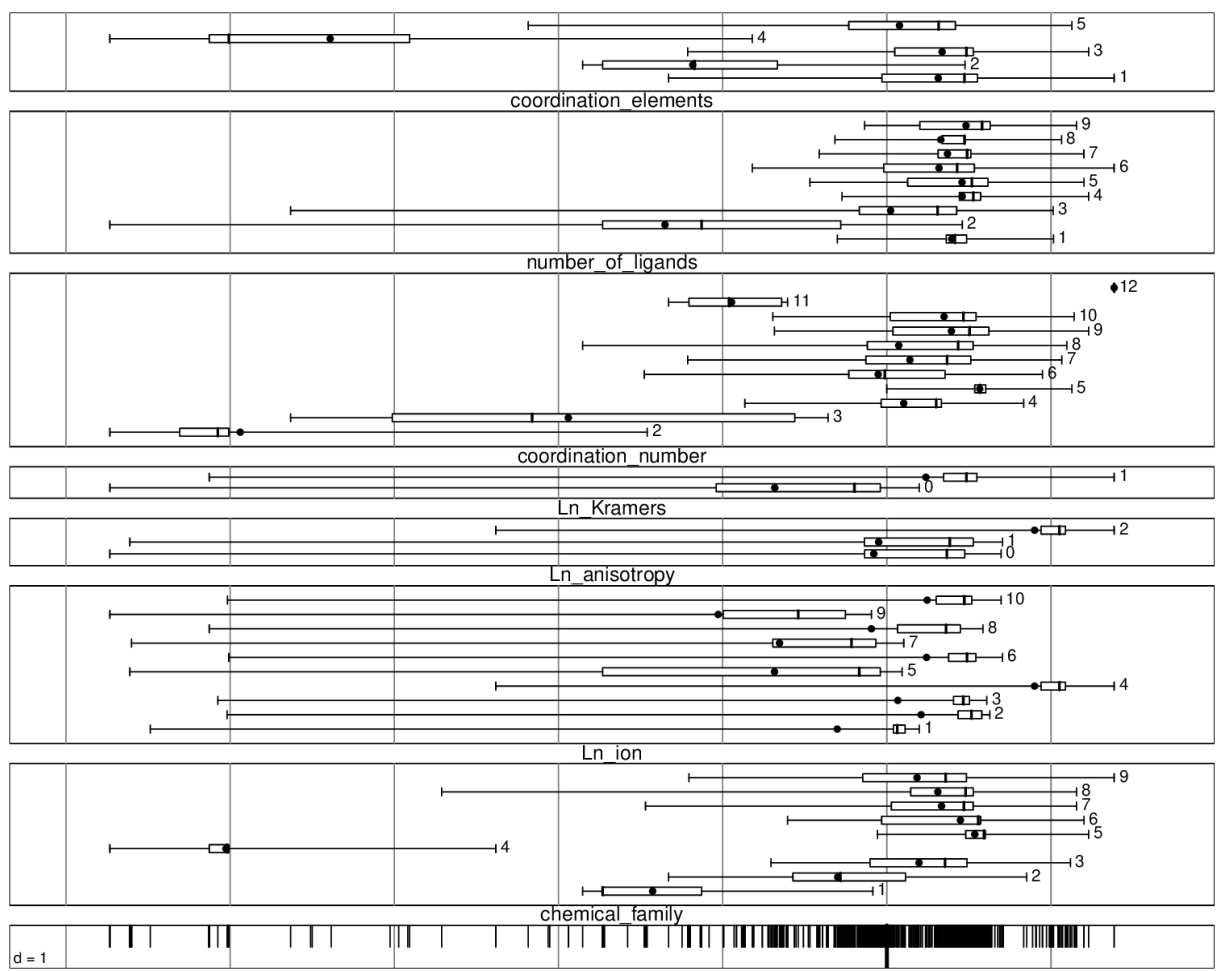

**Supplementary Figure 15 | Boxplots for the distribution along the first MCA dimension of the different categorical values for the chemical variables.** See the numbering convention for the categories of the variables in Supplementary Section 1.

A further alternative representation, depicting the distribution of values for the different variables in the dataset as subsets of points was also done (Supplementary Fig. 16). This allows to locate the different values for each chemical variable in terms of positive and negative value ranges for the MC dimensions 1 and 2 simultaneously. At the same time, it allows us to observe overlap between chemical variables, *e.g.* CN = 2 or 3 are in overlap with the metallocene family and with coordination by carbon. Similarly, there is a significant overlap between the LnPc$_2$ family, complexes with 2 ligands, Ln = Tb and a coordination sphere formed by nitrogen atoms. As we will see in Supplementary Section 4.2, while this kind of information points into a clear direction, specialised representations and statistical studies will allow us to study parameter association by clustering.

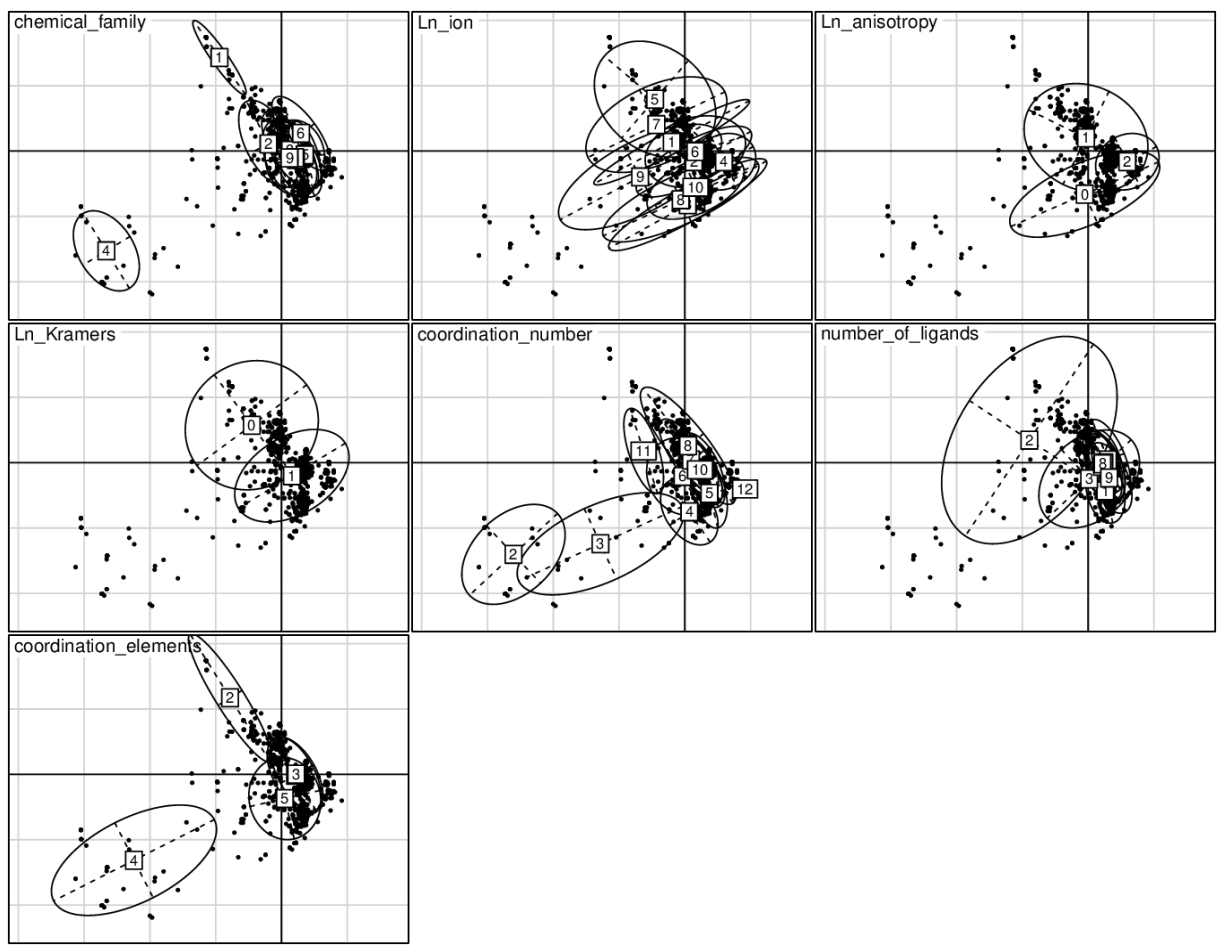

**Supplementary Figure 16 | Subsets for the different values of each of the seven categorical chemical variables in the plane of the two MCA dimensions.**

### 4.2. Clustering studies for the chemical variables

Clustering studies were performed as an independent test. In particular, we employ the package FactoMineR in R to perform the MCA.[57] This allows a graphical representation of the distances between individual samples and the relations between variables and their values. The goal in this case is to detect within the data types or profiles of data with similar characteristics. In other words, this procedure groups the values for the chemical parameter sets corresponding to individual measurements in families that present an overall similarity.

Due to the nature of our study, this is a crucial step since in practice chemical parameters are not homogeneously distributed as in a purely combinatorial approach. On the contrary, different research groups have chemical expertise in the preparation of different classes of compounds, and often, also evolving ideas on which design strategies would be more relevant for the desired physical property. This means that different research groups at different times focus on different chemical families and strategies for the molecular design of SIMs. As a result, the overall number of samples (~1400) can be judiciously divided in a small set of hierarchical clusters which share certain common features, as a kind of molecular taxonomy. Again, this will help us put in context our findings: it is to be expected that, when finding a pattern or a magneto-structural trend, we will actually be seeing the behaviour of a cluster rather than the influence of an isolated parameter. Supplementary Figs. 17 and 18 show two different perspectives on the dendrogram resulting from the clustering.

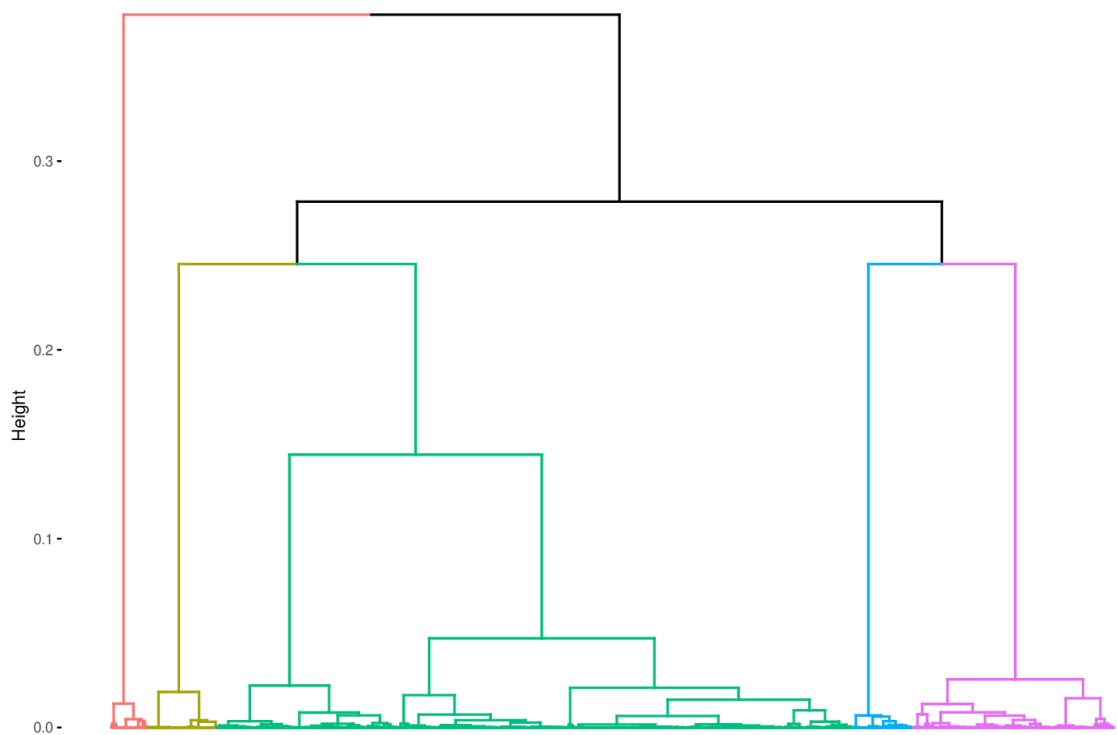

**Supplementary Figure 17 | Dendrogram visualisation of the calculated hierarchical clustering on the factor map.**

Depending on the height of the dendrogram cut in Supplementary Figures 17 and 18, one can obtain more or less fine-grained clusters. Our dataset contains a main categorisation with 5 tipologies A-E, described below, and we always offer an alternate, finer categorisation within the same hierarchical clustering (mol_cluster_2 in the dataset) with 7 tipologies A-F, which is also included in the Data tab of the the SIMDAVIS dashboard.

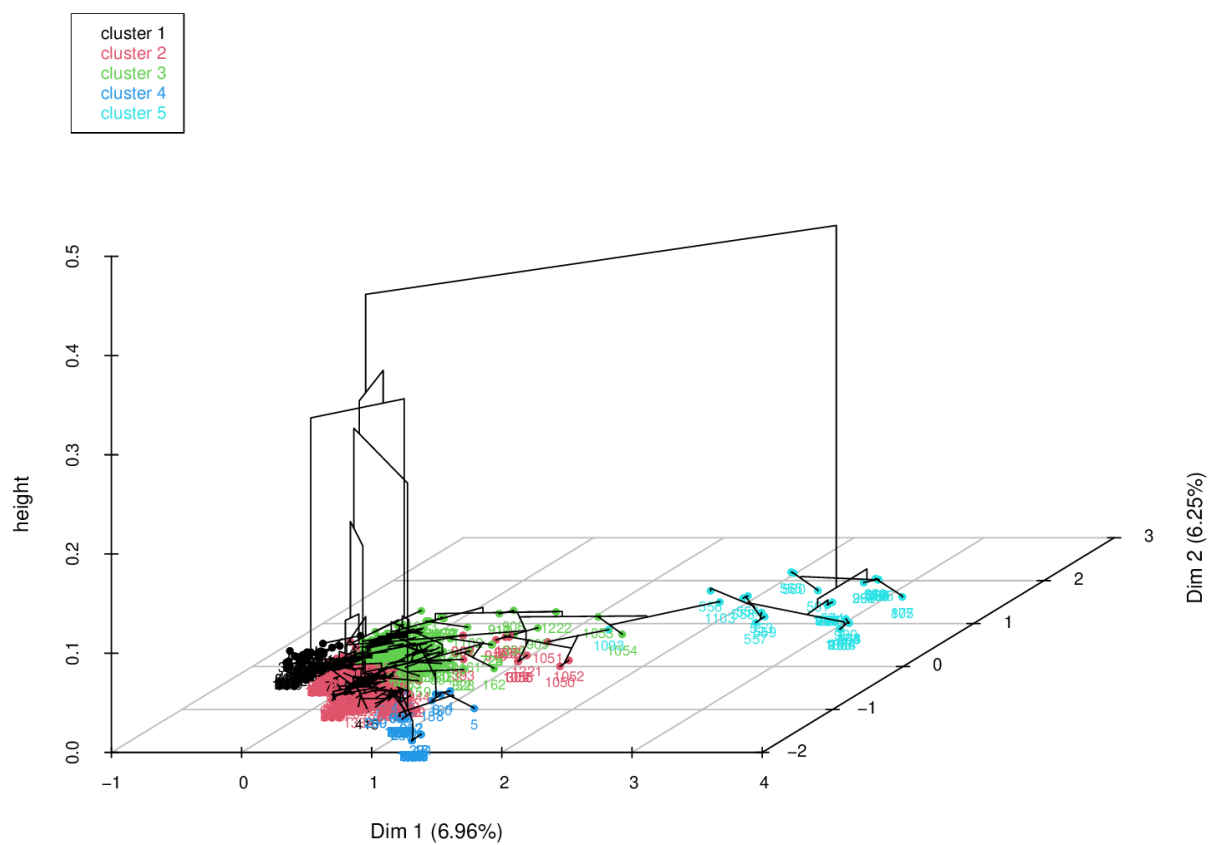

**Supplementary Figure 18 | Perspective view of the dendrograms visualisation of the calculated hierarchical clustering on the factor map.**

We found the following 5 tipologies (for details see Supplementary Figs. 19.1 and 19.2):

-Cluster A is small (6% of the samples) and corresponds almost perfectly with the set of $Gd^{3+}$ compounds, or, equivalently, of isotropic complexes. Properties that are strongly overrepresented in this cluster compared to the whole sample include belonging to the "radical" chemical family, number of ligands = 5 and a coordination sphere formed by oxygen only.

-Cluster B is the largest one (63% of the samples), and is composed entirely of oblate ions. 90% of the total of all $Dy^{3+}$ samples are inside Cluster B, however there is a large minority (33.3%) of samples inside Cluster B which are based on non-$Dy^{3+}$ ions. Cluster B also includes ~75% of all samples of "mixed ligands", of "other families" and where the coordination sphere is a mixture of Oxygens and Nitrogens, and 70% of the samples where the coordination sphere is all Oxygen.

-Cluster C is the second largest one (20% of the samples), and it is composed almost entirely of prolate Kramers ions. Above 50% are $Er^{3+}$, and almost 30% are $Yb^{3+}$. Additionally, coordination number = 8 is strongly overrepresented in this cluster.

-Cluster D is small (7.4% of the samples), includes practically all Pc double-deckers ($LnPc_2$), and indeed corresponds very well with this chemical family. The match is weaker in the case of the categories corresponding to all Nitrogen in the coordination sphere, 2 ligands and 8 atoms in the coordination sphere, in the sense that there is a minority of samples with these features which are outside this Cluster (Supplementary Fig. 19.2). 60% of the samples in Cluster D are based on $Tb^{3+}$, 30% are based on $Dy^{3+}$.

-Cluster E, which is the smallest (3.7% of the samples), corresponds almost perfectly to the metallocene family and, like in Cluster D, the match with all-Carbon coordination, coordination number = 2, and number of ligands = 2 is weaker since some samples with these features are outside this cluster. The samples in Cluster E are more often based on prolate than oblate ions (~40% prolate), but significantly less so than in the total set (~20% prolate). This can be related with the fact that, while ~30% of samples in Cluster D are based on $Dy^{3+}$, this is much less than for the total dataset (~50% Dy).

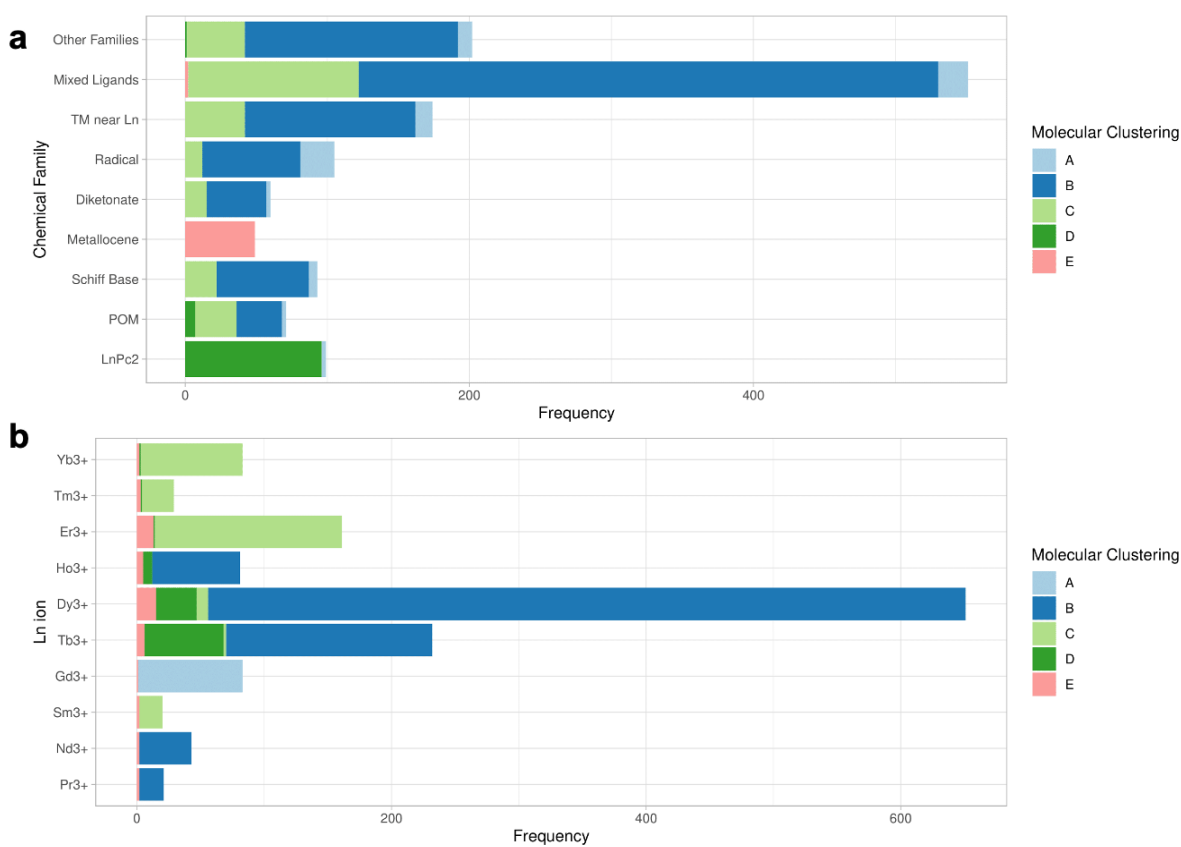

**Supplementary Figure 19.1 | Bar charts representing the frequencies of chemical variables, filled according to the molecular clustering described in this section. a,** chemical families, **b,** Ln ions.

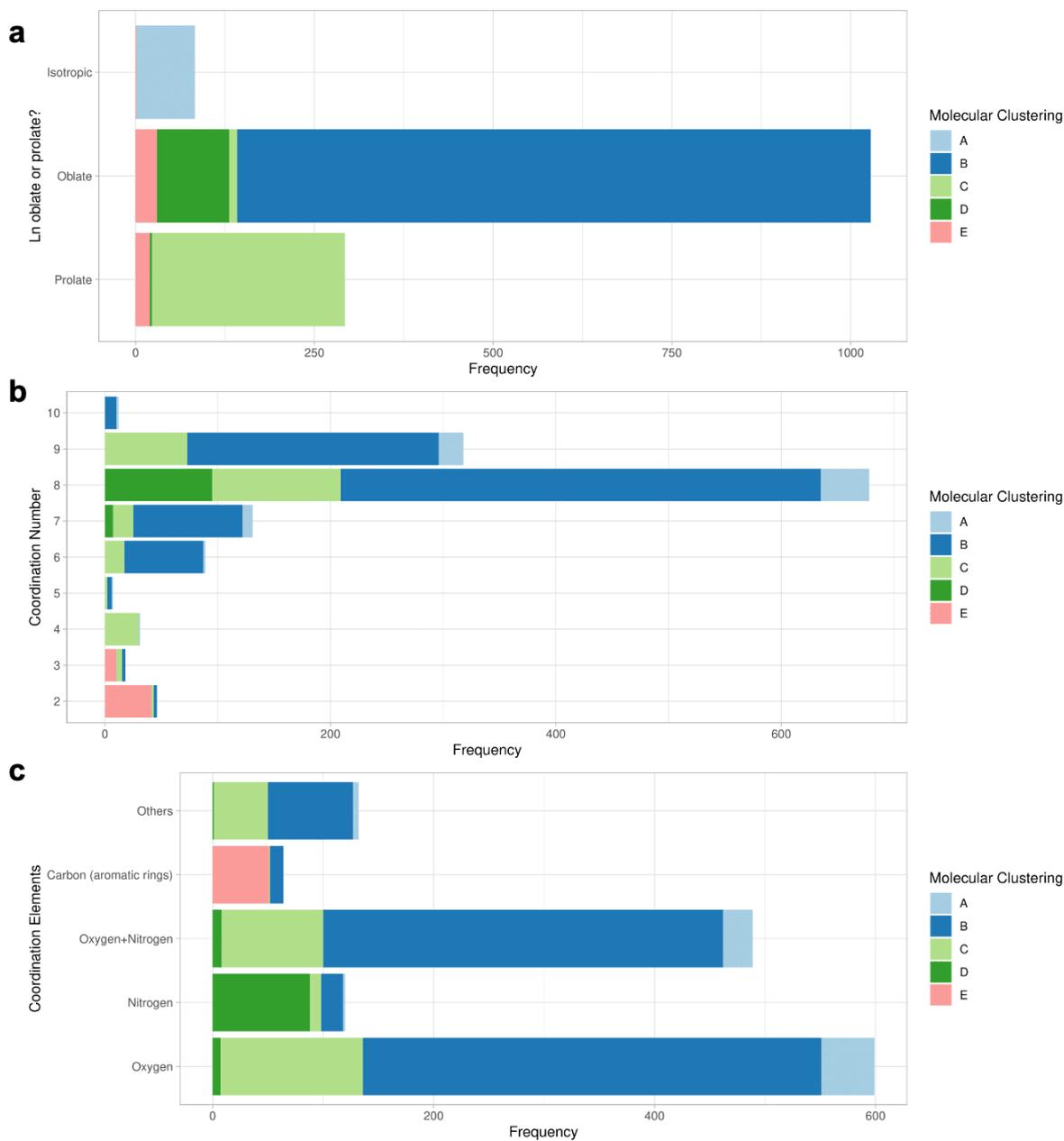

**Supplementary Figure 19.2 | Bar charts representing the frequencies of chemical variables, filled according to the molecular clustering described in this section. a**, oblate *vs* prolate nature of the ion, **b**, coordination numbers, **c**, coordination elements.

### 4.3. Lognormal modelling

Finally, we applied a lognormal model to detect possible association between variables. The different factors within the model indicate the different associations between the studied variables and parameters. This model works with the frequencies of each variable in the crosstabulation. These frequencies result from each combination of variables, so Poisson's distribution was employed. For any modelization one needs to declare the link function: a function of the expected value of the dependent variable, taking the form of a linear combination of the independent variables. For lognormal modelling, we do not distinguish between dependent and independent variables, since one is rather interested in associations between variables, but one does have a link function which unites the average frequency with the linear predictor, which in this case is the logarithmic function. Within the chemical variables we have many categorical ones, with many levels each. To avoid a useless complication of the analysis, we start by studying the relation between the lanthanide ion, its anisotropy and spin parity (Kramers or not). This serves solely as a test for the modelling and the integrity of our data, since we know that these are associated beforehand: for every lanthanide ion, its anisotropy type and Kramers nature is well defined, with no exceptions. As expected, our model found that it is sufficient to work with the lanthanide ion, and we discarded the other two variables for the subsequent clustering studies. We proceeded to use lognormal models, introducing the rest of the chemical variables, one by one, together with the lanthanide ion. We found that the chemical family, lanthanide ion and coordination elements are the main variables and sufficient to reasonably explain the frequencies of the rest of the variables.

### Supplementary Section 5. Statistical analysis of the physical variables

For all the statistical analyses of the physical variables that follow in Supplementary Sections 5.1 and 5.3, additionally to the analyses performed for the complete dataset (data from years 2003-2019), we repeated the study with the data subset in the timeframe 2003-2017 (~1000 samples instead of ~1400). All qualitative results presented here are robust whether one considers the whole dataset 2003-2019 or the 2003-2017 subset. Each of the weak numerical correlations indicated in Supplementary Fig. 20 are stable within a 0.2 window, and in particular all correlations higher than 0.9 are stable with deviations within 0.05 window between both subsets. All values for the Akaike Information Criterion (AIC) in Supplementary Tables 2, 3, 4 are higher on average by about a factor of 1.5-2, as expected, since they are roughly proportional to the number of samples.[58] We can therefore conclude that the quantitative and qualitative relations among the physical variables are stable, *i.e.* no new significant trends have been revealed recently.

## 5.1. Overview of the main statistical relationships

Here, our goal is to identify statistical relationships among the physical variables. At first, we want to confirm whether the simple model parameters $U_{eff}$, $U_{eff,2}$, $\tau_0$ are good statistical predictors (*i.e.* present a high correlation with) of the experimental behaviour. In particular, whether they are good predictors of $T_{B3}$, $T_{B3H}$, $T_{hyst}$, $H$, $\chi_{im,max}$ and "Hyst". Next, we want to determine whether the predictive power is different between $U_{eff}$ *vs* $U_{eff,ff}$, and $\tau_0$ *vs* $\tau_{0,ff}$. Supplementary Fig. 20 represents graphically all physical variables, with their relationships in pairs; $U_{eff,2}$ was eliminated from the graph since the scarcity of data produced errors in the correlation test.

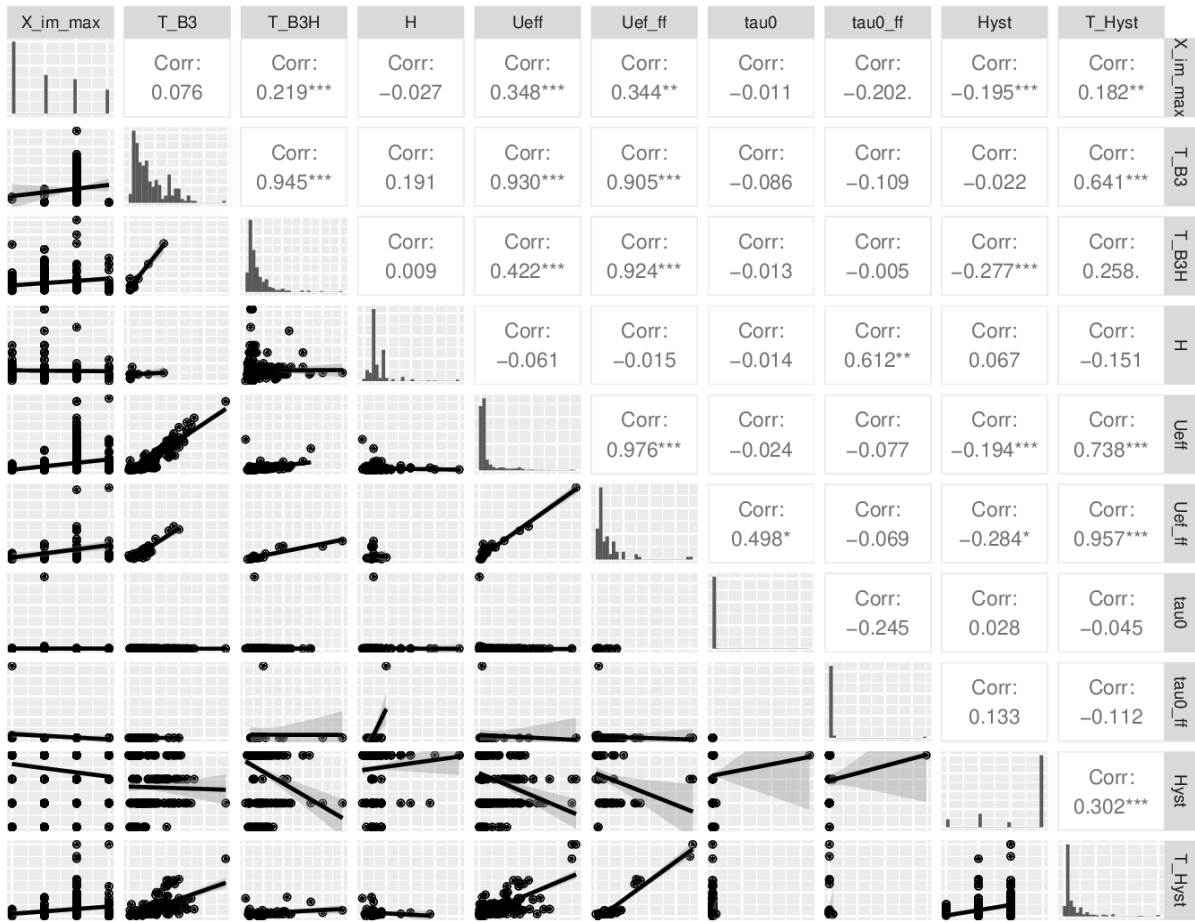

**Supplementary Figure 20 | Statistical relationships among the physical variables.** The diagonal shows the frequency of each value (range) for each of the physical variables. Below the diagonal we see graphical representations to visually show the relation between every pair of variables, and above the diagonal we find the quantification of each correlation, in the case of numerical variables, or an alternate boxplot, in the case of categorical variables.

Let us first go over this bidimensional array of statistical representations and analyses, and later on we will focus on the most significant pieces of information.

Keeping in mind the definition of $\chi_{im,max}$ (Supplementary Section 1), the properties as SIM are better when $\chi_{im,max}$ takes higher numbers ($0 < 1 < 2$), *e.g.* only the samples with $\chi_{im,max} = 2$ present values for $T_{B3}$, while $\chi_{im,max} = 3$ means having no information. The most frequent value is $\chi_{im,max} = 0$ (Freq-independent $\chi''$), with the rest being similarly abundant. For $T_{B3}$ and $T_{B3H}$, higher temperature values are associated with higher values of $\chi_{im,max}$. Higher effective energy barriers $U_{eff}$, $U_{eff,2}$, $U_{eff,ff}$ were estimated for $\chi_{im,max} = 3$ (corresponding to cases where the SIM properties are not necessarily bad, but just were not characterised via $\chi''$ vs T plots) and especially for $\chi_{im,max} = 2$ (as expected), compared with systems with worse properties, where almost no difference is found between $\chi_{im,max} = 0$ and $\chi_{im,max} = 1$. As in the case of the effective energy barriers, higher hysteresis temperatures $T_{hyst}$ were reported for $\chi_{im,max} = 3$, and especially for $\chi_{im,max} = 2$, as expected, compared with systems with worse properties.

The temperatures of the maximum in $\chi''$ at $10^3$ Hz are labelled as $T_{B3}$ (when no magnetic field is applied) and $T_{B3H}$ (when a magnetic field $H$ is applied). There are more cases reported in the presence of a magnetic field. In both cases, there is a distribution similar to an inverse exponential, meaning higher temperatures are less frequent, but this is more marked for $T_{B3H}$: when a magnetic field is applied to detect a maximum in $\chi''$, high temperature values for this maximum are rare. Higher values of the magnetic field $H$ do not correlate with higher values of $T_{B3H}$. There is a marked difference in the correlations of the temperatures $T_{B3}$, $T_{B3H}$ with the effective energy barriers $U_{eff}$, $U_{eff,ff}$. While there is a very strong correlation between $T_{B3}$ and $U_{eff}$, for $T_{B3H}$ the correlation only exists for $U_{eff,ff}$. This is consistent with the presence of other relaxation mechanisms in these cases, such as the QTM, which are quenched in presence of a strong field. In contrast with the effective energy barriers, there are no significant correlations between $T_{B3}$, $T_{B3H}$ and $\tau_0$, $\tau_{0,ff}$. This might be seen as puzzling, since each pair of effective energy barriers and preexponential times are extracted from a single fit. The reason for this behaviour can be understood by recalling that the search for correlation in Supplementary Figure 20 is linear, while the proper way to find the correlations in this case is via a logarithmic plot (Supplementary Section 5.3 and Supplementary Figs. 24.1 and 24.2). Since these values vary over several orders of magnitude, especially in the case of $\tau_0$, $\tau_{0,ff}$, a minimum square root analysis over the linear data is practically determined by the small and noisy cloud of data points with maximum values rather than the whole data range. Finally, there is a systematic qualitative improvement of the hysteretic behaviour with higher values of $T_{B3}$, $T_{B3H}$. In other words, high values of $T_{B3}$, meaning short-term (millisecond) magnetic memory up to a high temperature, are frequently associated with the presence of hysteresis. Moreover, within cases with hysteresis, short-term magnetic memory up to a high temperature is associated with the presence of full hysteresis rather than pinched (butterfly) hysteresis. While this cannot be clearly seen in Supplementary Fig. 20, it is evident in Supplementary Fig. 21. In contrast, the quantitative correlation of the hysteretic temperature $T_{hyst}$ with higher values of $T_{B3}$, $T_{B3H}$ is relatively weak in the case of $T_{B3}$, and nonexistent for $T_{B3H}$.

$H$ is the external magnetic field applied to measure $T_{B3H}$. It presents very weak (and negative) correlations with the effective energy barriers (both $U_{eff}$ and $U_{eff,ff}$). High values of $U_{eff}$ are visibly correlated with the sample presenting hysteresis (Hyst = 1), and within it, with samples presenting coercivity (Hyst = 2). These correlations are stronger in the case of $U_{eff,ff}$. The attempt times $\tau_0$ and $\tau_{0,ff}$ are however very poorly correlated, both with their corresponding barriers and with the other properties. As we will see in Supplementary

Section 5.3, this is probably due to the high dispersion in values: at least a qualitative correlation is visible when one works on a logarithmic scale.

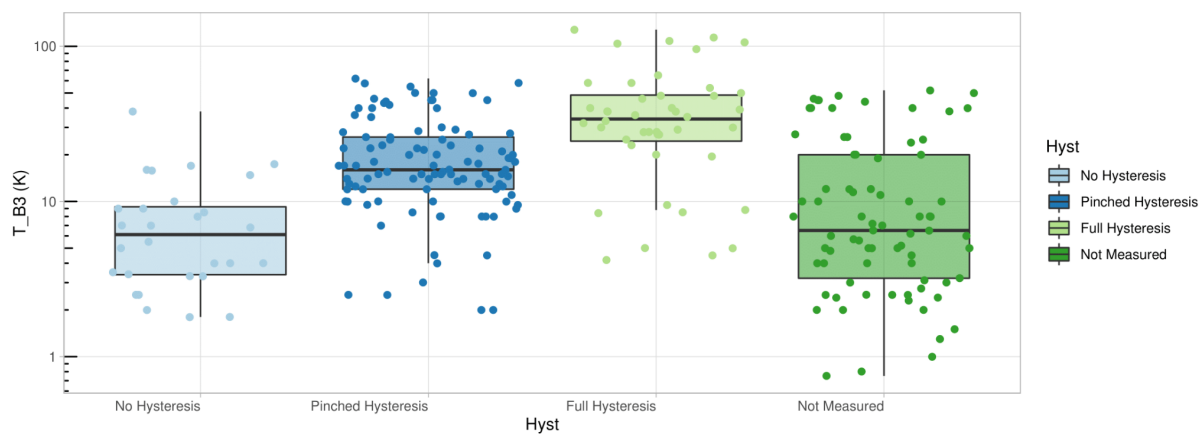

**Supplementary Figure 21 | Boxplots of $T_{B3}$ grouped by categories of hysteretic behaviour.**

### 5.2. Simple frequency distributions

Let us start by examining the individual frequency distributions of $T_{B3}$, $T_{B3H}$, $U_{eff}$, $U_{eff,ff}$ and $T_{hyst}$, *i.e.* the five single-variable bar chart representations of their distribution of values throughout the population (diagonal graphics in Supplementary Fig. 20). In all cases, an initial visual inspection evidences a roughly inverse exponential decay, or perhaps a gaussian distribution centred around very small values. This can be interpreted as a mostly random distribution of values, a signal that the studied population is large and mostly unbiased. An interesting exception can be found in $T_{B3}$, which takes a bimodal distribution. This has different possible interpretations, but likely just means that a fraction of the research in the field was focused on types of compounds where the median of $T_{B3}$ is at or above the maximum values of the other chemical families, *i.e.* the LnPc$_2$ family, see Supplementary Fig. 11.1, where it can be compared with the much less marked difference in the case of $U_{eff}$, where the distribution of the LnPc$_2$ family is less skewed to very high values.

### 5.3. Correlations between physical variables

As can be seen in Supplementary Fig. 20, the highest correlations involve $U_{eff}$ and $U_{eff,ff}$. $U_{eff}$ is highly correlated with $T_{B3}$ and with $T_{hyst}$. $U_{eff,ff}$ is highly correlated with $U_{eff}$, $T_{B3}$ and $T_{B3H}$. Furthermore, we apply Pearson's test to verify the correlation between $U_{eff}$ and $U_{eff,ff}$. We obtained a very robust correlation between the two variables, with a p-value $< 2.2 \cdot 10^{-16}$, a 95% confidence interval for the correlation of 0.951-0.989 and an estimated correlation of 0.976. The same procedure is applied to $U_{eff}$ and $U_{eff,2}$, obtaining results that are robust but substantially less so: p-value $< 1.113 \cdot 10^{-7}$, with a 95% confidence interval for the correlation of 0.701-0.941 and an estimated correlation of 0.864.

We apply the Akaike Information Criterion (AIC), a well-established method that evaluates how well a statistical model fits the data it was generated from. This method allows to compare the quality of a series of candidate models using the same data, so that the AIC estimates the quality of each of the models relative to the others. As the models are used to represent the process that generated the data, this representation will be losing some information because of the flaws of the model, and the AIC estimates the relative amount of information lost by each candidate model. This means, the preferred model will have the lowest AIC value in a given set of candidate models. To implement AIC, we employ the R functions lm, glm with family = binomial (stats package from R base)[59], and multinom (nnet package).[60] The results can be found in Supplementary Tables 2, 3 and 4.

**Supplementary Table 2 | AIC modelling experimental physical (response) variables as a function of modelling variables $U_{eff}$, $U_{eff,2}$, $\tau_0$.**

| Response variable | Data points | Variables included in the model | Significant variable | AIC |
|---|---|---|---|---|
| $\chi''_{max}$ | 23 | $U_{eff}$, $U_{eff,2}$, $\tau_0$ | - | 55.04 |
| | | $\boldsymbol{U_{eff}}$, $\boldsymbol{U_{eff,2}}$ | - | 49.04 |
| | | $U_{eff}$, $\tau_0$ | - | 65.52 |
| | | $U_{eff,2}$, $\tau_0$ | - | 55.62 |
| $T_{B3}$ | 4 | - | - | - |
| $T_{B3H}$ | 14 | $U_{eff}$, $U_{eff,2}$, $\tau_0$ | $\tau_0$ | 2.64 |
| | | $U_{eff}$, $U_{eff,2}$ | - | 35.35 |
| | | $U_{eff}$, $\tau_0$ | $\tau_0$ | 0.98 |
| | | $U_{eff,2}$, $\tau_0$ | $\tau_0$ | 4.25 |
| | | $\boldsymbol{\tau_0}$ | $\tau_0$ | 2.27 |
| Hyst | 23 | $U_{eff}$, $U_{eff,2}$, $\tau_0$ | - | 44.49 |
| | | $U_{eff}$, $U_{eff,2}$ | - | 40.49 |
| | | $U_{eff}$, $\tau_0$ | - | 40.56 |
| | | $\boldsymbol{U_{eff}}$ | - | 36.56 |
| | | $U_{eff,2}$, $\tau_0$ | - | 43.13 |
| $T_{hyst}$ | 4 | - | - | - |

Supplementary Table 2 contains the analysis based on the modelling variables $U_{eff}$, $U_{eff,2}$ and $\tau_0$. The analysis quantifies the models employing these three variables to explain the different response variables: {$\chi''_{max}$, $T_{B3}$, $T_{B3H}$, Hyst, $T_{hyst}$}. The results are heterogeneous. Depending on the response variable chosen, the best modelling variables can be either {$U_{eff}$, $U_{eff,2}$}, or $\tau_0$, or $U_{eff}$. Since very few samples are modelled considering two independent barriers ($U_{eff}$, $U_{eff,2}$), the data are scarce and the results are not statistically significant.

**Supplementary Table 3 | AIC modelling experimental physical (response) variables as a function of modelling variables $U_{\text{eff}}$, $\tau_0$.**

| Response variable | Data points | Variables included in the model | Significant variable | AIC |
|---|---|---|---|---|
| $\chi''_{\text{max}}$ | 608 | $U_{\text{eff}}$, $\tau_0$ | - | 1381.05 |
| | | $\boldsymbol{U_{\text{eff}}}$ | - | 1378.16 |
| | | $\tau_0$ | - | 1542.43 |
| $T_{\text{B3}}$ | 186 | $U_{\text{eff}}$, $\tau_0$ | $U_{\text{eff}}$ | 693.98 |
| | | $\boldsymbol{U_{\text{eff}}}$ | $U_{\text{eff}}$ | 692.98 |
| | | $\tau_0$ | - | 1051.61 |
| $T_{\text{B3H}}$ | 261 | $U_{\text{eff}}$, $\tau_0$ | $U_{\text{eff}}$ | 770.73 |
| | | $\boldsymbol{U_{\text{eff}}}$ | $U_{\text{eff}}$ | 768.74 |
| | | $\tau_0$ | - | 818.99 |
| Hyst | 601 | $U_{\text{eff}}$, $\tau_0$ | - | 40.56 |
| | | $\boldsymbol{U_{\text{eff}}}$ | $U_{\text{eff}}$ | 36.56 |
| | | $\tau_0$ | - | 42.04 |
| $T_{\text{hyst}}$ | 134 | $U_{\text{eff}}$, $\tau_0$ | $U_{\text{eff}}$ | 650.3 |
| | | $\boldsymbol{U_{\text{eff}}}$ | $U_{\text{eff}}$ | 648.32 |
| | | $\tau_0$ | - | 780.94 |

Supplementary Table 3 contains the analysis including $U_{\text{eff}}$ and $\tau_0$, *i.e.* when an Orbach-only model with a single energy barrier is considered, and quantifies the models employing these variables to explain different response variables: $\{\chi''_{\text{max}}, T_{\text{B3}}, T_{\text{B3H}}, Hyst, T_{\text{hyst}}\}$. The results are very robust in this case. $U_{\text{eff}}$ is consistently found to be the significant variable and the one producing the lowest AIC value.

**Supplementary Table 4 | AIC modelling experimental physical (response) variables as a function of modelling variables $U_{\text{eff,ff}}$, $\tau_{0,\text{ff}}$.**

| Response variable | Data points | Variables included in the model | Significant variable | AIC |
|---|---|---|---|---|
| $\chi''_{\text{max}}$ | 68 | $U_{\text{eff,ff}}$, $\tau_{0,\text{ff}}$ | - | 169.58 |
| | | $\boldsymbol{U_{\text{eff,ff}}}$ | - | 165.33 |
| | | $\tau_{0,\text{ff}}$ | - | 185.89 |
| $T_{\text{B3}}$ | 27 | $\boldsymbol{U_{\text{eff,ff}}, \tau_{0,\text{ff}}}$ | $U_{\text{eff,ff}}$, $\tau_{0,\text{ff}}$ | 88.5 |
| | | $U_{\text{eff,ff}}$ | $U_{\text{eff,ff}}$ | 91.51 |
| | | $\tau_{0,\text{ff}}$ | - | 137.25 |
| $T_{\text{B3H}}$ | 23 | $U_{\text{eff,ff}}$, $\tau_{0,\text{ff}}$ | $U_{\text{eff,ff}}$ | 74.62 |
| | | $\boldsymbol{U_{\text{eff,ff}}}$ | $U_{\text{eff,ff}}$ | 73.54 |
| | | $\tau_{0,\text{ff}}$ | - | 116.81 |
| Hyst | 67 | $U_{\text{eff,ff}}$, $\tau_{0,\text{ff}}$ | - | 137.57 |
| | | $\boldsymbol{U_{\text{eff,ff}}}$ | - | 131.57 |
| | | $\tau_{0,\text{ff}}$ | - | 156.68 |
| $T_{\text{hyst}}$ | 36 | $U_{\text{eff,ff}}$, $\tau_{0,\text{ff}}$ | $U_{\text{eff,ff}}$ | 102.08 |
| | | $\boldsymbol{U_{\text{eff,ff}}}$ | $U_{\text{eff,ff}}$ | 102.72 |
| | | $\tau_{0,\text{ff}}$ | - | 191.68 |

Supplementary Table 4 contains the same analysis but including only $U_{\text{eff,ff}}$ and $\tau_{0,\text{ff}}$, *i.e.* a more complete model that should be closer to reality. The modelling employing $U_{\text{eff,ff}}$ is almost as good as in the case of $U_{\text{eff}}$. However, since information about $U_{\text{eff,ff}}$ is available in fewer samples, $U_{\text{eff}}$ is statistically preferable.

To visualise these correlations, see Supplementary Figs. 22 and 23. The relations between $U_{\text{eff}}$, $\tau_0$, $\chi''_{\text{max}}$, $T_{\text{B3H}}$ are represented in Supplementary Fig. 22. Within a wide dispersion, a (log-log) inverse linear relation is apparent between (log-log) $U_{\text{eff}}$ and $\tau_0$ (more on this on Supplementary Figs. 24.1 and 24.2) and a (log-log) linear relation is apparent between $T_{\text{B3H}}$ and $U_{\text{eff}}$. These graphs also show how samples with higher values of $U_{\text{eff}}$ systematically present a maximum in $T_{\text{B3}}$, and in contrast samples where no frequency-dependent $\chi''$ is measured tend to display lower values of $U_{\text{eff}}$. To complete the picture, we represent the relations between $U_{\text{eff}}$, $Hyst$, $T_{\text{hyst}}$, $T_{\text{B3}}$. (Supplementary Fig. 23). A (log-log) linear tendency is apparent when plotting $T_{\text{B3}}$ *vs* $U_{\text{eff}}$; this is obscured in the case of $T_{\text{hyst}}$ *vs* $U_{\text{eff}}$ by the abundance of samples where the hysteresis was characterised only at 2 K. Like in the case of the ac susceptibility, there is a large dispersion of behaviours by samples with longer magnetic memory, in this case meaning the ones presenting whole hysteresis, tend to be grouped around higher values of $U_{\text{eff}}$, with samples presenting no hysteresis tend to present lower values of $U_{\text{eff}}$ and samples with pinched (butterfly) hysteresis presenting typically intermediate values.

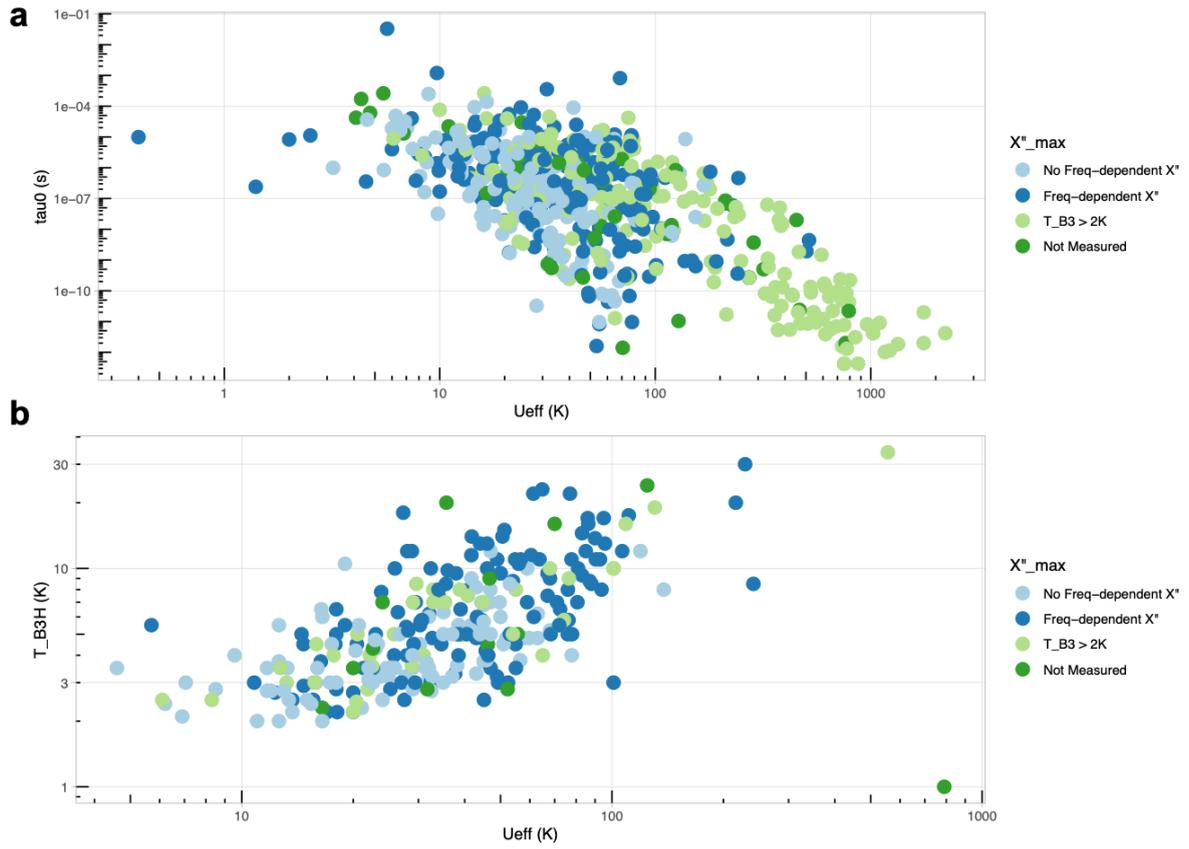

**Supplementary Figure 22 | Scatterplots depicting the relation of $U_{eff}$ with $\tau_0$, $\chi''_{max}$, $T_{B3H}$.** $\tau_0$ *vs* $U_{eff}$, colored by $\chi''_{max}$, **a**; and $T_{B3H}$ *vs* $U_{eff}$, colored by $\chi''_{max}$, **b**. Note that not all samples will be present in all graphs (see Supplementary Figure 1.1). As a consequence, an (x vs y) plot can only include samples for which x and y are simultaneously present in the dataset.

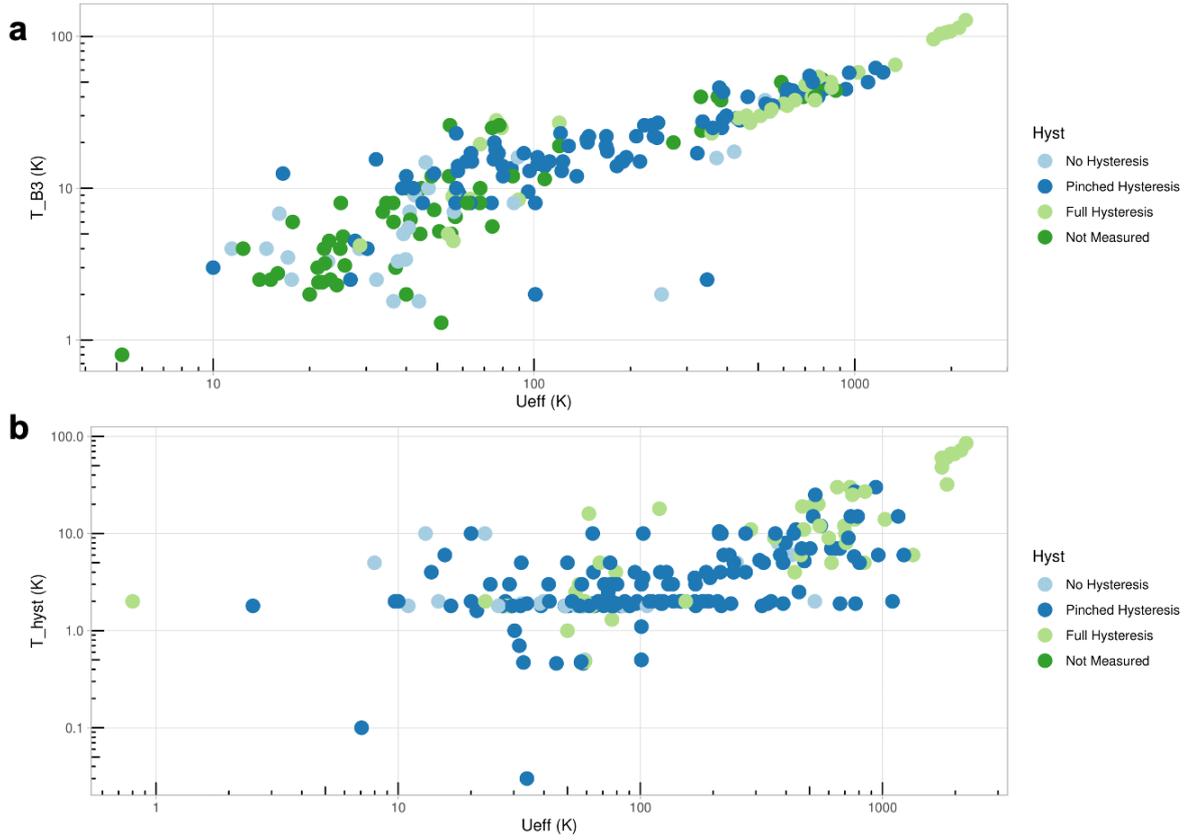

**Supplementary Figure 23 | Scatterplots depicting the relation of $U_{eff}$ with *Hyst*, $T_{hyst}$, $T_{B3}$.** $T_{B3}$ *vs* $U_{eff}$, colored by Hyst, **a**; $T_{hyst}$ *vs* $U_{eff}$, colored by Hyst, **b**. Note that not all samples will be present in all graphs (see Supplementary Figure 1.1). As a consequence, an (x vs y) plot can only include samples for which x and y are simultaneously present in the dataset.

The main conclusion of the study is that $U_{eff}$ derived from a simplistic Arrhenius plot is currently the best single predictor for physical behaviour. This means that, whether we are discussing in terms of the presence of maximum in out-of-phase component of the ac susceptibility ($\chi''_{max}$) or the temperature of said maximum ($T_{B3}$, $T_{B3H}$), $U_{eff}$ is a better predictor than $\tau_0$ or $U_{eff,2}$ (note that the number of studies with $U_{eff,2}$ is too small). Also the number of studies deriving $U_{eff,ff}$ from a full fit considering the other physical processes is very low. Furthermore, the correlation between $U_{eff,ff}$ and $U_{eff}$ is very high. The combination of the two facts mean that there is no statistical argument for the qualitative observation that $U_{eff,ff}$ from a full fit is a better predictor for $T_{hyst}$.

### 5.4 The question of $U_{eff}$ vs $U_{eff,ff}$

A crucial issue is to quantify up to what level the value of $U_{eff}$ and $\tau_0$ are well correlated with the slow relaxation of the magnetisation, or to determine whether one would need to employ $U_{eff,ff}$ instead. Let us proceed with increasing the order of complexity. A visual inspection in SIMDAVIS shows that, in a few cases where there is simultaneous information on $U_{eff}$ and $U_{eff,ff}$, their values are very similar (Fig. 4a in the main text). Furthermore, this partial information is corroborated by the very similar dependencies of $T_{B3}$ or $T_{hyst}$ vs either $U_{eff}$ or $U_{eff,ff}$, as well as in the numerical correlations (see Supplementary Sections 3.2 and 5.3). A categorical analysis (Figs. 4b, c) shows that the data dispersion is large, meaning that it is impossible to predict the experimental behaviour for an individual sample merely from its $U_{eff}$ value. However, it demonstrates that, as expected, samples which present a maximum in the out-of-phase susceptibility $\chi''$, or hysteresis, also present higher $U_{eff}$ values. A more thorough numerical analysis (see Supplementary Section 6) confirms these trends.

An in-depth statistical analysis of all physical parameters based on the Akaike Information Criterion (see Supplementary Section 5.3) concludes that $U_{eff}$ derived from a simple Arrhenius plot is the best single predictor for the magnetic behaviour in our dataset. This idea was previously proposed by Ding et al.[61] but studied with a much lower sample size. This means that, whether we are discussing in terms of the out-of-phase component of the ac susceptibility or magnetic hysteresis, $U_{eff}$ is a better predictor than $\tau_0$, $\tau_{0,ff}$, $U_{eff,2}$ and, in practice, than $U_{eff,ff}$. Factorial analysis of mixed data (see Supplementary Section 6) also reveals the predictive power of $U_{eff}$ compared with $\tau_0$. Note that this does not contradict previous studies which demonstrated that a variation in the Orbach barrier does not fully explain the differences in retention of magnetisation,[1] since we did not explicitly consider other relaxation mechanisms up to this phase of the work. Our observation could be due to the fact that, historically, the fit to the Orbach process has been applied to the relaxation times obtained at the highest temperatures, even in systems where at very low temperatures the relaxation times were indicating purely quantum tunnelling or Raman relaxation mechanism. To get further insights on this problem, we explored the available data on the two latter processes in the next section.

## 5.5 Dependence of $\tau_0$ vs $U_{\text{eff}}$

Let us address in more detail the remaining question of whether the effective energy barrier, despite being oversimplified, is meaningful.

In the classical text by Abragam and Bleaney (Published 1970, reprinted 2013, chapter 10, page 561, eq 10.55)[62] offered the following relation for the two-phonon Orbach process, assuming a Debye model for phonons:

$$\frac{1}{\tau_1} = \frac{3}{2\pi\hbar^4\rho\upsilon^5}\left|V^{(1)}\right|^2 U_{eff}^3 \frac{1}{exp\,(U_{eff}/kT)-1} \qquad (1)$$

In terms of notation, note that in the book $\Delta$ is employed for the energy difference between the starting state |b> and the excited state |c> in the two-phonon Orbach process, before relaxation to the final state |a>. Thus, $\Delta$ corresponds to $U_{\text{eff}}$ in our dataset.

Two major objections to the validity of this approximation are, (a) there is a consensus that the vast majority of studied compounds present relaxation mechanisms dominated by multiphonon processes i.e. high-order Orbach processes involving successive excitations to higher states followed by a cascade of de-excitations and (b) the Debye model is nowadays known to be a bad match for the local vibrations responsible for relaxation in molecular nanomagnets. Arguably, for our purposes this is still an interesting representation. Despite the first objection, one needs to consider that the two-phonon Orbach process with a single excited state corresponds well with the Arrhenius equation that has been widely employed in the literature to extract parameters $\tau_0$ and $U_{\text{eff}}$. About the second objection, one will just need to remember that part of the deviations of experimental data from this equation that one will find will be precisely due to the failing of the Debye model in magnetic molecules. For more on the repercussions on the failing of the Debye model, see the discussion in Supplementary Section 9.

From (1) we establish a relationship with the Arrhenius equation in the limit $U_{\text{eff}} >> kT$ (which is always the case in the experimental data, since at temperatures of the order of $U_{\text{eff}}$ there is no slow relaxation of the magnetization):

$$\frac{1}{\tau_1} = \frac{1}{\tau_0} \cdot exp\,(-\,U_{eff}/kT) \quad (2)$$

From combining (1) and (2) we extract the approximate relation:

$$\frac{1}{\tau_0} = R_{Or}\cdot(U_{eff})^3 \qquad (3)$$

, where we introduced an Orbach relaxation rate $R_{\text{Or}}$ as:

$$R_{Or} = \frac{3}{2\pi\hbar^4\rho\upsilon^5}\left|V^{(1)}\right|^2 (4)$$

We can rewrite Eq. (3) above more generally as:

$$\frac{1}{\tau_0} = R_{Or} \cdot (U_{eff})^n \qquad (5)$$

where $n = 3$. According to Abragam and Bleaney, reasonable parameters for rare earth elements resulted in an Orbach rate $R_{Or} \approx 10^4$ $K^{-3} \cdot s^{-1}$, and early experimental results were in the range $10^3$ $K^{-3} \cdot s^{-1} < R_{Or} < 10^5$ $K^{-3} \cdot s^{-1}$.

Plotting the data available in the dataset in terms of $\tau_0$ $vs$ $U_{eff}$ allows one to quantify the deviations, in practice, from the assumptions in eq. 10.55 in Abragam and Bleaney, as commented above. The results from the fits can be found in Supplementary Table 5.

**Supplementary Table 5 | Least squares fits of ln($U_{eff}$) $vs$ -ln($\tau_0$) and ln($U_{eff,ff}$) $vs$ -ln($\tau_{0,ff}$).**

| Data | Intercept ($R_{Or}$) | Slope ($n$) |
|---|---|---|
| All | 839.4 | 2.437 |
| Prolate | 2557.5 | 2.454 |
| Oblate | 504.3 | 2.506 |
| $Tb^{3+}$ | 700.5 | 2.415 |
| $Dy^{3+}$ | 504.0 | 2.515 |
| Full fit | 151.2 | 2.957 |

Note that the slopes are identical but the constant term is higher for oblate ions compared with prolate ions. We find $R_{Or}$(prolate) $\approx 5 \cdot R_{Or}$(oblate), meaning that, for comparable $U_{eff}$, $\tau_0$ for oblate ions is on average substantially greater, and relaxation substantially slower, than that for prolate ions (Supplementary Fig. 24.1). This is consistent with the observation that complexes of oblate ions present values of $T_{B3}$ higher than expected considering their $U_{eff}$ (see Supplementary Fig. 11.3). meaning an equivalent $U_{eff}$ relaxation will be substantially slower in oblate ions). Within the two main oblate ions ($Dy^{3+}$ and $Tb^{3+}$), the slope is slightly higher for $Dy^{3+}$, meaning a dramatic increase in $U_{eff}$ is somewhat more beneficial for $Tb^{3+}$ compared with $Dy^{3+}$ (Supplementary Fig. 24.2).

The limited (<100) data points of $U_{eff,ff}$, $\tau_{0,ff}$ pairs, where all relaxation processes were considered, present a better agreement on the exponent, with $n \approx 3$ but lower Orbach rates $R_{Or}$ $\approx$ 150 $K^{-3} s^{-1}$. For the rest of the dataset, given the limitations pointed out above, the coincidence with the expectation from the relationship in Eq. 5 is reasonable. The minor discrepancy with the expected exponent, a value of $n$ that is between 2.4 and 3 instead of $n=3$, and values of $R_{Or}$ of the order of $10^3$, around or below the expected lower limit of the range $10^3$-$10^4$, serves as an independent evaluation of the limitations of a two-phonon Orbach model that also assumes a Debye model for phonons.

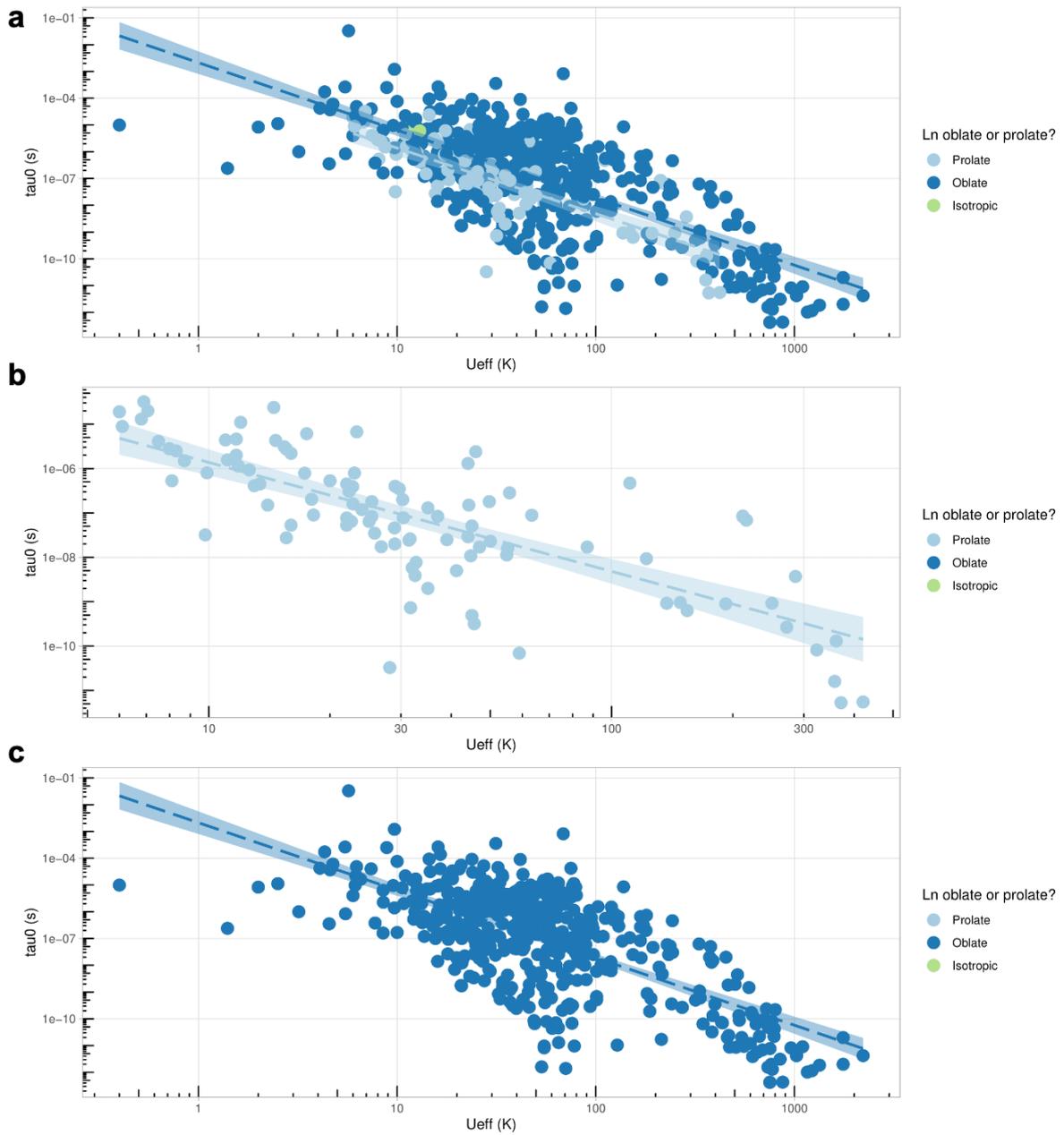

**Supplementary Figure 24.1 $\tau_0$ *vs* $U_{\text{eff}}$, for prolate and oblate ions. a,** Comparison between both. **b,** Only prolate ions. **c,** Only oblate ions. Note that not all samples will be present in all graphs (see Supplementary Figure 1.1). As a consequence, an (x vs y) plot can only include samples for which x and y are simultaneously present in the dataset.

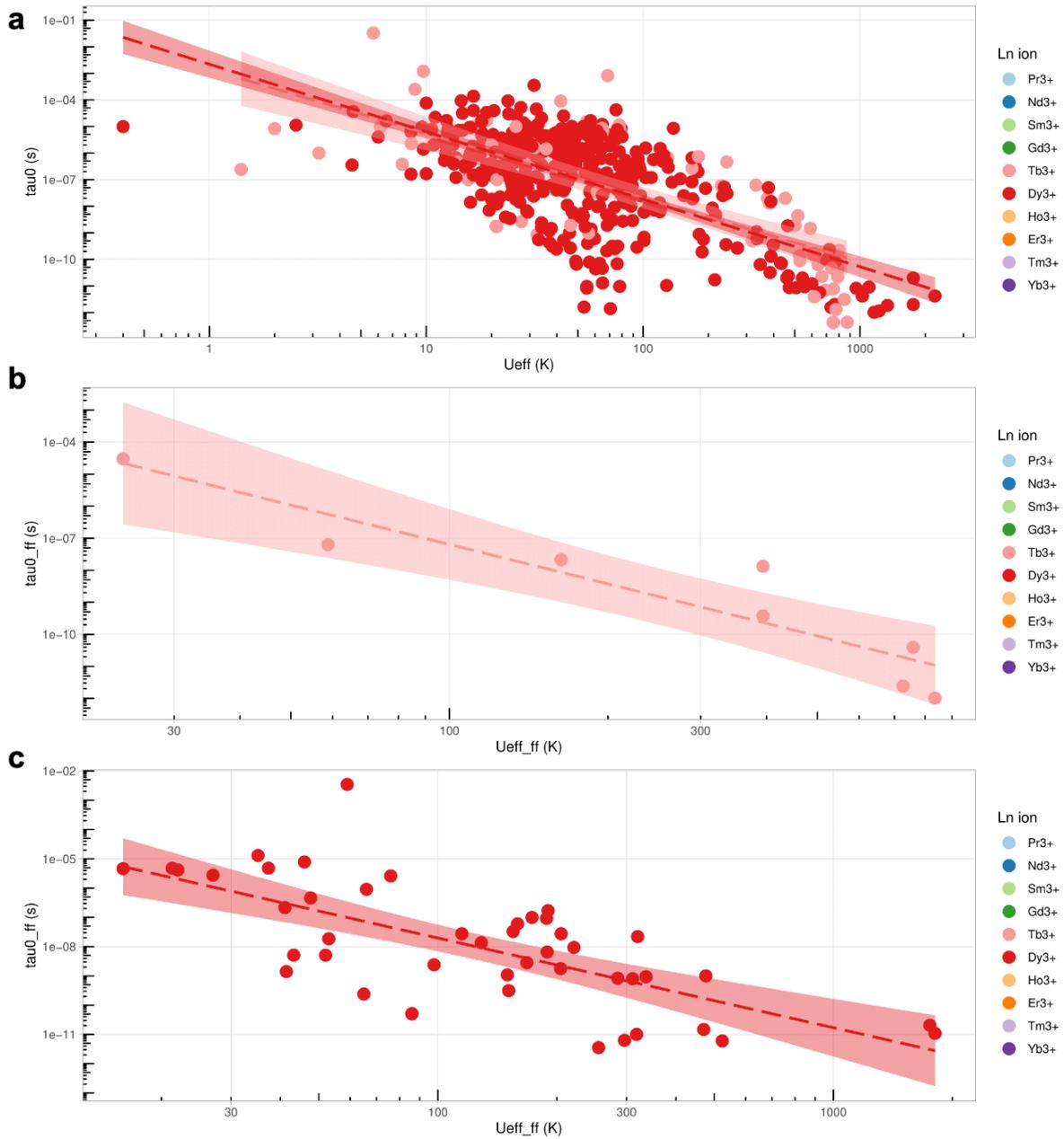

**Supplementary Figure 24.2 | $\tau_0$ *vs* $U_{\text{eff}}$ and $\tau_{0,\text{ff}}$ *vs* $U_{\text{eff,ff}}$ for Tb³⁺ and Dy³⁺. a,** Comparison between $\tau_0$ vs $U_{\text{eff}}$ for both ions. **b,** $\tau_{0,\text{ff}}$ vs $U_{\text{eff,ff}}$, only Tb³⁺ complexes. **c,** $\tau_{0,\text{ff}}$ vs $U_{\text{eff,ff}}$, only Dy³⁺ complexes. Note that not all samples will be present in all graphs (see Supplementary Figure 1.1). As a consequence, an (x vs y) plot can only include samples for which x and y are simultaneously present in the dataset.

**Supplementary Section 6. FAMD and magnetostructural clustering**

For all the statistical studies of the physical variables that follow in the present section, we performed the analysis two times, to check for consistency and robustness of the results. In particular, we performed the analysis of the full dataset (~1400 samples) and repeated it independently employing only the data subset in the timeframe 2003-2017 (~1000 samples). We found that all major qualitative results presented here are robust and independently obtained whether one considers the whole set 2003-2019 or the 2003-2017 subset. Furthermore, quantitative data were found to be within a 25% deviation, with a shift towards higher values of $U_{\mathrm{eff}}$, $T_{\mathrm{B3}}$ and $T_{\mathrm{hyst}}$ in the 2003-2019 set when compared with the 2003-2017 subset.

Factorial analysis of mixed data (FAMD) is a factorial method appropriate to analyse data containing both quantitative and qualitative variables. FAMD is a versatile method that acts as Principal Component Analysis for quantitative variables, and as Multiple Correspondence Analysis for qualitative variables. Qualitative and quantitative variables are normalised during the analysis to equilibrate the influence of each in the variable set. In this case this allows us a simultaneous analysis of physical and chemical properties, to perform a hierarchical clustering of samples with the goal of producing a magnetostructural taxonomy in our Ln-based SIMs catalogue. By grouping the samples by taking into consideration their molecular structure and their magnetic behaviour, we can aspire to obtain information on the main relation between form and function. To perform this analysis and data representation we employed R packages FactoMineR[57] and factoextra.[63]

We found that the chemical family, the lanthanide ion and the coordination elements are the best chemical predictors, as $U_{\mathrm{eff}}$ among the physical parameters. Only 608 samples in the dataset contain quantitative $U_{\mathrm{eff}}$ and $\tau_0$ data. We initially worked just with these 608 samples, and later repeated the analysis with all samples, obtaining the same result.

Let us start with representing the relation between the two main FAMD dimensions and the main physical and chemical variables (Supplementary Figs. 25, 26 and 27). Analysing the contribution from the main chemical and physical variables to the main FAMD dimensions (Supplementary Figs. 25 and 26), one can see that the distinctive traits for both dimension 1 and dimension 2 are the variables "chemical family" and "coordination elements", with the "Ln ion" choice appears in a distant third place, while "$U_{\mathrm{eff}}$" participates only in dimension 1. This is similar to what was seen in Supplementary Section 4.2. The FAMD factor map (Supplementary Fig. 27) provides additional information on the actual values of the variables presented by the samples and their relation to the two main dimensions.

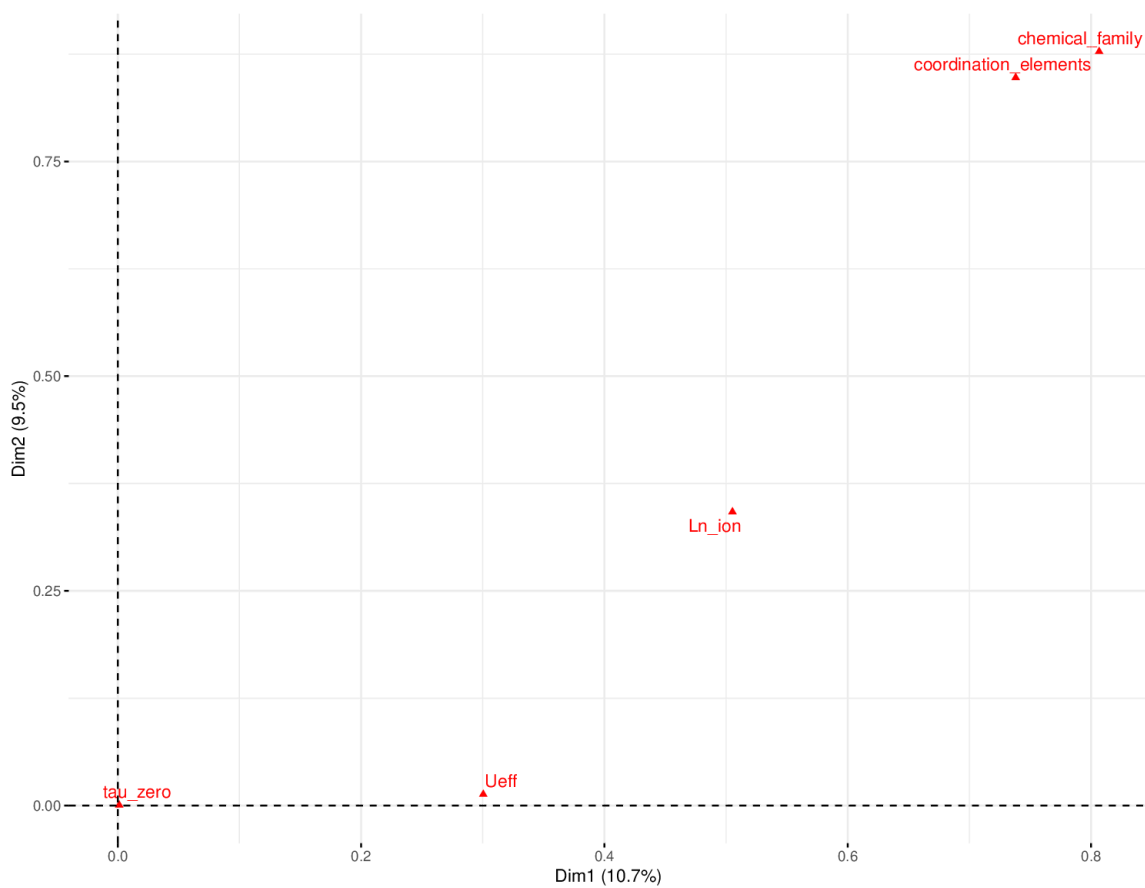

**Supplementary Figure 25 | Representation of the physical and chemical variables according to a FAMD method.**

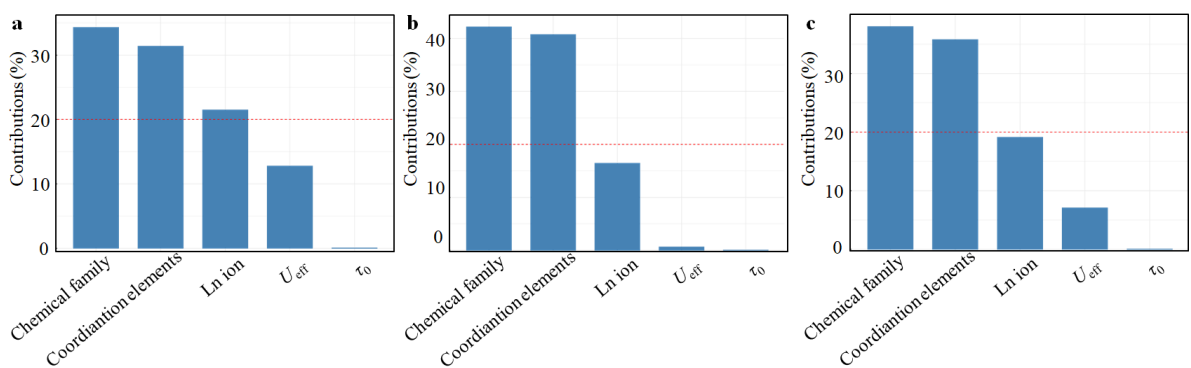

**Supplementary Figure 26 | Contribution from the main chemical and physical variables to the main FAMD dimensions. a**, Contributions to dimension 1. **b**, Contributions to dimension 2. **c**, Combined contributions to dimensions 1 and 2.

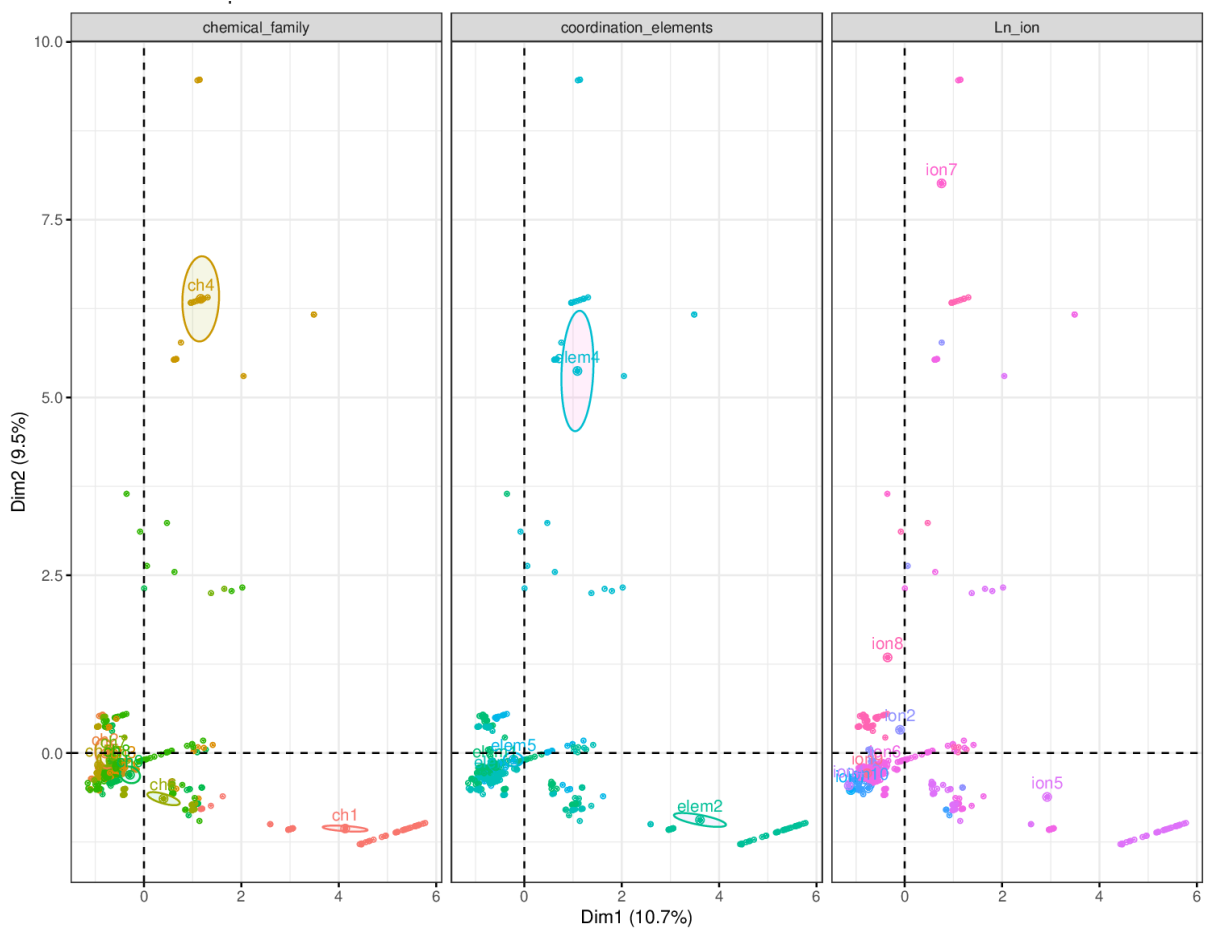

**Supplementary Figure 27 | FAMD factor map.** The groupings of the different values for the three main chemical variables is shown. See numbering convention for the categories of variables in Supplementary Section 1.

## 6.1. Magnetostructural clusters

We proceed to analyse the magnetostructural hierarchical clustering. This is comparable with the molecular clustering presented in Supplementary Section 4.2, but considering both physical and chemical variables. Dendrograms are represented in Supplementary Figs. 28 and 29, and a description of the different clusters follows.

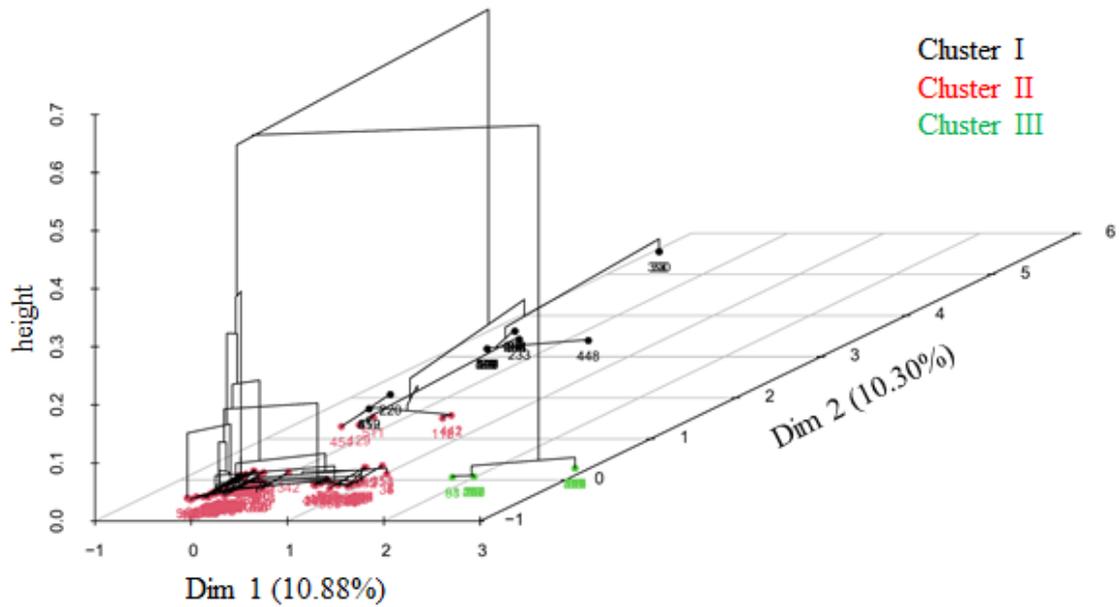

**Supplementary Figure 28 | Dendrogram depicting the hierarchical clustering on the factor map.** The numbers in the plot represent sample_IDs.

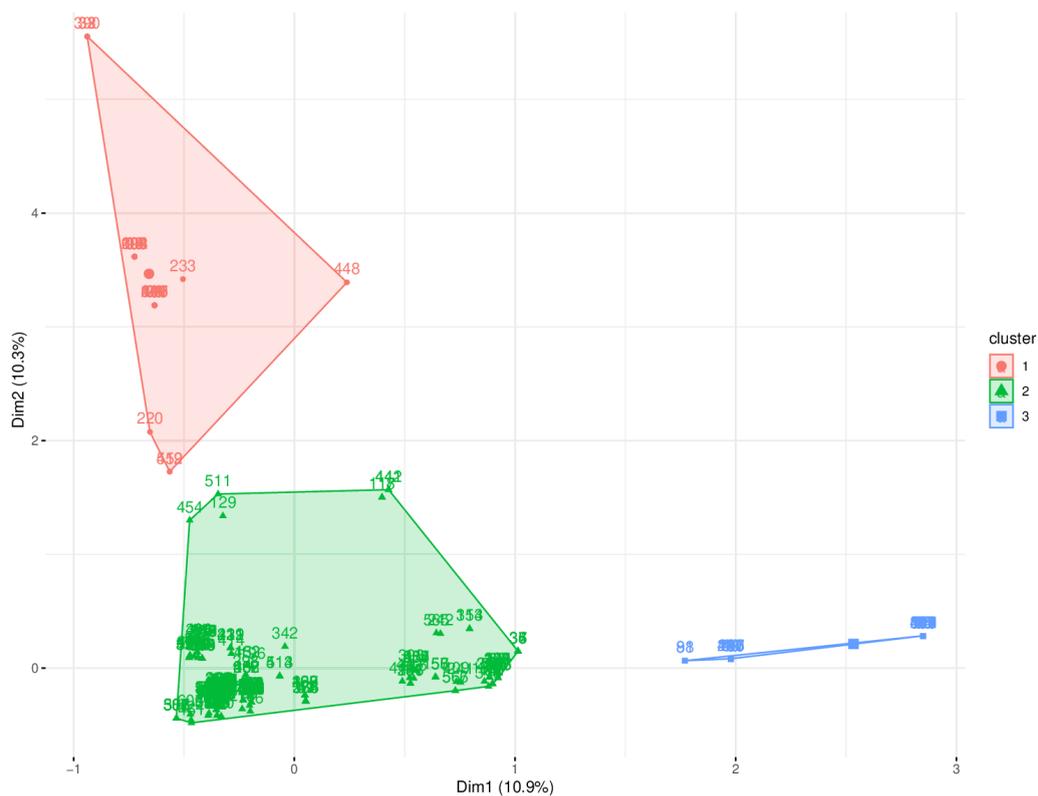

**Supplementary Figure 29 | Alternate view of the calculated hierarchical clustering on the factor map.** The numbers in the plot represent sample_IDs.

The main results from these analysis are as follows:

-Cluster I: $U_{eff}$ is the most associated variable in cluster I. The average value of $U_{eff}$ (252 K) in cluster I is considerably higher than the general average of $U_{eff}$ (117 K). In addition, the average value of $T_{hyst}$ in cluster I (9.9 K) is also significantly above the average value of $T_{hyst}$ (5.5 K). Indeed, cluster I is characterised by higher-than-average values of $U_{eff}$ and $T_{hyst}$.

-Cluster II is characterised by close-to-average values of $T_{B3}$ (18.5 K < 19.2 K) and $T_{hyst}$ (5.2 K < 5.5 K), and lower-than average $U_{eff}$ (87 K < 117 K).

-Cluster III, like cluster I but less intensely, is characterised by higher-than-average values for $U_{eff}$ (353 K > 117 K) and $T_{B3}$ (26 K > 19 K).

A partial clustering taking into account of the first 1000 data points (*i.e.* discarding data from 2018 and 2019) results in a very similar classification, but primarily characterises cluster I by higher-than-average values for $U_{eff}$ and $T_{B3}$ and cluster III by higher-than-average values for $T_{hyst}$ and $U_{eff}$. As a notable difference, discarding recent data results in a significant decrease in the average value for cluster III down to $U_{eff}$ = 199 K.

This general "magnetostructural" clustering classification, when described strictly from the point of view of the chemical variables, is depicted in Supplementary Fig. 30 and can be simplified to:

·cluster I: metallocene-type sandwiches, carbon as donor atoms, with $Ho^{3+}$ and $Er^{3+}$ as the most abundant ions.

·cluster II: predominantly mixed ligands, *i.e.* a mixture of different coordination ligands, predominantly $Dy^{3+}$ ion, and either only oxygens or a mixture of nitrogen and oxygen as donor atoms. This is by far the most abundant class of compounds with reported $U_{eff}$.

·cluster III: $Tb^{3+}$ ion (followed by $Dy^{3+}$), LnPc$_2$ family, nitrogens as donor atoms. These values for these variables are partially overlapping.

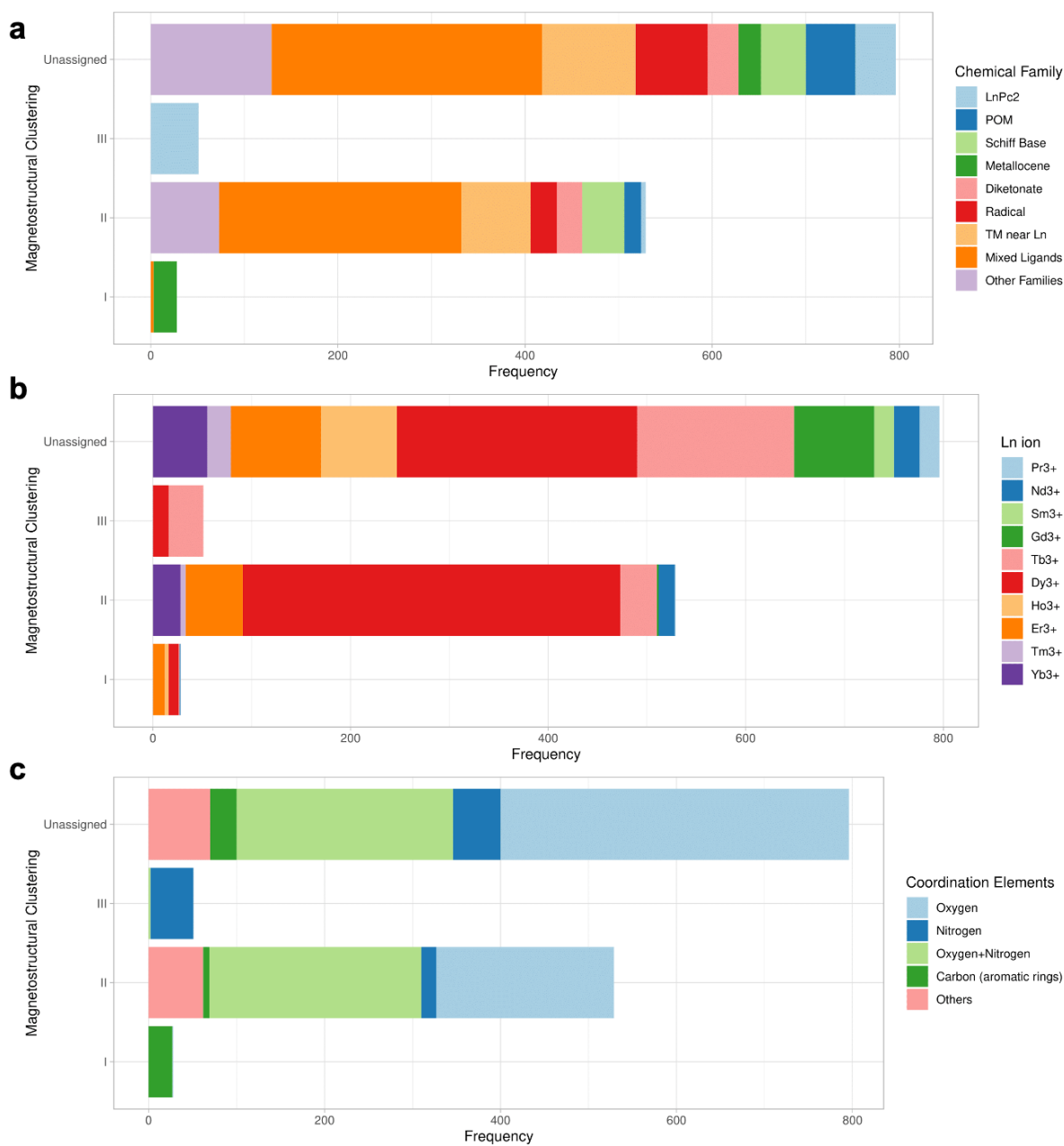

**Supplementary Figure 30 | Bar charts for magnetostructural clusters and their relation with the main chemical variables.** From top to bottom, the bar charts are filled according to: **a,** chemical family; **b,** lanthanide ion and **c,** coordination elements in the coordination sphere.

The correlation between the magnetostructural clusters I-II-III and the chemical clusters A-B-C-D-E is depicted in Supplementary Fig. 31, and can be summarised as follows; for readability we add the main feature of each cluster in parentheses:

-cluster I is mostly composed of samples from cluster E (metallocene)

-cluster II is a mixture of samples from cluster B (oblate) and C (prolate), and as well some from cluster D, but mostly from B

-cluster III is mainly samples of cluster D (LnPc$_2$)

-samples outside the I-II-III classification (not assigned), meaning samples with no recorded value of $U_{eff}$ are a mixture of all of the A-B-C-D-E cases, notably including all of the cases of cluster A ($Gd^{3+}$ complexes), and are, in order of relative abundance: B, C, A, D, E.

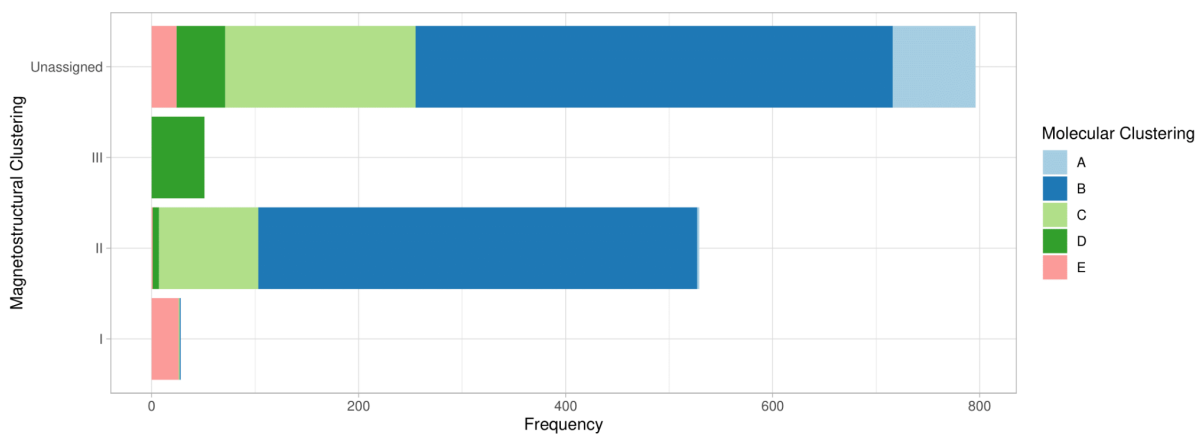

**Supplementary Figure 31 | Bar charts for magnetostructural clusters and their relation with the molecular (chemical) clusters.** NA stands for not assigned samples, *i.e.* samples that do not belong in any of the 3 magnetostructural clusters I-III.

As in the molecular clustering, depending on the height of the dendrogram cut in Supplementary Figure 28, one can obtain more or less fine-grained clusters. Our dataset contains a main categorisation with 3 tipologies I-III (+ "Unassigned"), described below, and we always offer an alternate, finer categorisation within the same hierarchical clustering (mag_struct_cluster_2 in the dataset) with 8 tipologies I-VIII (+ "Unassigned"), which is also included in the Data tab of the the SIMDAVIS dashboard.

**Supplementary Section 7. Extended SHAPE analysis, comparison with reference polyhedra**

### 7.1 Methodology

In order to define the ligand environment to analyse the reference polyhedra, we used PyCrystalField[64] to extract the nearest neighbour ligands to define the reference polyhedra from the material crystallographic information files.[65] PyCrystalField is a software designed to calculate the ligand and crystal field Hamiltonian of a single ion using a point charge model. Prior to this study, PyCrystalField considered only the space-group symmetry of the crystal, which is generally enough for conventional solid-state materials. Due to the diverse and low-symmetry environments that are generally found in molecules, and thus in the SIMDAVIS dataset, we added the ability to identify near-symmetries using continuous symmetry measures.[66] In the present work, this allows analysis of the correlations between the shape of the coordination environment and other variables. In future works, it will also allow intelligent predictions of the single-ion states and quantization axes of low-symmetry crystals and molecules (including single-ion magnets).

Equipped with this new functionality, we wrote a script to batch-process the ligand environments of all materials in the SIMDAVIS database, using the following routine. In cases where the coordination number CN was identified in the original study, we took the $n$=CN nearest atoms to the central magnetic ion as the coordination sphere. However, in some cases CN was not identified in the literature, for a variety of reasons. In these cases we used the following automated procedure: starting with $n$=7, we took the nearest $n$ atoms to be the coordination sphere, and calculated their geometrical centroid. If the central magnetic ion was off from the centroid by more than 25% of the greatest bond length, we added another ligand (or group of ligands if the next nearest ligands are all the same ion) and recalculated. This procedure was followed up to $n$=20. After identifying the appropriate ligand sphere, we used the SHAPE software[67] to identify the closest reference polyhedra, as well as the continuous shape measure (CSM) to it, and added both to the dataset. Note that in the specific case of aromatic ligands, the formal coordination number CN in our dataset may be much less than the number of atomic ligands $n$ because aromatic rings are counted as a single coordination ligand. The script for batch-processing can be found at:

https://github.com/asche1/PyCrystalField/tree/master/Publications/SIM_BatchProcess

Although we do not report the results here, PyCrystalField allows us to estimate the single-ion ground state wavefunction of each material using the point charge model. There are significant challenges in appropriately assigning effective charges to the ligands, and in orienting the coordination sphere in extreme low-symmetry cases. Because solving these issues is an ongoing task, we leave the wavefunction calculations for a future iteration of this project.

**7.2 Results**

Let us analyse the most frequent coordination polyhedra, and in particular the polyhedra that are most frequent for the most common coordination numbers, namely CN = 7-9 (see Supplementary Figure 32.1). And in this analysis, let us answer the question: are the different coordination polyhedra equally good, in terms of frequently resulting in magnetic memory effects at relatively high temperatures? The results, which we shall discuss below, are in Supplementary Figure 32.2.

First note that, since the data are relatively scarce compared with analyses of previous sections, we were merely looking for differences that can be striking when the results of a polyhedron type are compared with the results of the whole dataset. Moreover, it is important to allow the reader to keep in mind the scarcity of this kind of information, and the possibility of the insights to be overly influenced by a few outlier compounds. For this purpose, when the number of samples < 50 for a given type of polyhedron, in the discussion we include in parentheses the total samples as "ts", and the number of unique doi (articles) associated with this polyhedron as "ud".

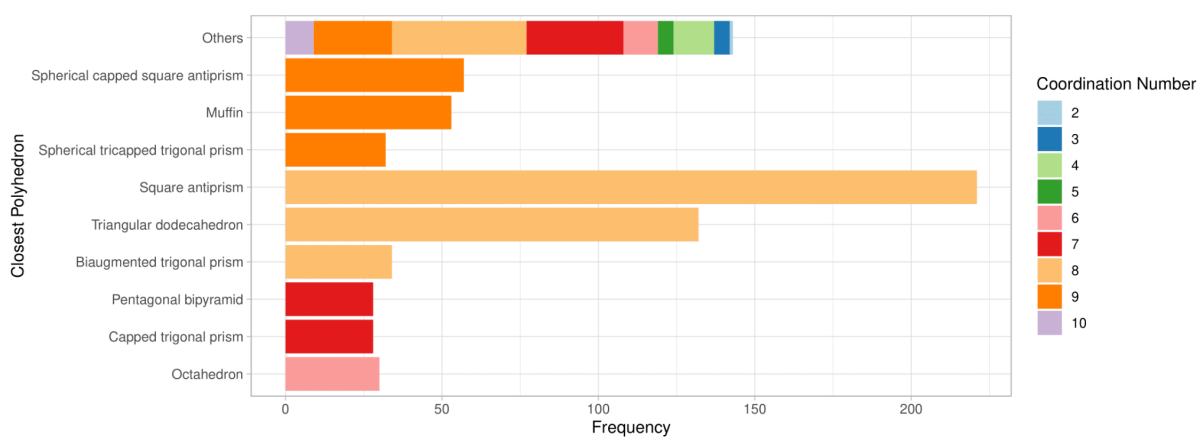

**Supplementary Figure 32.1 | Bar charts for closest polyhedra and their relation with the coordination number.** One can see the most abundant values for CN, and the most abundant polyhedra.

Let us start with CN=7, where, as we shall see, the sharpest contrast can be found between the different coordination shapes. The most common polyhedra with CN=7 are pentagonal bipyramids and capped trigonal prisms. Capped trigonal prisms (28 ts, 13 ud) tend to present no frequency dependent x'', much less an ac peak or hysteresis. In this sense, capped trigonal prisms presents among the worst magnetic results for any shape in the present dataset, at least when it comes to direct experimental results (its distribution of $U_{eff}$, $\tau_0$ values, while following this tendency, i.e. slightly lower $U_{eff}$, higher $\tau_0$, are not markedly different from the rest of the dataset). In contrast with the capped trigonal prism shape, pentagonal bipyramid (29 ts, 18 ud) is apparently the shape with the highest tendency to present an ac peak at T > 2K and f = 1000 Hz and to present hysteresis (see Supplementary Figure 32.2).

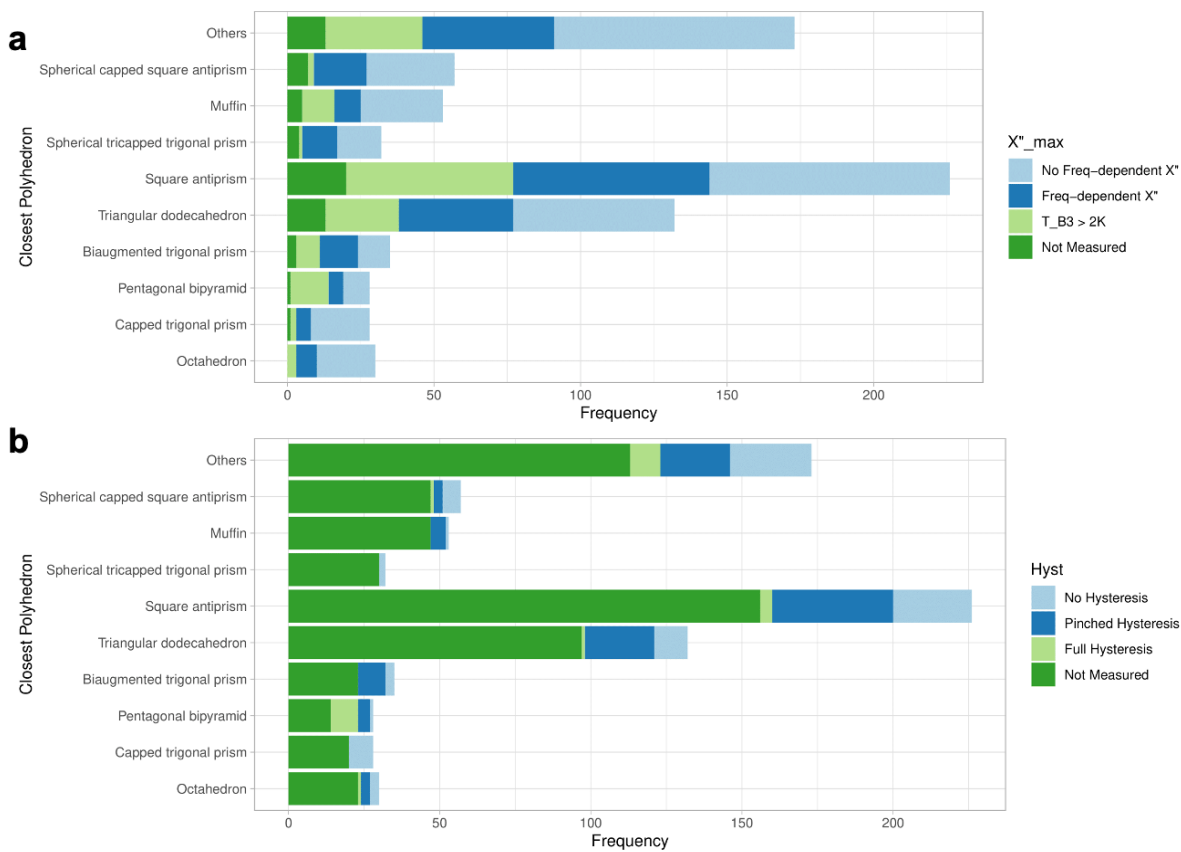

**Supplementary Figure 32.2 | Bar charts for closest polyhedra and their relation with the magnetic behaviour. a,** out of phase ac susceptibility response and **b,** hysteresis.

Moreover, said ac peak tends to be at very high temperatures compared with other shapes (median $T_{B3} > 30$ K), with the same observation being true for $T_{Hys}$ (median $T_{Hys} > 10$ K, when most other shapes present median $T_{Hys}$ in the window 2K< median $T_{Hys} < 4$K). Both of these observations can be quantified in Supplementary Figure 32.3. As can be seen in Figure 6, in terms of parameterized Arrhenius behaviour, pentagonal bipyramids in our dataset present outstanding values of $U_{eff}$, with a median value $U_{eff}$ median ~ 400 K that is an order of magnitude above the usual for other coordination shapes ($U_{eff}$ median ~ 40 K). $\tau_0$, as we see elsewhere in the present analysis, is strongly correlated with $U_{eff}$, and in this case pentagonal bipyramids tend to present values of tau0 that are much below the usual range for other polyhedra (median $\tau_0 \sim 10^{-11}$ s, when most other shapes present median $10^{-8}$ s $< \tau_0 < 10^{-7}$ s).

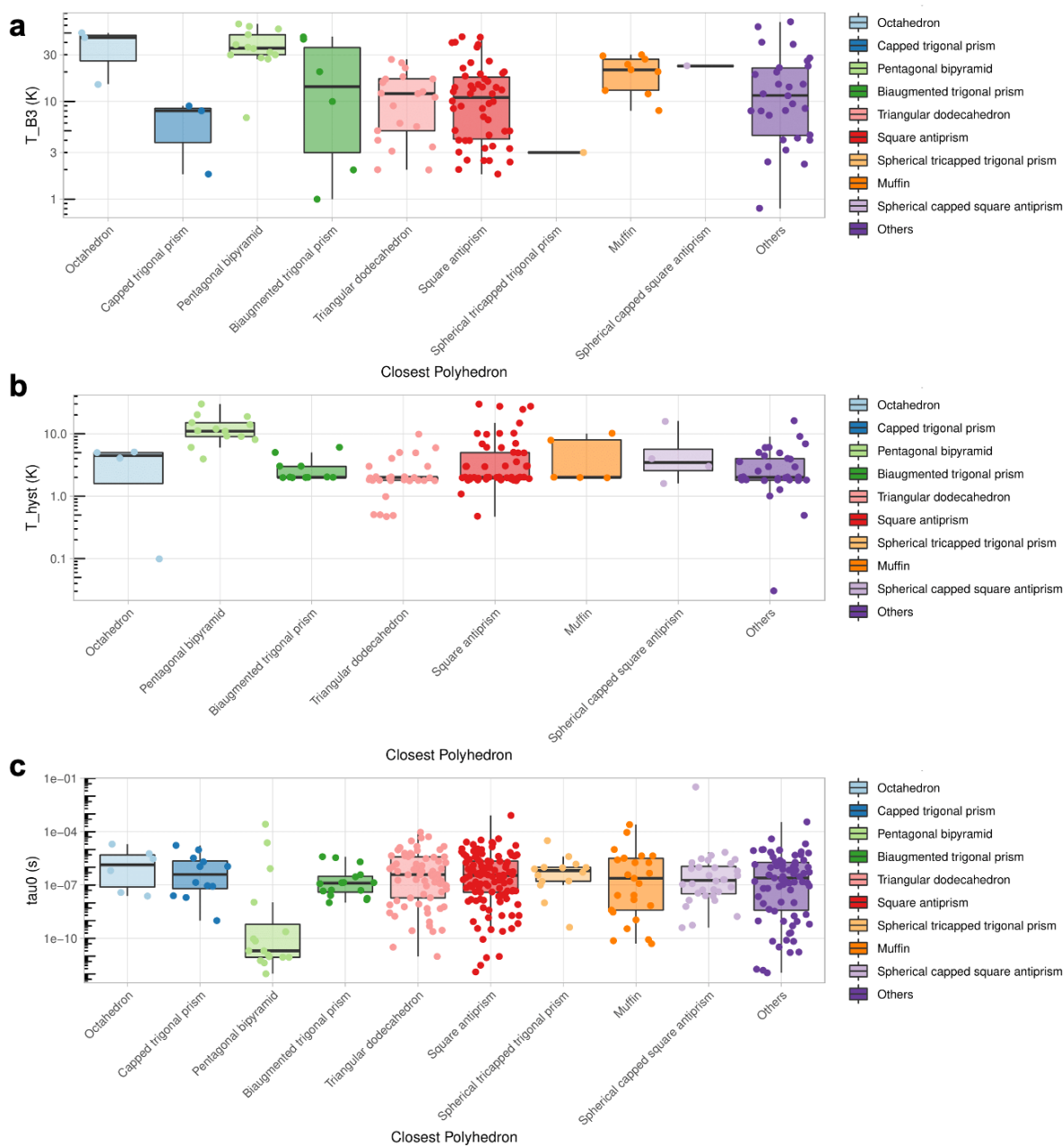

**Supplementary Figure 32.3 | Box plots for closest polyhedra and their relation with the magnetic behaviour.** From top to bottom: $T_{B3}$, $T_{hyst}$, $\tau_0$.

Let us investigate this unusual behaviour presented by the pentagonal bipyramid (PBPY-7) and try to rationalise it with the help of the spherical harmonics (see Supplementary Figure 32.4). First one needs to note that a regular PBPY-7 polyhedron, much like a regular square antiprism, presents only diagonal components in the crystal field Hamiltonian, with all other contributions being zero, either because the ligands are located on a node or due to symmetry cancellation. This has been very often argued to be a positive feature, since it facilitates pure crystal field states, minimising the mixing and thus lowering the transition probabilities between states. However, this cannot be the only reason behind the success of PBPY-7 complexes, since we have seen that square antiprisms are not as markedly good as SIMs. We need to note, additionally, that a PBPY-7 presents only two types of ligands: perfectly axial ligands and perfectly equatorial ligands. In both cases, this means that the position of each donating atom coincides with an angular maximum of the diagonal spherical harmonics, for $B_2^0$, $B_4^0$ and $B_6^0$. In this context, let us consider vibronic coupling. Vibrations that alter metal-ligand bond distance tend to be high-energy and thus, in practice, are not the ones limiting the working temperature for slow relaxation of the magnetisation. In contrast, vibrations where metal-ligand bond distance is kept constant and only the angles change can be low frequency. It is against these distortions where being exactly at an angular maximum is important, since it means that the first derivative of the spherical harmonics, and thus, the change in any diagonal term in the crystal field Hamiltonian, is zero. We postulate that this special geometrical correspondence allows pentagonal bipyramids to be specially resilient to angular (twisting, wagging, bending) vibrations which, being the lowest in energy, can be present even at low temperatures and facilitate magnetic relaxation most often.

| l: | | $P_\ell^m(\cos\theta)\,\cos(m\varphi)$ | | | | | | | $P_\ell^{|m|}(\cos\theta)\,\sin(|m|\varphi)$ | | | | | |
|----|---|---|---|---|---|---|---|---|---|---|---|---|---|---|
| 0 | s | | | | | | | | | | | | | |
| 1 | p | | | | | | | | | | | | | |
| 2 | d | | | | | | | | | | | | | |
| 3 | f | | | | | | | | | | | | | |
| 4 | g | | | | | | | | | | | | | |
| 5 | h | | | | | | | | | | | | | |
| 6 | i | | | | | | | | | | | | | |
| | m: | 6 | 5 | 4 | 3 | 2 | 1 | 0 | -1 | -2 | -3 | -4 | -5 | -6 |

**Supplementary Figure 32.4 | Spherical harmonics.** Only the shapes corresponding to $l$ = 2,4,6 are relevant for Crystal Field in lanthanides.

In the case of CN=8, the most common polyhedra are square antiprisms, triangular dodecahedra and biaugmented trigonal prisms (35 ts, 21 ud), which not only come in a far third place but also consistently present higher distortions CSM. Between square antiprisms and triangular dodecahedra in our dataset, there is no marked difference neither in the magnetic performance nor in the parameterization. This is notable, since proximity to square antiprisms has historically very often been invoked as a promising geometry to obtain SMM behavior. Biaugmented trigonal prisms tend to present higher values of $U_{eff}$ and lower values of $\tau_0$ compared with other polyhedra, but there is no marked difference in their magnetic performance.

Finally, in the case of CN=9, the most common polyhedra are spherically capped square antiprisms, muffins and spherically tricapped trigonal prisms (36 ts, 20 ud). The differences between the three shapes are more marked than in the case of CN=8 but less than in the case of CN=7. Among CN=9, muffin polyhedra, which present a high degree of distortion compared with any other polyhedra in our dataset, present an ac peak at T > 2K at f = 1000 Hz and also present hysteresis most often, and in both regards spherically capped square antiprisms come in second place and spherically tricapped trigonal prisms comes last. In terms of maintaining magnetic memory up to high temperatures, CN=9 muffins are not as exceptional as the CN=7 pentagonal bipyramids we discussed above, but they do perform markedly better than any CN=8 shape, again, shockingly, given the popularity of square antiprisms.

Also deserving further theoretical study (in the next section) is the fact that $U_{eff}$, $T_{B3}$, $T_{Hys}$ vs CSM present overall positive slopes, meaning that higher distortion tends to produce higher $U_{eff}$, $T_{B3}$, $T_{Hys}$. This can be initially surprising, if one is thinking in terms of extradiagonal parameters in the crystal field Hamiltonian. It has often been argued precisely that ideal geometries are preferable to avoid mixing in the spin states. However, it makes more sense if one considers that reference polyhedra in SHAPE are as spherical as possible, meaning $B_2^0$ tends to cancel.

**Supplementary Section 8. Extended SHAPE analysis, axial distortions**

**8.1 Methodology**

For the analysis of the effect of structural distortions, we aimed to quantify the elongation or contraction of each coordination sphere employing SHAPE. Note that the reference polyhedra in SHAPE are as spherical as possible, in the sense that all vertices are at the same distance from the centre. So, after determining the overall distortion as CSM for the coordination sphere of each complex, our goal in this step was to get some insight into the character of said distortion: is it systematically elongated, or contracted, or are the distortions of a rather "isotropic" character? This is however not a standard feature of the program, so we developed an auxiliary methodology, which we explain here.

We first obtained a set of reference structures for the 9 most represented families of coordination polyhedra within our dataset (see Supplementary Table 6): capped square antiprism (CSAPR-9), square antiprism (SAPR-8), triangular dodecahedron (TDD-8), pentagonal bipyramid (PBPY-7), tricapped trigonal prism (TCTPR-9), muffin (MFF-9), biaugmented trigonal prism (BTPR-8), octahedron (OC-6) and capped trigonal prism (CTPR-7). The coordinates for the reference polyhedra were obtained from the SHAPE program by employing the keyword %test . We fixed the resulting orientation, which puts the z axis as a maximum symmetry axis. This z axis was the one we employed as a reference for axial compression or axial elongation. Note that in principle all results in the present analysis are conditioned by the choice of the elongation/compression axis.

Supplementary Table 6: Frequencies of the 9 most abundant closest polyhedra within the data set.

| # | Closest polyhedron | Counts |
|---|---|---|
| 1 | Square antiprism | 226 |
| 2 | Triangular dodecahedron | 132 |
| 3 | Spherical capped square antiprism | 58 |
| 4 | Muffin | 54 |
| 5 | Spherical tricapped trigonal prism | 36 |
| 6 | Biaugmented trigonal prism | 35 |
| 7 | Octahedron | 31 |
| 8 | Pentagonal bipyramid | 29 |
| 9 | Capped trigonal prism | 28 |

Path to repository with the sampling script:

https://github.com/silsgs/extended_SHAPE_analysis/blob/main/simpre_sampling_sph.sh

Example of SHAPE script generating an ideal 6-vertices octahedron coordination structure ("OC-6").

```
$ Generate X-vertices reference eg: 6-vertices refs OC-6; 7th atom is the metal

%test

  6 7

  3
```

Output file:

```
  7

H         0.00000000    0.00000000   -1.00000000
H         1.00000000    0.00000000    0.00000000
H         0.00000000    1.00000000    0.00000000
H        -1.00000000    0.00000000    0.00000000
H         0.00000000   -1.00000000    0.00000000
H         0.00000000    0.00000000    1.00000000
N         0.00000000    0.00000000    0.00000000
```

In a second step, and starting from the ideal polyhedron structure of each family (obtained from SHAPE), we propose to standardise a methodology to quantify non-regular structures assigning a value representative of their 'elongated' or 'compressed' character versus the ideal structure. For that, ideal structures were either axially compressed or elongated up to 10%, 20% and 50%. In practice, an extension/compression factor between 0.5-1.5 over the z coordinate of each ideal polyhedron was applied. For each of the 8 new distorted structures (varying z' from z'=z·0.5 to z'=z·1.5) for each reference polyhedron in Supplementary Table 6, we prepared a file of user-defined reference polyhedra. We run SHAPE in batches employing the code 0 to command SHAPE to read a user-defined "ideal" (.ide) reference file. Following this procedure, we obtained, first, a series of CSM(z') values, measuring how much each coordination polyhedron differs, not from the reference "spherical" polyhedra, but from each of the elongated or compressed references. In turn, this allowed us to obtain a series of new axial distortion quantifications Δ by difference of continuous shape measures CSM:

axial distortion $\Delta_{CSM(z')}$ = CSM(z')-CSM(original)

Let us briefly discuss what this axial distortion $\Delta_{CSM(z')}$ means in practice. If (for a given coordination sphere, reference polyhedron and axial distortion) $\Delta_{CSM(z')}$ is negative it means that this coordination sphere is closer to an axially distorted reference (smaller value of CSM(z')) rather than to the original one (larger value of CSM(original)). For example, let us say we are examining a hypothetical square antiprism (SAPR) where the CSM with respect to the reference SAPR is CSM(z·1) = 3. The question we want to answer is: is this strong distortion mostly axial, mostly isotropic, or somewhere in between? An example of a mostly axial distortion case would be an elongated SAPR which, other than that, is a rather perfect D4d. An example of a mostly isotropic distortion would be a distorted SAPR where all spatial coordinates of the coordinating atoms have suffered random changes. An intermediate example, of course, would be a mixture of random distortions and axially directed ones.

A way to distinguish between these three situations is calculating $\Delta_{CSM(z')}$ for different values of z'. We can see this as a sampling of axial distortions. The values of $\Delta_{CSM(z')}$, plotted against z', look like a local minimum. Thus, the value of z' where one finds the most negative value of $\Delta_{CSM(z')}$, i.e. the minimum value of CSM, corresponds to the reference polyhedron with the elongation/compression that most closely resembles the one of the real coordination sphere. So with this set of calculations we would have an (admittedly rough) categorisation corresponding to this sampling of $(z'/z) \in \{0.5, 0.75, 0.9, 1, 1.1, 1.25, 1.5\}$.

However, our goal is slightly more ambitious, so additionally we prepared a minimally elongated structure (elongated by 1% i.e. multiplying z·1.01) to establish a distortion slope, which is a description of how much is CSM varying when certain distortion is applied, and it is defined as

axial distortion CSM slope s = 100*(CSM(z·1.01)-CSM(original))

This axial distortion CSM slope served as an intermediate step towards obtaining our goal "axial distortion" continuous metric. We compared the $\Delta_{CSM(z')}$ values of all structures with their axial distortion CSM slopes, and found some pretty robust correspondence, not just qualitative but even quantitative. That is, not only the sign of the slope corresponds in every case to the sign in $\Delta_{CSM(z')}$, but also larger slopes consistently correspond to negative $\Delta_{CSM(z')}$ for higher values of z'. Let us see it with some examples. If $\Delta_{CSM(1.5)} \lesssim 0$, it would mean that the structure is extremely elongated, because the complex is a bit closer to a SAPR with z' = z·1.5 than to a "spherical" SAPR. One could assume that it's axial distortion is of the order of z' = z·1.25 (the elongation point where a structure starts being to a reference with z' = z·1.5 than to the original reference with z' = z). In these cases, we tend to find a threshold value of s ≈ -10. Or, if $\Delta_{CSM(z')} > 0$ for all values of z', it means that the complex is not at all elongated or compressed, but rather distorted in another way. Thus, any elongation or compression to the reference polyhedron can only increase the CSM value. In these cases, we tend to find a threshold value of |s| < 2. Finally, if $\Delta_{CSM(0.9)} \lesssim 0$, it would means that this structure is at least slightly contracted. One could assume that its axial distortion corresponds approximately to z' = z·0.95 (the compression point where a structure starts being to a reference with z' = z·0.9 than to the original reference with z' = z). In these cases, we tend to find a threshold value of s ≈ 2.5

We were able to extract the following approximate relation to estimate the axial distortion:

axial distortion = s² -2.5s +1

## 8.2 Results

In the Supplementary Fig. 33 a correlation of three magnetic parameters $U_{eff}$, $U_{eff,ff}$, $T_{B3}$, with the elongation / contraction of two most important polyhedra (PBPY-7 and SAPR-8) is presented.

In the top graph we observe that in the case of the pentagonal bipyramid, most of the data is restricted to a 'compressed' range of [0.92-0.94] with a dispersion of $U_{eff}$ values from approximately 5K up to 900K. In contrast, for the case of the square antiprism, a high number of real structures are considered as 'elongated', with values around [1.00-1.20]. In all these cases, $U_{eff}$ values still show some dispersion with a soft dependence showing a $U_{eff}$ increase with the elongation of the structure. In both cases, a small but relevant correlation of the two parameters can be extracted being a positive slope for the case of the square antiprism and negative for the pentagonal bipyramid.

In the centre panel, the full-fit effective barrier $U_{eff,ff}$ is represented versus the elongation / compression of the square antiprism structure which is the only coordination geometry where there is sufficient $U_{eff,ff}$ data to show the dependence between these two values. In this case the number of points is very scarce because very few studies performed a full-fit study considering Direct, Orbach and Raman relaxation paths with a crystal structure solved. In concordance with that happening in the case of the Orbach-only fit, a weak positive correlation is found, with the barrier height increasing with the structural elongation.

Finally, regarding the bottom graph, the $10^3$ Hz AC blocking temperature ($T_{B3}$) is shown versus the structural distortion. In this case, the pentagonal bipyramid shows the more relevant correlation with the structural compression/ elongation, again showing a negative slope. For the square antiprism, the dependence is noisy but positive, in agreement with that observed in the previous plots.

Let us expose a factible explanation for this behaviour based on the axial term $B_0^2$. In the case of the pentagonal bipyramid, a compression of the axial ligands produces an increment of the axial strain and, in consequence, a higher $B_0^2$ value, thus increasing the effective barrier and the blocking temperature. In apparent contrast, the regular square antiprism turns more axial when the ligands are distorted by an axial elongation, with the same consequences as the former. Note that these observations are only valid for oblate ions ($Tb^{3+}$, $Ho^{3+}$, $Dy^{3+}$) which are among the most usual in the dataset. In the case of prolate ions, ($Er^{3+}$, $Tm^{3+}$, $Yb^{3+}$) the expected behaviour would be the opposite.

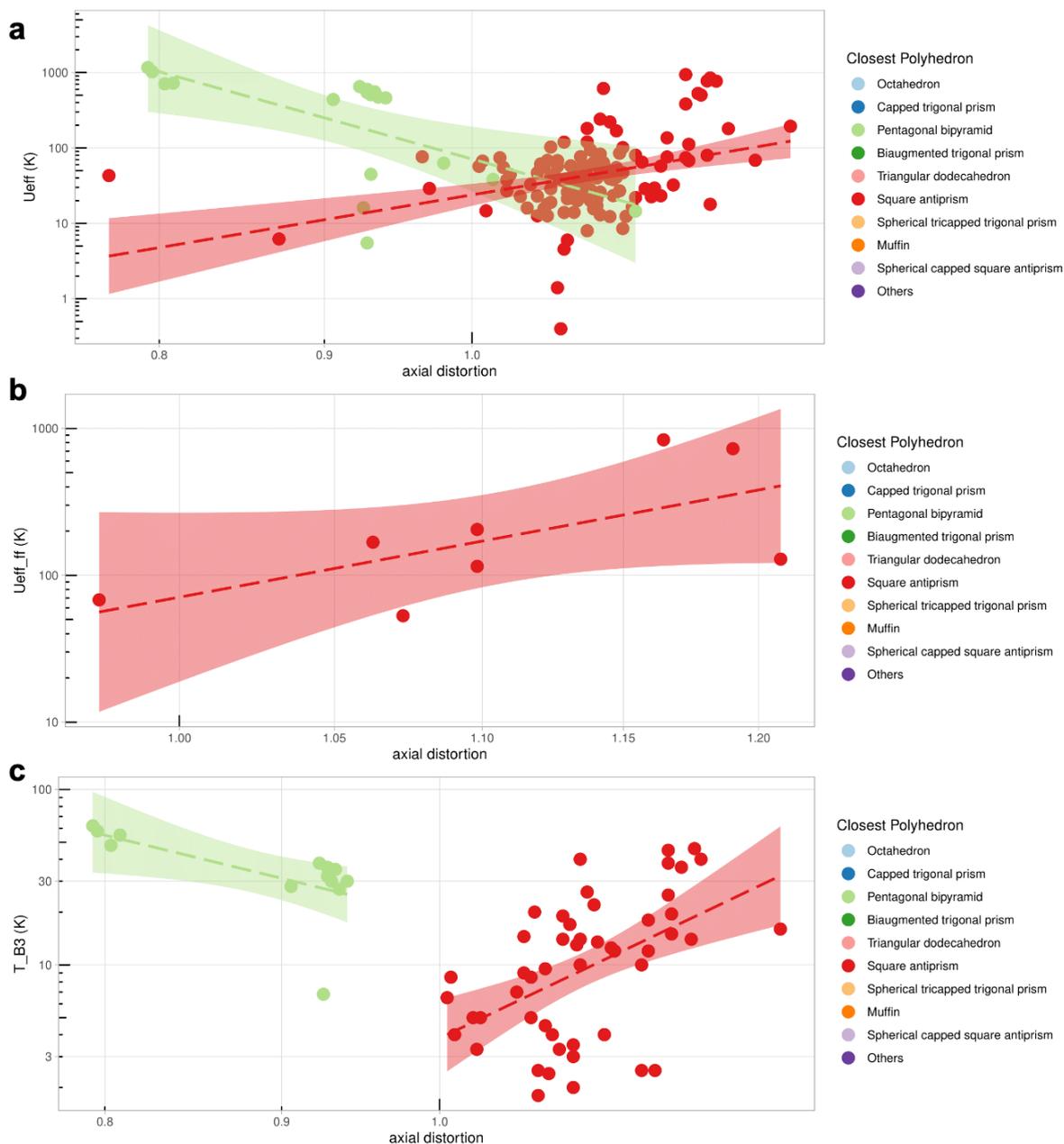

**Supplementary Figure 33 | Scatter plot showing the variation vs axial distortion for $U_{eff}$ (top), $U_{eff,ff}$ (centre), $T_{B3}$ (bottom).** The only displayed polyhedra are the ones that present a more clear dependence: square antiprism and pentagonal bipyramid. Note that not all samples will be present in all graphs (see Supplementary Figure 1.1). As a consequence, an (x vs y) plot can only include samples for which x and y are simultaneously present in the dataset.

**Supplementary Section 9. Extended figures and discussion about $U_{eff}$ vs Raman vs QTM**

**9.1 Investigation of the relation between different relaxation parameters**

In Figure 5 of the main text we have seen an apparent correlation between $U_{eff,ff}$ and Raman relaxation prefactor $C$: faster relaxation via an Orbach mechanism (lower $U_{eff}$) happens together with faster mechanism via a Raman mechanism (higher $C$). As we can see in Supplementary Figure 34, the correlation is of the same sign, but more noisy, in the case of $U_{eff,}$. Also interestingly, the correlation extends to the Raman exponent $n$, again with the same sign: faster relaxation at high temperature via an Orbach mechanism (lower $U_{eff}$) happens together with faster mechanism at high temperature via a Raman mechanism (higher $n$).

Let us start by addressing the fact of the so-called "anomalous" Raman exponents $n \neq \{7,9\}$ which are the norm rather than the exception in our dataset, and how this may be related to the meaning of the effective energy barrier $U_{eff}$ considering an Orbach relaxation process. The matter of the anomalously low Raman exponents was recently the subject of a theoretical work by Gu *et al.*[68] First, it is important to recall that considering a pure Raman relaxation process the standard exponent at low temperatures should be either $n = 7$ (non-Kramers ions) or $n = 9$ (Kramers ions). Gu *et al.* explained why anomalies where the exponent n is in the range $3 < n < 5$ are frequently obtained by a simultaneous fitting of the relaxation rates by a 'full fit' procedure considering Raman, Orbach and direct processes, but just a single path for each mechanism. As it is known, the maximum value of the vibration energies (Debye energy) assigned to a magnetic relaxation process limits the temperature where the pure Raman process is applicable. Gu argued that, as one analyses the system above such temperature, multiple Orbach processes increase their contributions to the total rate. Indeed, the highly discrete nature of the vibrational DOS in magnetic molecules means that a sum including a single Orbach barrier $U_{eff}$ and a single Raman contribution is an oversimplification. In particular, the low-energy nature of SIMs intra / intermolecular vibrations results in most of the magnetic measurements to be performed in the limit of the Debye energy and thus most of the measurements may be contaminated both by multiple Raman rates based on different phonons and by non-Raman rates. In a previous work[69] it was hypothesized that these values could be explained by using an optical-acoustic mechanism. However, Gu *et al.* argued that this explanation is not realistic for SIMs with a high magnetic anisotropy and a small Zeeman splitting in the ground doublet.

In any case, a plausible connection between the parameters assigned to different relaxation mechanisms can have precisely this source: the so-called "full" fit is still an oversimplification, up to the point where different parameters contaminate each other.

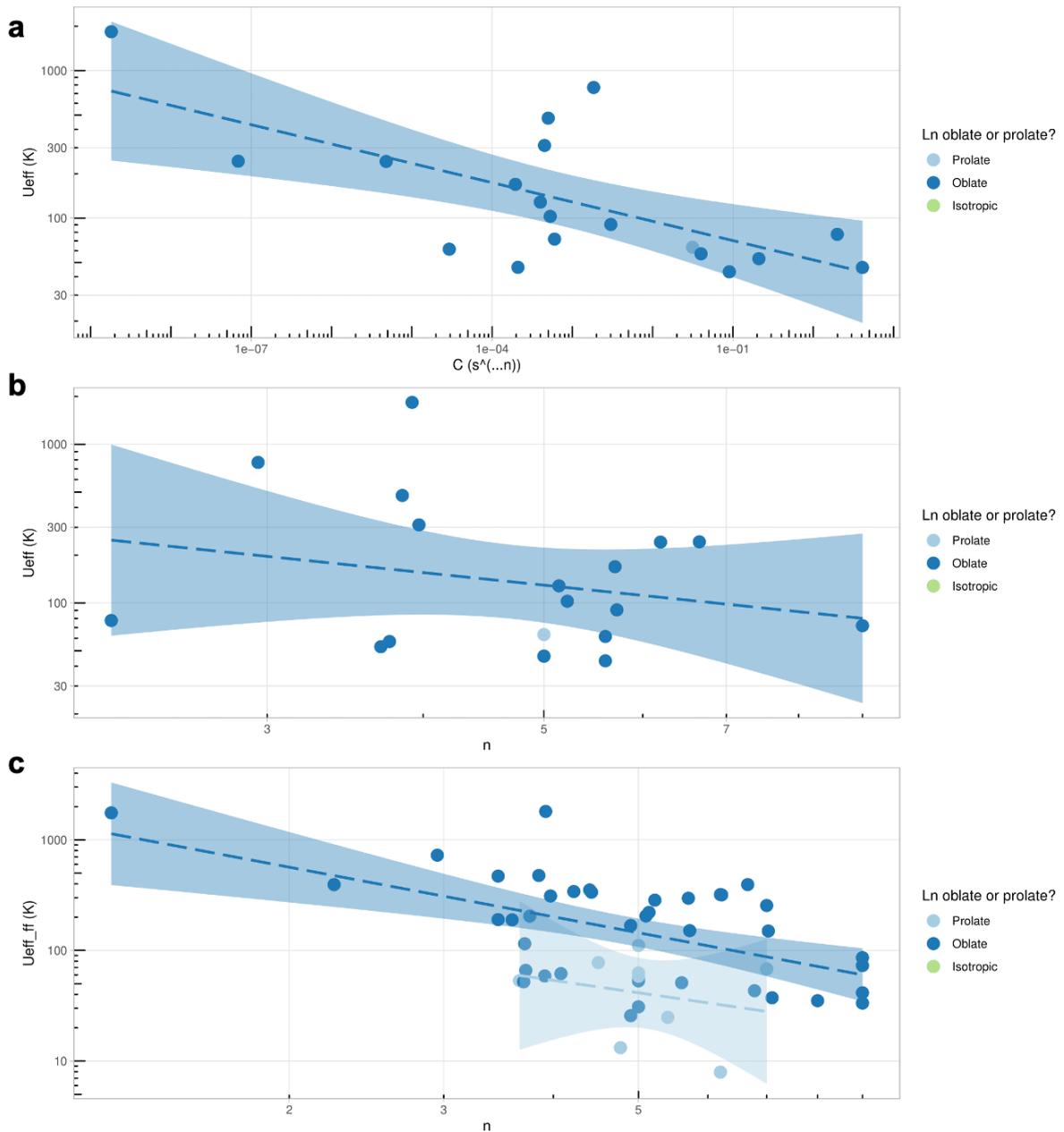

**Supplementary Figure 34 | Scatter plots showing the dependence between $U_{eff}$ and C (a), $U_{eff}$ and n (b), $U_{eff,ff}$ and n (c).** The correlation with n is less clear than in the case of C. Note that not all samples will be present in all graphs (see Supplementary Figure 1.1). As a consequence, an (x vs y) plot can only include samples for which x and y are simultaneously present in the dataset.

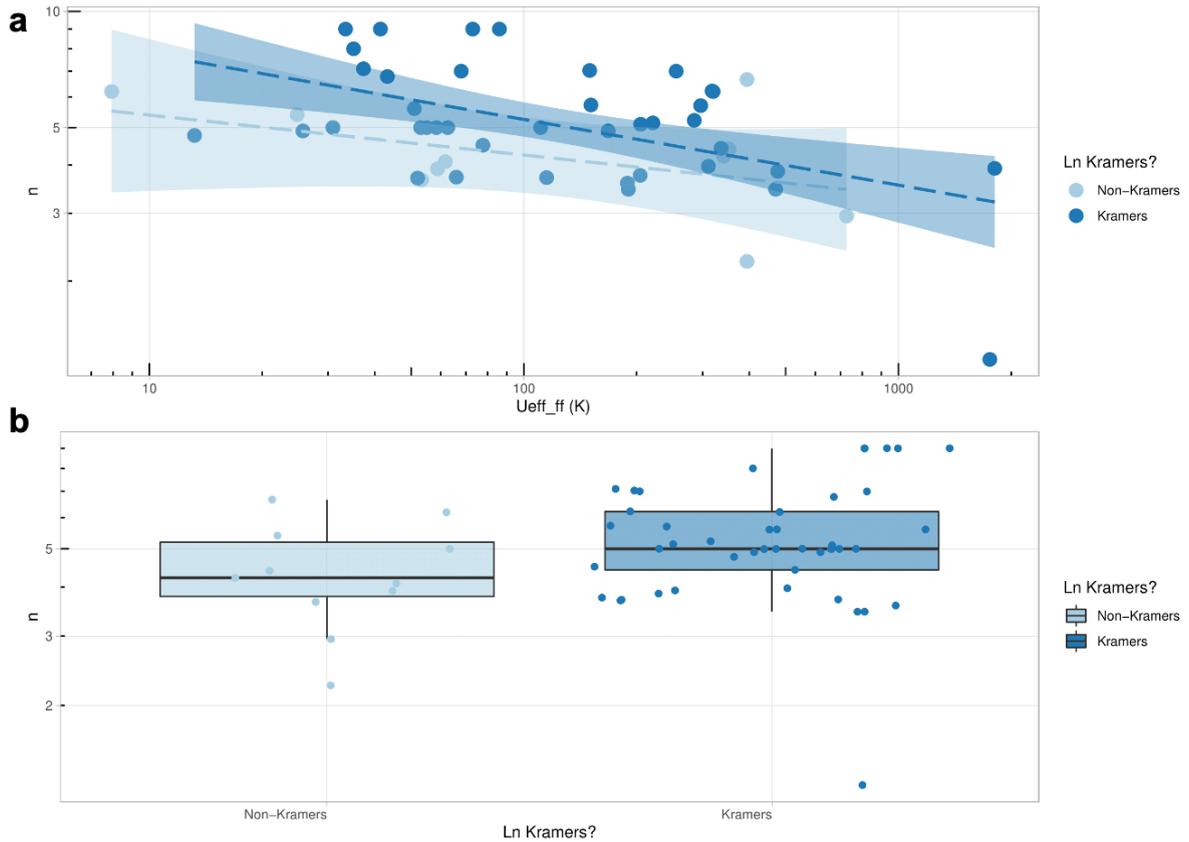

**Supplementary Figure 35 | Scatter plot showing the dependence between *n* and $U_{\text{eff}}$ (top); Boxplot showing values of *n* for Kramers and non-Kramers ions (bottom).**

In Supplementary Figure 35, we represent the tendencies for both Kramers / non-Kramers ions between the Orbach barrier ($U_{\text{eff}}$) and the Raman exponent (*n*), as well as the distribution of *n* for Kramers vs non-Kramers ions. It is worth noting that for lower $U_{\text{eff}}$ values, the corresponding Raman coefficients tend to the expected values 7 or 9 (9 for Kramers ions, 7 for non-Kramers). On the other hand, higher $U_{\text{eff}}$ values are expected to produce anomalous coefficients in the range $3 < n < 5$. Again, this can be rationalised considering that for high barriers, the measuring temperatures are well over Debye energies thus producing an incorrect consideration of the Raman coefficients.

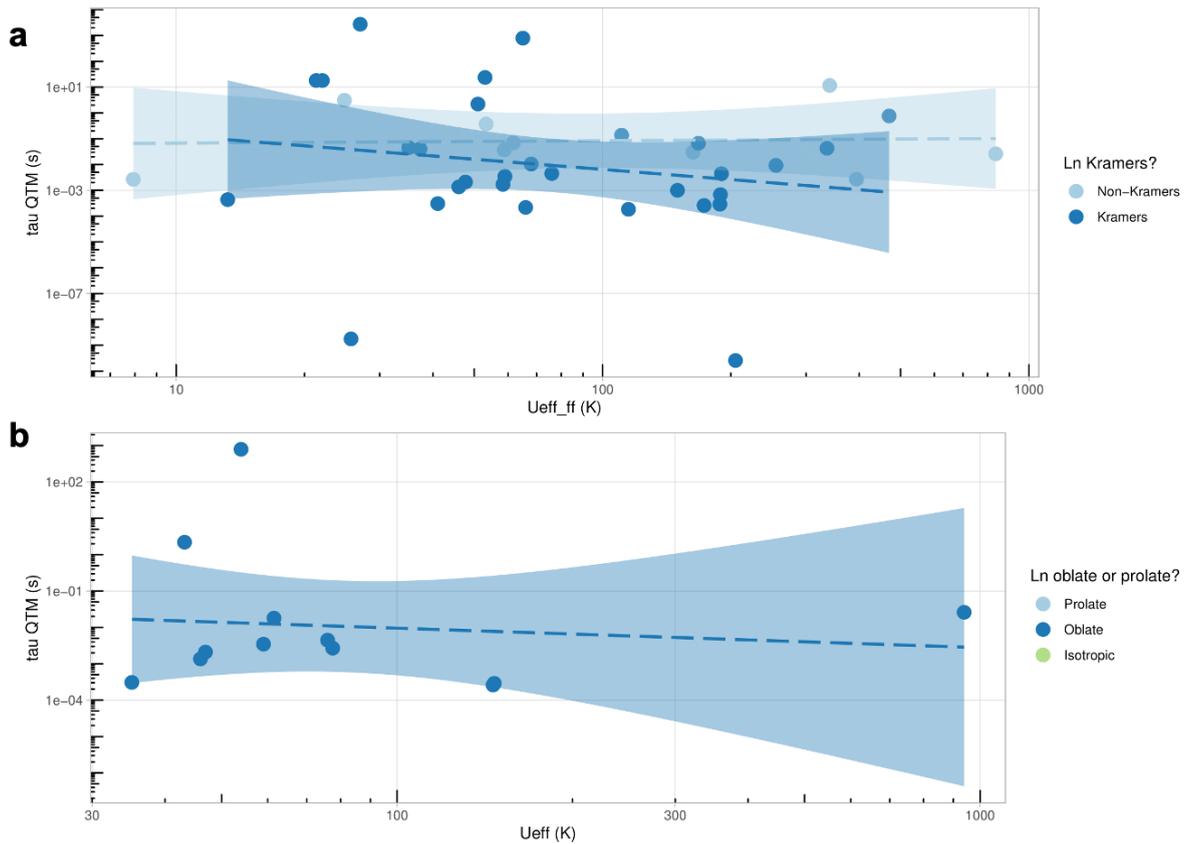

**Supplementary Figure 36 | Scatter plots showing the (lack of) dependence between $\tau_{\mathrm{QTM}}$ and $U_{\mathrm{eff}}$ (a), $U_{\mathrm{eff,ff}}$ (b).** There seems to be no correlation in this case.

In Supplementary Figure 36, we see the relation between $\tau_{\mathrm{QTM}}$ and $U_{\mathrm{eff}}$ (top), $U_{\mathrm{eff,ff}}$ (bottom), or rather, the lack thereof. This could be naïvely expected, of course, since Orbach and QTM are two physically independent mechanisms, but at this point it is somewhat surprising. If, as we have hypothesised, the correlation between Raman and Orbach parameters is due to so-called "full" fits being oversimplifications up to the point where different parameters contaminate each other, why should QTM be exempted from this mixing? This supports the hypothesis we expose in the main text of high values of $U_{\mathrm{eff}}$ being a witness for weak vibronic couplings, which also translate into low values of $C$, $n$, but do not affect QTM since phonons are not involved in QTM.

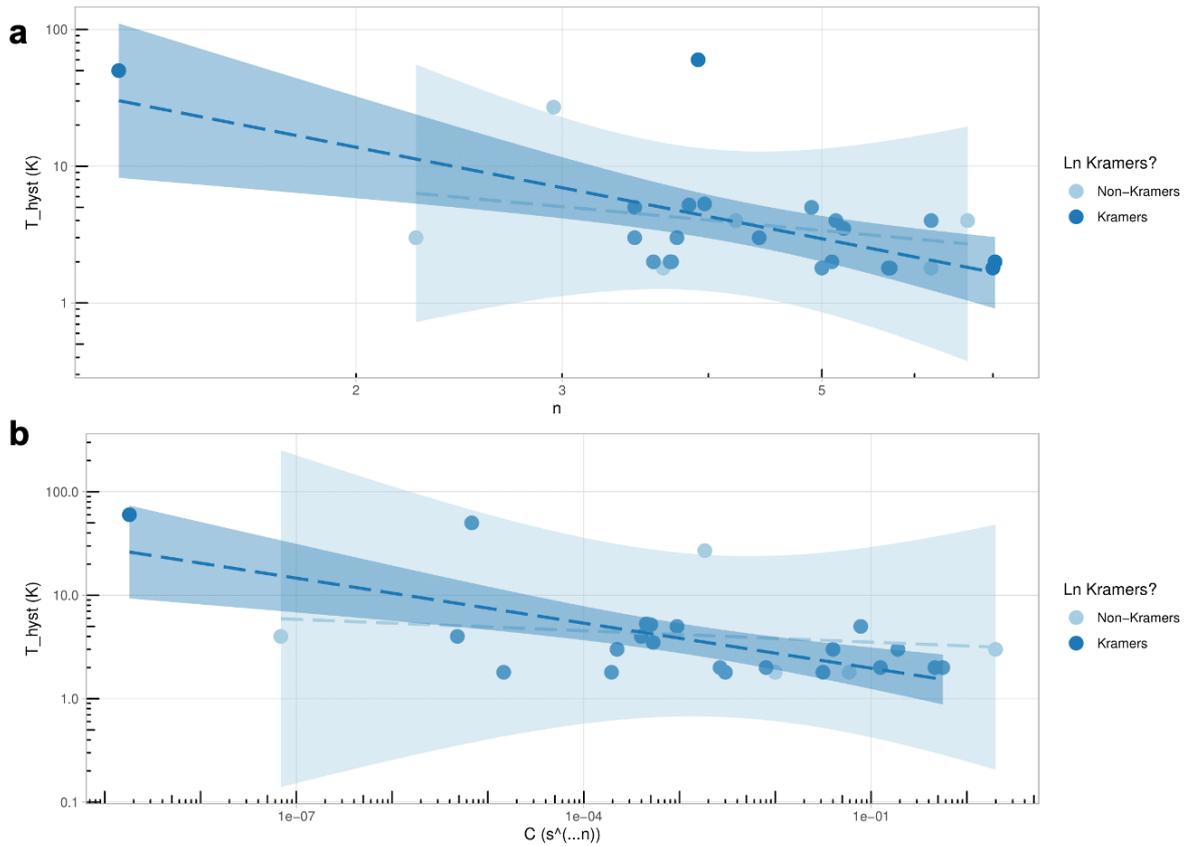

**Supplementary Figure 37 | Scatter plots showing the dependence between n and $T_{hyst}$ (a), C and $T_{hyst}$ (b).** The correlation is less clear than in the case of C. Note that not all samples will be present in all graphs (see Supplementary Figure 1.1). As a consequence, an (x vs y) plot can only include samples for which x and y are simultaneously present in the dataset.

As the last plots among physical parameters, let us discuss the relation between $T_{hyst}$ and Raman parameters in Supplementary Figure 37. The lowering tendency of $T_{hyst}$ with both *n* and *C* is expected, qualitatively meaning Raman relaxation mechanism needs to be weak for high hysteresis temperatures. This also supports the idea above that $U_{eff}$ is tied to Raman and both of them are linked to magnetic behaviour.

## 9.2 Influence of CN and the number of ligands to optimise Raman and $U_{eff}$

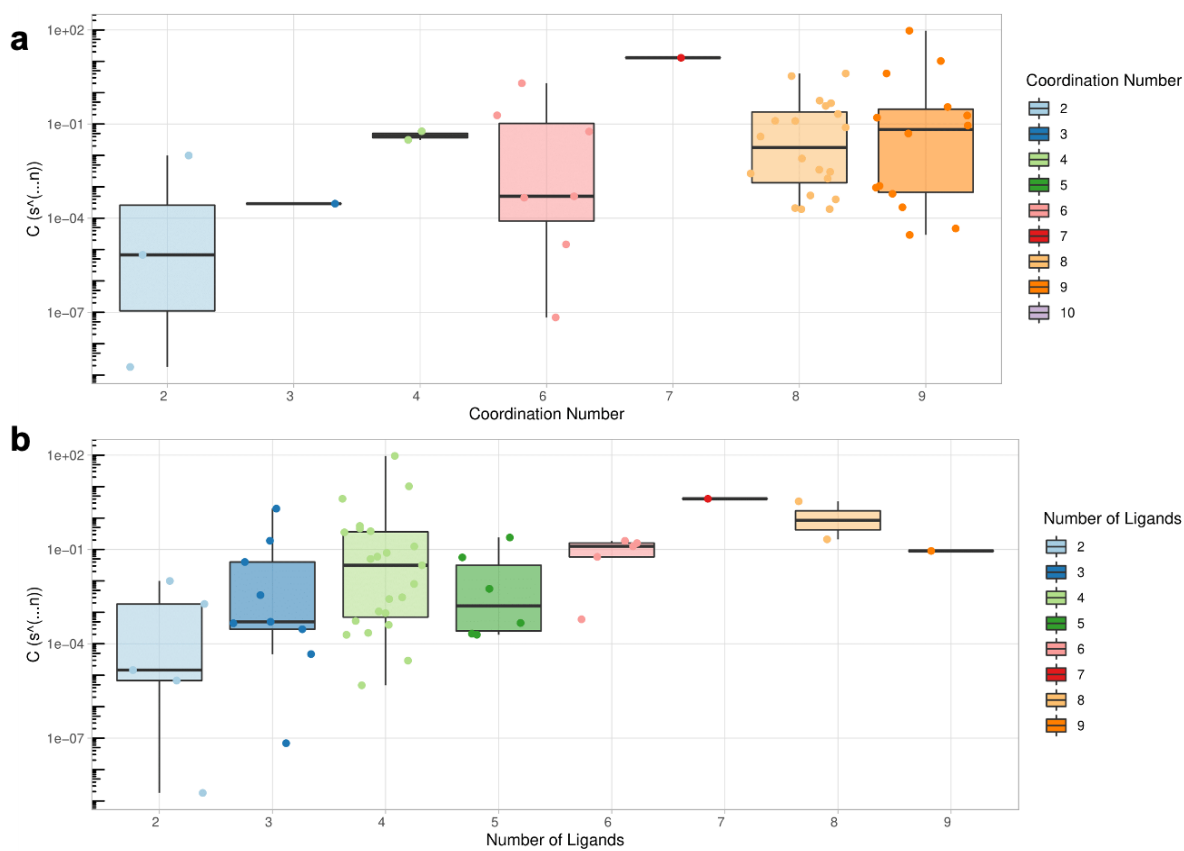

**Supplementary Figure 38 | Boxplots showing the values of C for different coordination numbers (top) and numbers of ligands (bottom).** Although the data is limited, in both cases the higher values of $C_{Ra}$ tend to correspond to higher values of CN or numbers of ligands in the coordination sphere.

Since the Raman mechanism is governed by vibronic coupling, let us represent the evolution of C vs the CN and the number of ligands (see Supplementary Figure 38).

Even within very scarce data, there is an apparent tendency towards higher values of C vs the number of ligands and the coordination number. In the same sense but with more abundant data, there is also a tendency towards lower values of $U_{eff}$, with the coordination number, where CN=7 and number of ligands=7 constituting an anomaly to the general trend (Supplementary Figure 39).

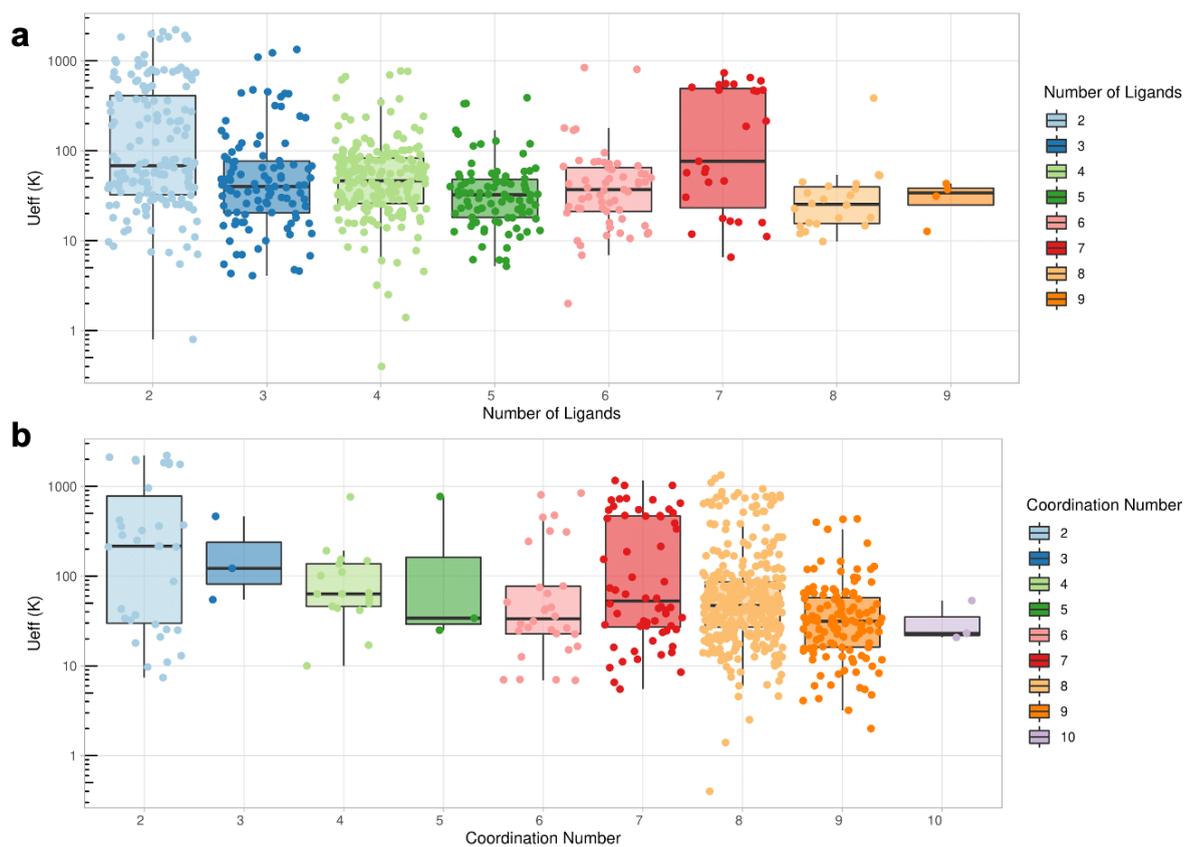

**Supplementary Figure 39 | Boxplots showing the values of $U_{\text{eff}}$ for different coordination numbers (top) and numbers of ligands (bottom).** The data are more abundant, and the trend is consistent with the previous case, but less marked: in both cases the higher values of $U_{\text{eff}}$ tend to correspond to lower values of CN or numbers of ligands in the coordination sphere, with 7 (either as CN or as number of ligands) seemingly marking an exception..

**Supplementary Section 10. Evidence for/against the main tested hypotheses**

**Supplementary Table 7 | Main hypotheses in the present studies and location of the evidence for or against each of them.**

| hypothesis | answer | evidence |
|---|---|---|
| LnPc$_2$ distinctly promising as SIMs? | yes | ·Fig. 3 a,b<br>·SI Figs 11.1, 11.6, 12.1, 12.2<br>·SI Sect. 6.1 (magnetostructural clustering analysis) |
| metallocenes distinctly promising as SIMs? | yes | |
| any other family among {Schiff bases, polyoxometalates, diketonates, radicals, TM near Ln} promising as SIMs? | no | |
| oblate (Dy,Tb…) > prolate (Er…) ? (ac, $U_{eff}$) | yes | ·Fig 3c<br>·SI Fig 11.3 |
| oblate (Dy,Tb…) > prolate (Er…) ? (hysteresis) | no | ·Fig 3d<br>·SI Fig 11.5 |
| $U_{eff}$ an excellent predictor? | yes | ·Fig 5<br>·SI Sect. 5 |
| $U_{eff}$ correlated with $\tau_0$? | yes | ·SI Figs. 24.1, 24.2<br>·SI Tabl. 5 |
| $U_{eff}$ independent of Raman? | no ! | ·Fig 5<br>·SI Sect 9 |
| any promising underexplored coordination polyhedron? | yes ! (compressed) pentagonal bipyramid | ·Fig 6b,6c,6d<br>·SI Sect 7<br>·SI Sect 8 |

## Supplementary References